\documentclass[leqno,12pt]{article}
\usepackage{latexsym}
\usepackage{amssymb}

\usepackage{graphicx}
\usepackage{graphicx}
\usepackage{epsfig}
\usepackage{rotating}
 \usepackage{enumitem}

\usepackage[authoryear]{natbib}
 0

\def \R{I \!\! R}

\newcommand{\su}{\sum_{i=1}^n}

\newcommand{\Pkonv}{\stackrel{P}{\longrightarrow}}

\newcommand{\Kern}[4]{K_{#1}\left(\frac{#2-#3}{#4}\right)}

\newcommand{\E}{\mbox{E}}

\newcommand{\abs}[1]{\left|#1\right|}
\newcommand{\Pro}[1]{\mbox{P}\Bigl(#1\Bigr)}

\newtheorem{theorem}{Theorem}[section]
\newtheorem{lemma}[theorem]{Lemma}

\newtheorem{remark}[theorem]{Remark}

\newcommand{\pn}{\stackrel{P}{\longrightarrow}}

\renewcommand{\theequation}{\thesection.\arabic{equation}}
\makeatletter\@addtoreset{equation}{section}\makeatother

\newcommand{\ba}{\begin{array}}
\newcommand{\ea}{\end{array}}
\newcommand{\beqohne}{\begin{eqnarray*}}
\newcommand{\eeqohne}{\end{eqnarray*}}
\newcommand{\beohne}{\begin{equation*}}
\newcommand{\eeohne}{\end{equation*}}

\renewcommand{\epsilon}{\varepsilon}
\newcommand{\eps}{\varepsilon}

\def\3{\ss}

\def\en{\mathbb{N}}

\def \nti {n\rightarrow\infty}
\def \knti {\stackrel{n\rightarrow\infty}{\longrightarrow}}

\newcommand{\bea}{\begin{eqnarray*}}
\newcommand{\eea}{\end{eqnarray*}}
\newcommand{\beq}{\begin{equation}}
\newcommand{\eeq}{\end{equation}}

\newcommand{\Ind}[1]{I_{\left\{ #1 \right\}}}

\newcommand{\alle}{\forall \,}

\begin{document}

\author{ Stanislav Volgushev$^{\normalsize\rm a,b}$\thanks{Supported by the Sonderforschungsbereich ``Statistical modeling of nonlinear dynamic processes" (SFB~823) of the Deutsche Forschungsgemeinschaft.}, Holger Dette$^{\normalsize\rm a*}$,  
\\
$^{\normalsize\rm a}$ Ruhr-Universit\" at Bochum \vspace{1mm}
\\
$^{\normalsize\rm b}$ University of Illinois at Urbana-Champaign. 
$\;$\vspace{-7mm}\\
}
\date{}

\title{Nonparametric quantile regression for twice censored data}

\maketitle

\begin{abstract}
We consider the problem of nonparametric quantile regression for twice censored data.
Two new estimates are presented, which are constructed by applying  concepts of
monotone rearrangements to estimates of the  conditional distribution function. The
proposed methods  avoid the problem of crossing quantile curves.
Weak uniform consistency and weak convergence is established for both estimates
and their finite sample properties are investigated by means of a simulation study.
As a by-product, we obtain  a new result regarding
 the weak convergence of the Beran estimator for right censored data
 on the maximal possible domain, which is of its own interest.

\end{abstract}

AMS Subject Classification: 62G08, 62N02, 62E20

Keywords and Phrases:  quantile regression, crossing quantile curves, censored data, monotone rearrangements, survival analysis, Beran estimator

\section{Introduction}
Quantile regression offers great flexibility in assessing covariate effects on event times.
 The method was introduced by  \cite{koebas1978}  as a
supplement to least squares methods focussing on the estimation of the conditional mean function and since this seminal  work it has found
numerous applications in different fields [see \cite{koenker2005}]. Recently \cite{koegel2001}   have proposed quantile regression techniques
as an alternative to the classical Cox model for analyzing survival times. These  authors argued that  quantile regression methods offer an
interesting alternative, in particular if there is heteroscedasticity in the data or inhomogeneity in the  population, which is a common
phenomenon in survival analysis [see \cite{portnoy2003}]. Unfortunately the ``classical'' quantile regression techniques cannot be directly
extended to survival analysis, because  for the estimation of a quantile one has to estimate the censoring distribution for each observation.
As a consequence rather stringent assumptions are required in censored regression settings. Early work by \cite{powell1984,powell1986},
requires that the censoring times are always observed. Moreover, even under this rather restrictive and -- in many cases -- not realistic
assumption the objective function is not convex, which results in some computational problems [see for example \cite{fitzenberger1997}]. Even worse, recent research indicates that using the information contained in the observed censored data actually reduces the estimation accuracy [see \cite{koenker2008}]. \\
Because in most survival settings the information regarding the censoring times is incomplete several authors have tried to address this
problem by making restrictive assumptions on the censoring mechanism. For example, \cite{yinjunwei1995}  assumed that the responses and censoring times are independent, which is stronger than the usual assumption of conditional independence. \cite{yang1999}  proposed a method for median regression
under the assumption of i.i.d. errors, which is computationally difficult to evaluate and cannot be directly generalized to the
heteroscedastic case. Recently, \cite{portnoy2003}  suggested a recursively re-weighted quantile regression estimate under the assumption that
the censoring times and responses are independent conditionally on the predictor. This estimate adopts the principle of self consistency for
the Kaplan-Meier statistic [see \cite{efron1967}] and can be considered as a direct generalization of this classical estimate in survival analysis.
\cite{penghuan2008} pointed out that the large sample properties of this recursively defined
estimate  are still not completely understood and proposed an alternative approach, which is based on martingale
estimating equations. In particular, they proved consistency and asymptotic normality of their estimate.

While all of the cited literature considers the classical linear quantile regression model with right censoring, less results are available for quantile regression in a nonparametric context. Some results on nonparametric quantile regression when no censoring is present can be found in \cite{chaudhuri1991} and \cite{yujon1997,yujon1998}. \cite{chefergal2006} and \cite{detvol2008} pointed out that many of the commonly proposed parametric or nonparametric estimates lead to possibly crossing quantile curves and modified some of these estimates to avoid this problem.
Results regarding the estimation of the conditional distribution function from right censored data
can be found in \cite{dabrowska1987, dabrowska1989} or \cite{lidoss1995}. The estimation of conditional quantile functions in the same setting is briefly stressed in \cite{dabrowska1987} and further elaborated in \cite{dabrowska1992a}, while \cite{ghouchkeilegom2008} proposed a quantile regression procedure for right censored and dependent data.
On the other hand, the problem of nonparametric quantile regression for censored data where the observations can be censored from either left or right
does not seem to have been considered in the literature. 

This gap can partially be explained by the difficulties arising in the estimation of the conditional distribution function with two-sided censored data. The problem of estimating the (unconditional) distribution function for data that may be censored from above and below has been considered by several authors. For an early reference see \cite{turnbull1974}. More recent references are \cite{changyang1987, chang1990, guzhang1993} and \cite{patiroli2006}. On the other hand- to their best knowledge- the authors are not
aware of literature on nonparametric conditional quantile regression, or estimation of a conditional distribution function, for left and right censored data when the censoring is not always observed and only the conditional independence of censoring and lifetime variables is assumed.

In the present paper we consider the problem of nonparametric quantile regression for twice censored data. We consider a censoring mechanism
introduced by \cite{patiroli2006} and propose an estimate of the conditional distribution function in several steps. On the basis of this
estimate and the preliminary statistics which are used for its definition, we construct two quantile regression estimates using the concept of
simultaneous inversion and isotonization [see \cite{detneupil2005}] and monotone rearrangements [see \cite{detneupil2006},
\cite{chefergal2006} or \cite{anevfoug2007} among others]. In Section \ref{prel} we introduce the model and the two estimates, while Section \ref{main} contains our main results.
In particular, we prove uniform consistency and weak convergence of the estimates of the conditional distribution function and its quantile
function. As a by-product we obtain
a new result on the weak convergence of the Beran estimator on the maximal possible interval,
which is of independent interest. In Section 4 we illustrate the finite sample properties of the proposed estimates by means of a simulation study. Finally, all proofs
and technical details are deferred to an Appendix.

\section{Model and estimates} \label{prel}

We consider  independent identically distributed random vectors $(T_i,L_i,R_i,X_i)$, $i=1,\dots,n$, where $T_i$ are the variables of
interest, $L_i$ and $R_i$ are left and right censoring variables, respectively, and the $\R^d$-valued random variables $X_i$ denote the covariates. We assume that the distributions of the
random variables $L_i,R_i$ and $T_i$ depend on $X_i$ and denote by $F_L(t|x):=P(L\leq t|X=x)$
the conditional distribution function of $L$ given $X=x$. The conditional distribution functions  $F_R(.|x)$ and $F_T(.|x)$ are defined analogously.

Additionally, we assume that the random variables $T_i, L_i, R_i$ are almost surely nonnegative and independent conditionally on  the covariate $X_i$. Our aim is to estimate the
 conditional quantile function
$F^{-1}_T(.|x)$. However, due to the censoring, we can only observe the triples $(Y_i,X_i,\delta_i)$ where $Y_i = \max(\min(T_i,R_i),L_i)$ and the
indicator variables $\delta_i$ are defined by
\begin{eqnarray}\label{cens}
\delta_i := \left\{
\begin{array}{ccl}
0 &,& L_i<T_i\leq R_i\\
1 &,& L_i<R_i<T_i\\
2 &,&   T_i\leq L_i < R_i\ \mbox{or} \ R_i\leq L_i.
\end{array}
\right.
\end{eqnarray}

\begin{remark}{\rm 
An unconditional version of this censoring mechanism was introduced by \cite{patiroli2006}. Examples of situations where this kinds of data occur can for example be found in chapter $15$ of \cite{meekesco1998}. This model also is closely related to the double censoring model, see \cite{turnbull1974}
for the case without covariates. In that setting, the assumption of independence between the random variables
$L,R,T$ is replaced by the assumption that $T$ is independent of the pair $(R,L)$ and additionally $P(L<R) = 1$. Note that none of the two assumptions is strictly more or less restrictive then the other. Rather the two models describe different situations. Moreover, since $L,T,R$ are never observed simultaneously, it is not possible to decide which of the models is most approriate. Instead, an understanding of the underlying data generation process is crucial to identify  the right model. A more detailed comparison of the two models can be found  in \cite{patiroli2001} and \cite{patiroli2006} for the case without covariates.}
\end{remark}

Roughly speaking, the construction of an estimate for the conditional quantile function of $T$ can be accomplished in three steps. First, we define the variables $S_i := \min(T_i,R_i)$ and consider the model $Y_i = \max(S_i,L_i)$, which is a classical right censoring model. In this model we estimate the conditional distribution $F_L(.|x)$ of $L$. In a second step, we use this information to reconstruct the conditional distribution of $T$ [see Section 2.1]. Finally, the concept of simultaneous isotonization and inversion [see \cite{detneupil2005}] and the monotone rearrangements, which was recently introduced by \cite{detneupil2006} in the context of monotone estimation of a regression function, are used to obtain two estimates of the conditional quantile function [see Section 2.2].

\subsection{Estimation of the conditional distribution function} \label{subsecestcdf}

To be more precise, let $H$ denote  the conditional distribution of $Y$. We introduce the notation $H_k(A|x)=\Pro{A\cap\{\delta=k\}|X=x}$ and
obtain the decomposition $H = H_0+H_1+H_2$ for the conditional distribution of $Y_i$. The sub-distribution functions $H_k$ $(k=0,1,2)$ can be represented as follows
\begin{eqnarray} \label{E0.1}
H_0(dt|x) &=& F_L(t-|x)(1 - F_R(t-|x))F_T(dt|x) \\
H_1(dt|x) &=& F_L(t-|x)(1-F_T(t|x))F_R(dt|x)  \label{E0.2} \\
H_2(dt|x) &=& \left\{1-(1-F_T(t|x))(1-F_R(t|x))\right\}F_L(dt|x) = F_S(t|x)F_L(dt|x). \label{E0.3}
\end{eqnarray}
Note that the conditional (sub-)distribution functions $H_k$ and $H$ can easily be estimated from the observed data by
\begin{eqnarray}
H_{k,n}(t|x) := \sum_{i=1}^n W_i(x)\Ind{Y_i\leq t, \delta_i=k}, \label{DS1} \quad H_n(t|x) := \sum_{i=1}^n W_i(x)\Ind{Y_i\leq t},
\end{eqnarray}
where the quantities $W_i(x)$ denote local weights depending on the covariates $X_1,...,X_n$,  which will be specified below. We will use the
representations (\ref{E0.1}) - (\ref{E0.3}) to obtain an expression for $F_T$ in terms of  the functions $H, H_k$ and then replace the
distribution functions
 $H, H_k$ by their empirical counterparts $H_n, H_{k,n}$, respectively.
We   begin with the reconstruction of $F_{L}$. First note that
\beq
M_2^-(dt|x) := \frac{H_2(dt|x)}{H(t|x)} = \frac{F_S(t|x)F_L(dt|x)}{F_L(t|x)F_S(t|x)} = \frac{F_L(dt|x)}{F_L(t|x)} \label{E.1}
\eeq
is the predictable reverse hazard measure corresponding to $F_L$  and hence we can reconstruct $F_L$  using the product-limit representation \beq F_L(t|x)
= \prod_{(t,\infty]}(1-M_2^-(ds|x))   \label{E.2} \eeq

[see e.g.\ \cite{patiroli2006}]. Now having a representation for the conditional distribution function $F_L$ we can define in a second step
\begin{eqnarray}
\Lambda_T^-(dt|x) &:=& \frac{H_0(dt|x)}{F_L(t-|x)-H(t-|x)} =  \frac{H_0(dt|x)}{F_L(t-|x)(1-F_S(t-|x))}\label{E.3}
\\
&=&  = \frac{H_0(dt|x)}{F_L(t-|x)(1-F_R(t-|x))(1-F_T(t-|x))} \nonumber
\\
&=& \frac{F_L(t-|x)(1-F_R(t-|x))F_T(dt|x)}{F_L(t-|x)(1-F_R(t-|x))(1-F_T(t-|x))} = \frac{F_T(dt|x)}{1-F_T(t-|x)},
\nonumber
\end{eqnarray}
which yields an expression for the predictable hazard measure of $F_T$. Finally, $F_T$ can be reconstructed by using the product-limit representation \beq 1 -
F_T(t|x) = \prod_{[0,t]}(1-\Lambda_T^-(ds|x))   \label{E.4} \eeq [see e.g.\ \cite{gilljoha1990}]. Note that formula (\ref{E.4}) yields an explicit representation of the conditional distribution function $F_T(.|x)$ in terms of the quantities $H_0, H_1, H_2, H$, which can  be estimated from the data [see equation (\ref{DS1})].
The estimate of the conditional distribution function is now defined as follows. First, we use the representation (\ref{E.2}) to obtain an estimate of $F_L(.|x)$, that is
\beq
\label{es1} F_{L,n} (t | x)= \prod_{(t, \infty]} (1 - M^-_{2,n} (ds|x)),
\eeq
where
\beq
\label{es2} M^-_{2,n} (ds|x)= \frac {H_{2,n}(ds|x)}{H_n(s|x)}.
\eeq
Second, after observing (\ref{E.3}) and (\ref{E.4}), we define
\beq
\label{es3} F_{T,n} (t| x)= 1 - \prod_{[0,t]} (1 - \Lambda^-_{T,n} (ds|x)),
\eeq
where
\beq
\label{es4} \Lambda^-_{T,n} (ds|x)= \frac {H_{0,n}(ds|x)}{F_{L,n}(s-|x) - H_n(s-|x)}.
\eeq \\
In Section 3 we    will analyse the
asymptotic  properties of these estimates, while in the following Section 2.2 these estimates are used to construct nonparametric and noncrossing quantile curve estimates.

\begin{remark}
{\rm Throughout this paper, we will adopt the convention $'0/0=0'$. This means that if, for example, $H_{0,n}(dt|x)=0$ and
$F_{L,n}(t-|x)-H_n(t-|x)=0$, the contribution of $$\frac{H_{0,n}(dt|x)}{F_{L,n}(t-|x)-H_n(t-|x)}$$ in (\ref{es4}) will be interpreted as
zero.}
\end{remark}

\subsection{Non-crossing quantile estimates by monotone rearrangements}
\label{S:concondquant}
In practice, nonparametric estimators of a conditional distribution function $F(.|x)$ are not necessarily increasing for finite sample sizes [see e.g. Yu, Jones (1998)]. Although this problem often vanishes asymptotically, it still is of great practical relevance, because in a concrete application it is not completely obvious how to invert a non-increasing function. Trying to naively invert such estimators may lead to the well-known problem of quantile crossing [see  \cite{koenker2005} or \cite{yujon1998}] which poses some difficulties in the interpretation of the results. In this paper we will discuss the following two possibilities to deal with this problem

\begin{enumerate}
\item Use a procedure developed by \cite{detvol2008} which is based on a simultaneous isotononization  and inversion of a nonincreasing distribution function. As a by-product this method yields non-crossing  quantile  estimates. To be precise, we consider the operator
\begin{eqnarray} \label{op1}
\Psi: \left\{
\begin{array}{l}
L^\infty(J) \rightarrow L^\infty(\R)
\\
f \mapsto \left( y \mapsto \int_J \Ind{f(u) \leq y} du \right)
\end{array}
\right.
\end{eqnarray}
where $L^\infty(I)$ denotes the set of bounded, measurable functions on  the set $I$ and $J$ denotes a bounded interval.
Note that for a strictly increasing function $f$ this operator yields the right continuous inverse of $f$, that is
$\Psi(f)=f^{-1}$ [here and in what follows, $f^{-1}$ will denote the generalized inverse, i.e. $f^{-1}(t) := \sup\{s:f(s)\leq t \}$]. On the other hand, $\Psi(f)$ is always isotone, even in the case where $f$ does not have this property.
Consequently, if $\hat f$ is a not necessarily isotone estimate of an isotone function $f$, the function $\Psi(\hat f)$ could be
regarded as an isotone estimate of the function $f^{-1}$.
Therefore, the first idea to construct an estimate of the conditional quantile function consists in the application of the operator $\Psi$ to the estimate $F_{T,n}$ defined in (\ref{es3}), i.e.
\beq
\hat q (\tau|x)= \Psi(F_{T,n} (.|x))(\tau).
\eeq
However, note that formally the mapping $\Psi$ operates on functions defined on bounded intervals. More care
is necessary if the operator has to be applied to a function with an unbounded support. A detailed discussion and a solution of this problem can be found in \cite{detvol2008}. In the present paper we use different approach which is a slightly modified version of the ideas from \cite{anevfoug2007}.
To be precise note that estimators of the conditional distribution function $F(.|x)$ [in particular those  of the form (\ref{DS1}), which will be used later] often are constant outside of the compact interval
$J := [j_1,j_2] =[\min_i Y_i, \max_i Y_i]$. Now the structure of the estimator $F_{T,n} (.|x)$ implies that $F_{T,n} (.|x)$ will also be constant outside of $J$.
We thus propose to consider the modified operator $\tilde \Psi_J$ defined as
\begin{eqnarray} 
\tilde \Psi_J: \left\{
\begin{array}{l}
L^\infty(\R) \rightarrow L^\infty(\R)
\\
f \mapsto \left( y \mapsto
j_1 + \int_J \Ind{f(u) \leq y} du
\right).
\end{array}
\right.
\end{eqnarray}
Consequently the first estimator of the conditional quantile function is given by
\beq
\label{qes1} \hat q (\tau|x)= \tilde \Psi_{J}(F_{T,n} (.|x))(\tau).
\eeq

\item Use the concept of increasing rearrangements [see \cite{detneupil2006} and  \cite{chefergal2006} for details] to
construct an increasing estimate of the conditional distribution function, which is then inverted in a second step.
More precisely, we define the operator
\begin{eqnarray} \label{op2}
\Phi: \left\{
\begin{array}{l}
L^\infty(J) \rightarrow L^\infty(\R)
\\
f \mapsto \left( y \mapsto (\Psi f(.))^{-1}(y) \right)
\end{array}
\right.
\end{eqnarray}
where $\Psi$ is introduced in (\ref{op1}). Note that for a strictly increasing right continuous function $f$ this operator reproduces $f$, i.e.\ $\Phi(f)=f$. On the other hand, if $f$ is not isotone, $\Phi(f)$ is an isotone function and
the operator preserves the $\mbox{L}^p$-norm, i.e.
$$
\int_J | \Phi(f(u))|^p \ du = \int_J | f(u)|^p \ du.
$$
Moreover, the operator also defines a contraction, i.e.
$$
\int_J | \Phi (f_1) (u) - \Phi (f_2)(u) |^p \ du  \leq \int_J |f_1 - f_2|^2 \ du \quad \alle p\geq 1
$$
[see \cite{hardylittlewoodpolya} or \cite{lorentz1953}].
This means if $\hat f (=f_1)$ is a not necessarily isotone estimate of the isotone function $f(=f_2)$, then the isotonized estimate $\Phi(\hat f)$
is a better approximation of the isotone function $f$ than the original estimate $\hat f$ with respect to any $\mbox{L}^p$-norm [note that $\Phi(f)=f$ because $f$ is assumed to be
isotone].  For a general discussion of monotone rearrangements and the operators (\ref{op1}) and (\ref{op2}) we refer to \cite{bennshar1988}, while some statistical applications can be found in \cite{detneupil2006} and \cite{chefergal2006}.\\
The idea is now to use rearranged estimators of $H_{i}(.|x)$ and $H(.|x)$ in the representations (\ref{E.1})-(\ref{E.4}). For this purpose we need to modify the operator $\Phi$ so that it can be applied to functions of unbounded support. We propose to proceed as follows
\begin{itemize}
\item Define the operator $\tilde \Phi_J$ indexed by the compact interval $J=[j_1,j_2]$ as
\begin{eqnarray} 
\quad \quad \quad \tilde \Phi_J: \left\{
\begin{array}{l}
L^\infty(\R) \rightarrow L^\infty(\R)
\\
f \mapsto \left( y \mapsto \Ind{y<j_1}f(j_1-) + (\tilde \Psi_J f(.))^{-1}(y)\Ind{j_1\leq y\leq j_2} + \Ind{y>j_2}f(j_2) \right)
\end{array}
\right.
\end{eqnarray}
\item Truncate the estimator $H_n(\cdot|x)$ for values outside of the interval $[0,1]$, i.e.
\[
\tilde H_n(t|x) := H_n(t|x)\Ind{H_n(t|x)\in[0,1]} + \Ind{H_n(t|x)>1}
\]
[note that in general estimators of the form (\ref{DS1}) do not necessarily have values in the interval $[0,1]$ since the weights $W_i(x)$ might be negative]
\item Use the statistic $H_n^{IP}(t|x) := \tilde \Phi_{J_Y}(\tilde H_n(\cdot|x))(t)$ as estimator for $H(t|x)$.
\item Observe that the estimator $H_n^{IP}(t|x)$ is by construction an increasing step function which can only jump in the points $t=Y_i$, i.e. it admits the representation
\beq \label{ra1}
H_n^{IP}(t|x) = \sum_i W_i^{IP}(x)\Ind{Y_i\leq t}
\eeq
with weights $W_i^{IP}(x) \geq 0$.
Based on this statistic, we define estimators $H^{IP}_{k,n}$ of the subdistribution functions $H_{k}$ as follows
\beq \label{ra2}
H^{IP}_{k,n}(t|x) = \sum_i W_i^{IP}(x) \Ind{Y_i\leq t} \Ind{\delta_i=k}, \quad k=0,1,2
\eeq
In particular, such a definition ensures that $H^{IP}(t|x) = H^{IP}_{0,n}(t|x)+H^{IP}_{1,n}(t|x)+H^{IP}_{2,n}(t|x)$.
\end{itemize}
So far we have obtained increasing estimators of the quantities $H$ and $H_i$. The next step in our construction is to plug these estimates in representation (\ref{E.1}) to obtain:
\begin{equation} \label{IP1}
\tilde M^-_{2,n} (dt|x) = \frac {H^{IP}_{2,n}(dt|x)}{H_n^{IP}(t|x)},
\end{equation}
which defines an increasing function with jumps of size less or equal to one. This implies that $\tilde F_{L,n}(t|x)= \prod_{(t, \infty]} (1- \tilde M^-_{2,n}(ds|x))$ is also increasing.
For the rest of the construction, observe the following Lemma which will be proved at the end of this section.
\begin{lemma} \label{le:LambdaT}
Assume that $Y_i \neq Y_j$ for $i \neq j$. Then the function
\begin{equation}\label{IP2}
\tilde\Lambda^-_{T,n} (dt|x):= \frac {H^{IP}_{0,n}(dt|x)}{\tilde F_{L,n}(t-|x) - H_n^{IP}(t-|x)}
\end{equation}
is nonnegative, increasing and has jumps of size less or equal to one.
\end{lemma}
This in turn yields the estimate
\begin{equation} \label{IP3}
F^{IP}_{T,n} (t|x)= 1 - \prod_{[0,t]} (1 - \tilde \Lambda^-_{T,n}(ds|x)).
\end{equation}
In the final step we now simply invert the resulting estimate of the conditional distribution function $F^{IP}_{T,n}$ since it is increasing by construction. We denote this estimator of the conditional quantile function by
\begin{equation} \label{IP4}
\hat q^{IP}(t|x) := \sup \left\{s: F^{IP}_{T,n}(s|x) \leq t \right\}.
\end{equation}
\end{enumerate}
In the next section, we will discuss asymptotic properties of the two proposed estimates $\hat q$ and $\hat q^{IP}$ of the conditional quantile curve.
\begin{remark} \label{undef}
\rm
In the classical right censoring case, there is no uniformly good
 way to define the Kaplan-Meier estimator beyond the largest uncensored
 observation [see e.g. \cite{flemharr1991}, page 105]. Typical approaches
 include setting it to unity, to the value at the largest uncensored observation,
 or to consider it unobservable within certain bounds [for more details,
 see the discussion in \cite{flemharr1991}, page 105 and \cite{andborgilkei1993}, page 260]. When censoring is light, the first of the above mentioned approaches seems to yield the best results [see \cite{andborgilkei1993}, page 260]. \\
When the data can be censored from either left or right, the situation becomes even more complicated since now we also have to find a reasonable definition below the smallest uncensored observation.
From definitions (\ref{E.1})-(\ref{E.4}) it is easy to see that $F_{T,n}$ equals zero below the smallest uncensored observation with non-vanishing weight and is constant at the largest uncensored observation and above.
In practice, the latter implies that the estimators $\hat q (\tau|x)$  and $\hat q^{IP}(\tau|x)$ are
 not defined as soon as $\sup_t F_{T,n}(t|x) < \tau$ or  $\sup_t F_{T,n}^{IP}(t|x) < \tau$, respectively.
 A simple ad-hoc solution to this problem
is to define  the estimator $F_{T,n}$ or $F_{T,n}^{IP}$  as $1$  beyond the last
 observation with non-vanishing weight or to locally increase the bandwidth.
 A detailed investigation of this problem is postponed
 to future research.
\end{remark}

We conclude this section with the proof of Lemma \ref{le:LambdaT}.\\
\textbf{Proof of Lemma \ref{le:LambdaT}} In order to see that $\tilde\Lambda^-_{T,n} (dt|x)$ is increasing, we note that
\bea
H_n^{IP}(t-|x) &=& \prod_{[t,\infty)}
\Bigl(1-\frac{H_n^{IP}(ds|x)}{H_n^{IP}(s|x)} \Bigr)
= \prod_{[t,\infty)}\Bigl( 1-\frac{H_{2,n}^{IP}(ds|x)}{H_n^{IP}(s|x)} - \frac{H_{0,n}^{IP}(ds|x) + H_{1,n}^{IP}(ds|x)}{H_n^{IP}(s|x)}\Bigr)
\\
&\leq& \prod_{[t,\infty)}\Bigl( 1-\frac{H_{2,n}^{IP}(ds|x)}{H_n^{IP}(s|x)}\Bigr) = \tilde F_{L,n}(t-|x).
\\
\eea
Thus $\tilde F_{L,n}(t-|x) - H_n^{IP}(t-|x) \geq 0$ and the nonnegativity of $\tilde\Lambda^-_{T,n} (dt|x)$ is established. In order to prove the inequality $\tilde\Lambda^-_{T,n} (dt|x) \leq 1$ we assume without loss of generality that $Y_1<Y_2<\cdots<Y_n$. Observe that as soon as $\delta_k=0$ we have for $k\geq 2$
\bea
&&\tilde F_{L,n}(Y_k-|x) - H_n^{IP}(Y_k-|x)
\\
&=& \Bigl[ 1- \prod_{[Y_k,\infty)}\Bigl( 1- \frac{H_{0,n}^{IP}(ds|x) + H_{1,n}^{IP}(ds|x)}{H_n^{IP}(s|x)}\Bigr)\Bigr]
\prod_{[Y_k,\infty)}\Bigl( 1-\frac{H_{2,n}^{IP}(ds|x)}{H_n^{IP}(s|x)}\Bigr)
\\
&\stackrel{(*)}{=}& \Bigl[ 1- \prod_{j\geq k, \delta_j\neq 2}\Bigl( 1- \frac{\Delta H_{0,n}^{IP}(Y_j|x) + \Delta H_{1,n}^{IP}(Y_j|x)}{H_n^{IP}(Y_j|x)}\Bigr)\Bigr]
\prod_{j\geq k+1, \delta_j=2}\Bigl( 1-\frac{\Delta H_{2,n}^{IP}(Y_j|x)}{H_n^{IP}(Y_j|x)}\Bigr)
\\
&=& \Bigl[ 1- \prod_{j\geq k, \delta_j\neq 2}\Bigl( \frac{H_{n}^{IP}(Y_{j-1}|x)}{H_n^{IP}(Y_j|x)} \Bigr)\Bigr]
\prod_{j\geq k+1, \delta_j=2}\Bigl( \frac{H_{n}^{IP}(Y_{j-1}|x)}{H_n^{IP}(Y_j|x)}\Bigr)
\\
&\stackrel{(**)}{=}& \Bigl[ 1- \frac{H_n^{IP}(Y_{k-1}|x)}{H_n^{IP}(Y_k|x)}\prod_{j\geq k+1, \delta_j\neq 2}\Bigl( \frac{H_{n}^{IP}(Y_{j-1}|x)}{H_n^{IP}(Y_j|x)} \Bigr)\Bigr]
\prod_{j\geq k+1, \delta_j=2}\Bigl( \frac{H_{n}^{IP}(Y_{j-1}|x)}{H_n^{IP}(Y_j|x)}\Bigr)
\\
&\geq& \Bigl[ 1 - \frac{H_n^{IP}(Y_{k-1}|x)}{H_n^{IP}(Y_k|x)}\Bigr]
\prod_{j\geq k+1}\Bigl( \frac{H_{n}^{IP}(Y_{j-1}|x)}{H_n^{IP}(Y_j|x)}\Bigr)
\\
&=& \Bigl[ \frac{H_n^{IP}(Y_k|x)-H_n^{IP}(Y_{k-1}|x)}{H_n^{IP}(Y_k|x)}\Bigr] \frac{H_{n}^{IP}(Y_k|x)}{H_n^{IP}(Y_n|x)}
\\
&=& \Delta H_n^{IP}(Y_k|x),
\eea
where the equalities $(*)$ and $(**)$ follow from $\delta_k=0$. An analogous result for $k=1$ follows by simple algebra. Hence we have established that for $\delta_k=0$ we have $\Delta \tilde\Lambda^-_{T,n} (Y_k|x) \leq 1$, and all the other cases need not be considered since we adopted the convention '0/0=0'. Thus the proof is complete. \hfill$\Box$

\section{Main results} \label{main}
The results stated in this section describe the asymptotic properties of the proposed  estimators.
In particular, we investigate weak convergence of the processes $\{H_{k,n}(t|x)\}_t, \{ F_{T,n}(t|x)\}_t$, etc. where  the predictor $x$ is
 fixed. Our main results deal with the weak uniform consistency and the weak convergence of the process $\{F_{T,n}(t|x) - F_T(t|x) \}_t$ and the
corresponding quantile processes obtained in Section \ref{prel}.  In order to derive the process convergence, we will assume that it holds for the initial estimates $H_n,H_{k,n}$ and give sufficient conditions for this property in Lemma \ref{le:g1}. In a next step we apply the delta method [see \cite{gill1989}] to the map $(H,H_2) \mapsto M_2^-$ defined in $(\ref{E.1})$ and the product-limit maps defined in $(\ref{E.2})$ and
$(\ref{E.4})$. Note that the product limit maps are Hadamard differentiable on the set of cadlag functions with total variation bounded by a constant
[see Lemma A.1 on page 42 in \cite{patiroli2001}], and hence the process convergence of $M^-_{2,n}$ and $\Lambda^-_{T,n}$ will directly entail
the weak convergence results for $F_{L,n}$  and   $F_{T,n}$, respectively. However, the Hadamard differentiability of the map $(H_2,H) \mapsto M_2^-$ only holds on domains where $H(t) > \epsilon > 0$, and hence more work is necessary to obtain the corresponding weak convergence results
on the interval $[t_{00},\infty]$ if $H(t_{00}|x)=0$, where
\begin{equation}\label{t0}
t_{00} := \inf\left\{t:H_0(t|x)>0 \right\}.
\end{equation} This situation occurs for example if $F_R(t_{00}|x)=0$, which is quite natural in the context considered in this paper because $R$ is the right censoring
variable.

For the sake of a clear representation and for later reference, we present all required technical conditions for the asymptotic results at the beginning of this section.
We
assume that the estimators of the conditional subdistribution functions are of the form (\ref{DS1})
with weights $W_j(x)$ depending on the covariates $X_1,...,X_n$ but not on $Y_1,...,Y_n$ or $\delta_1,...,\delta_n$. The first set of conditions concerns the weights that are used in the representation (\ref{DS1}). Throughout this paper, denote by $\|\cdot\|$ the maximum norm on $\R^d$.

\begin{enumerate}[label=(W\arabic{*})]
\item \label{W1} With probability tending to one, the weights in (\ref{DS1}) can be written in the form $$W_i(x) = \frac{V_i(x)}{\sum_{j=1}^n V_j(x)},$$
where the real-valued functions $V_j$ $(j=1,\dots,n)$ have the following properties:
\begin{enumerate}[label=(\arabic{*}), ref=\ref{W1}(\arabic{*})]
\item \label{W1a} There exist constants $0<\underline{c}<\overline{c}<\infty$ such that for all $n\in\en$ and all $x$ we  have either $V_j(x) = 0$ or $\underline{c}/nh^d \leq V_j(x) \leq \overline{c}/nh^d$
\item \label{W1b} If $\|x-X_j\|\leq Ch$ for some constant $C<\infty$, then  $V_j(x) \neq 0$  and $V_j(x) = 0$ for $\|x-X_j\|\geq c_n$
for some sequence $(c_n)_{n \in \mathbb{N}}$ such that $c_n=O(h)$. Without loss of generality, we will assume that $C=1$ throughout this paper.
\item \label{W1c} $\sum_i V_i(x) = C(x)(1+o_P(1))$ for some positive function $C$.
\item \label{W1d} $\sup_t \Big\|\sum_i V_i(x)(x-X_i)\Ind{Y_i\leq t}\Big\| = o_P(1/\sqrt{nh^d})$.
\end{enumerate}
Here [and throughout this paper] $h$ denotes a smoothing parameter converging to 0 with increasing sample size.
\item \label{W2} We assume that the weak convergence
\[
\sqrt{nh^d}(H_{0,n}(.|x) - H_0(.|x),H_{2,n}(.|x) - H_2(.|x),H_n(.|x) - H(.|x)) \Rightarrow (G_0,G_2,G)
\]
holds in   $D^3[0,\infty]$, where the limit denotes a centered Gaussian process which has a version with a.s. continuous sample paths and a covariance structure of the
form
\bea
\mbox{Cov}(G_i(s|x),G_i(t|x)) &=& b(x)(H_i(s\wedge t|x) - H_i(s|x)H_i(t|x))\\
\mbox{Cov}(G(s|x),G(t|x)) &=& b(x)(H(s\wedge t|x) - H(s|x)H(t|x))\\
\mbox{Cov}(G_i(s|x),G(t|x)) &=& b(x)(H_i(s\wedge t|x) - H_i(s|x)H(t|x))
\eea
for some function $b(x)$. Here and throughout this paper weak
convergence is understood as convergence with respect to the sigma algebra generated by the closed balls in the supremum norm [see \cite{pollard1984}].
\item \label{W3} The estimators $H_{k,n}(.|x)$ $(k=0,1,2)$ and $H_n(.|x)$ are weakly uniformly consistent on the interval $[0,\infty)$
\end{enumerate}

\begin{remark} \label{nwll}
{\rm It will be shown in Lemma \ref{le:g1} below that, under suitable assumptions on the smoothing parameter $h$, important examples for weights satisfying conditions \ref{W1}-\ref{W3} are given by the Nadaraya-Watson weights
\begin{equation}\label{nw}
W_i^{NW}(x) = \frac{\frac{1}{nh^d}\prod_{k=1}^d K_h((x-X_i)_k)}{\frac{1}{nh^d}\sum_j \prod_{k=1}^d K_h((x-X_i)_k)} =: \frac{V_i^{NW}(x)}{\sum_j V_j^{NW}(x)},
\end{equation}
or (in one dimension) by the local linear weights
\begin{eqnarray}\label{ll}
W_i^{LL}(x) &=& \frac{\frac{1}{nh}K_h(x-X_i)\left(S_{n,2} - (x-X_i)S_{n,1} \right)}{S_{n,2}S_{n,0}-S_{n,1}^2} \\
&=& \frac{\frac{1}{nh}K_h(x-X_i)\left(1
- (x-X_i)S_{n,1}/S_{n,2} \right)}{\frac{1}{nh}\sum_j K_h(x-X_j)\left(1 - (x-X_j)S_{n,1}/S_{n,2} \right)} =: \frac{V_i^{LL}(x)}{\sum_j
V_j^{LL}(x)}, \nonumber
\end{eqnarray}
where $K_h(.) := K(./h)$,  $S_{n,k} := \frac{1}{nh}\sum_j K_h(x-X_j)(x-X_j)^k$ and the  kernel satisfies
the following condition.
\begin{enumerate}
\item[(K1)] The kernel $K$ in (\ref{nw}) and (\ref{ll}) is a symmetric density of bounded total variation with compact support, say $[-1,1]$,
which satisfies $c_1 \leq K(x) \leq c_2$
 for all $x$ with $K(x)\neq 0$ for some constants $0<c_1\leq c_2<\infty$.
\end{enumerate}
}
\end{remark}

For the distributions of the random variables $(T_i,L_i,R_i,X_i)$ we assume that for some $\eps>0$ with $U_\eps(x):=\{y:|y-x|<\eps\}$
\begin{enumerate}[label=(D\arabic{*})]
\item \label{D0000} The conditional distribution function $F_R$ fulfills $F_R(t_{00}|x)<1$
\item \label{D00} For $i = 0,1,2$ we have $\lim_{y\rightarrow x} \sup_t |H_i(t|y)-H_i(t|x)| = 0$
\item \label{D0} The conditional distribution functions $F_L(.|x), F_R(.|x), F_T(.|x)$ have densities,\\
say $f_L(.|x), f_R(.|x), f_T(.|x)$, with respect to the Lebesque measure
\item \label{D1} $\int_{t_{00}}^{\infty} \frac{f_L(u|x)}{F_L^2(u|x)F_S(u|x)} du < \infty$ \label{Th1.1}
\item \label{D2} $\sup_{k=1,...,d}\int_{t_{00}}^{\infty} \frac{1}{F_L(u|x)F_S(u|x)}\left|\partial_{x_k} \frac{f_L(u|x)}{F_L(u|x)}\right| du < \infty$  \label{Th1.2}
\item \label{D3} $\sup_{k,j=1,...,d}\sup_{(t,z)\in (t_{00},\infty)\times U_{\epsilon}(x)}\left| \partial_{z_k}\partial_{z_j} \frac{f_L(t|z)}{F_L(t|z)}\right| < \infty$ \label{Th1.3}
\item \label{D4} The functions $H_k(t|x)$ $(k=0,1,2)$ are twice continuously differentiable with respect to the second component
in some neighborhood $U_\eps(x)$ of $x$ and for $k=0,1,2$ we have $$\sup_{k,j=1,..,d}\sup_t \sup_{|y-x|<\eps} |\partial_{y_k}\partial_{y_j} H_k(t|y)| < \infty $$
\item \label{D5} The distribution function $F_X$ of the covariates $X_i$ is twice continuously differentiable in $U_\eps(x)$. Moreover, $F_X$ has a uniformly continuous density $f_X$ such with $f_X(x)\neq 0$. 
\item \label{D6} There exists a constant $C>0$ such that $H(t|y)\geq CH(t|x)$ for all $(t,y)\in [t_{00},t_{00}+\eps)\times I$ where $I$ is a set with the property $\int_{I \cap U_\delta(x)} f_X(s) ds \geq c\delta^d$ for some $c>0$ and all $0<\delta\leq\eps$. 
\item \label{D7} $\frac{f_L(t|y)}{F_L(t|y)} = \frac{f_L(t|x)}{F_L(t|x)}(1+o(1))$ uniformly in $t\in [t_{00},t_{00}+\eps)$ as $y \rightarrow x$  \label{Th1.7}
\item \label{D8} For $\tau_{T,0}(x) := \inf\{t:F_T(t|x)>0\}$ we have $\inf_{y\in U_\eps(x)} F_L(\tau_{T,0}(y)|y) > 0.$
\end{enumerate}

\begin{remark} \label{rm:t00} \rm{
From the definition of $t_{00}$ and $H_0$ we immediately see that under condition $\ref{D0000}$ we have $t_{00} = \tau_{T,0}(x)\vee \tau_{L,0}(x)$ where we use the notation $\tau_{L,0}(x) := \inf\{t:F_L(t|x)>0\}$. In particular, this implies that under either of the assumptions \ref{D1} or \ref{D8} the equality $t_{00} = \tau_{T,0}(x)$ holds.
}\end{remark}

Finally, we make some assumptions for the smoothing parameter
\begin{enumerate}[label=(B\arabic{*})]
\item \label{B1} $nh^{d+4}\log n=o(1)$ and $nh\longrightarrow\infty$.
\item \label{B2} $h \rightarrow 0$ and $nh^d/\log n\longrightarrow\infty$.
\end{enumerate}

Some important practical examples for weights satisfying conditions \ref{W1} - \ref{W3} include Nadaraya-Watson and local linear weights. This is the assertion of the next Lemma.

\begin{lemma} \label{le:g1} \
\begin{enumerate}
\item Conditions \ref{W1a} and \ref{W1b} are fulfilled for the Nadaraya-Watson weights $W_i^{NW}$ with a Kernel $K$ satisfying condition (K1). If the density $f_X$ is continuous at the point $x$, condition \ref{W1c} also holds. Finally, if the function $x\mapsto f_X(x)F_Y(t|x)$ is continuously differentiable in a neighborhood of $x$ for every $t$ with uniformly (in $t$) bounded first derivative and (B1) is fulfilled, condition \ref{W1d} holds.\\
If additionally to these assumptions $d=1$ and the density $f_X$ of the covariates $X$ is continuously differentiable at $x$ with
bounded derivative, condition \ref{W1} also holds for the local linear and rearranged local linear weights $W_i^{LL}$ and $W_i^{LLI}$ defined in (\ref{ll}) and (\ref{ra1}), (\ref{ra2}) respectively, provided that the corresponding kernel fulfills condition (K1) .
\item If under assumptions \ref{D4}, \ref{D5} and \ref{B1} the density $f_X$ is twice continuously differentiable with uniformly bounded derivative, condition \ref{W2} holds for the Nadaraya-Watson ($d$ arbitrary), local linear ($d=1$) or rearranged local linear ($d=1$) weights based on a positive, symmetric kernel with compact support.
\item If under assumptions \ref{B2}, \ref{D00}, \ref{D0} the density $f_X$ is twice continuously differentiable with uniformly bounded derivative, condition \ref{W3} holds for the Nadaraya-Watson weights $W_i$ based on a positive, symmetric
kernel with compact support ($d$ arbitrary). If additionally $d=1$ and the density $f_X$ of the covariates $X$ is continuously differentiable at $x$ with bounded derivative, condition \ref{W3} also holds for local linear or rearranged local linear weights.
\end{enumerate}
\end{lemma}
The proof of this Lemma is standard, a sketch can be found in the Appendix. \\
\\
Note that the assumption \ref{B1} does not allow to choose $h \sim n^{-1/(d+4)}$, which would be the MSE-optimal rate
for Nadaraya-Watson or local linear weights and functions with two continuous derivatives with respect to the predictor.
This assumption has been made for the sake of a transparent presentation and implies that the bias of the estimates is negligible compared to the stochastic part. Such an approach is standard in nonparametric estimation for censored data, see \cite{dabrowska1987}  or \cite{lidoss1995}.
In principle, most results of the present paper can be extended to bandwidths $h \sim n^{-1/(d+4)}$ if a corresponding bias term is subtracted.

Another useful  property of estimators constructed from weights satisfying condition \ref{W1} is that they are increasing with probability
tending to one.
\begin{lemma} \label{le:pitto}
Under condition \ref{W1a} we have
\[
\Pro{\mbox{``The estimates} \ (H_n(.|x),H_{0n}(.|x),H_{1n}(.|x),H_{2n}(.|x) \ \mbox{are increasing''}} \stackrel{\nti}{\longrightarrow} 1.
\]
\end{lemma}
The Lemma follows from the relation
\[
\{ \mbox{``The estimates} \ H_n(.|x),H_{0n}(.|x),H_{1n}(.|x),H_{2n}(.|x) \ \mbox{are increasing''}\} \supseteq \{W_i(x)\geq 0\ \forall\  i\}
\]
and the fact that under assumption (W1) the probability of the event on the right hand side converges to one. We will use Lemma \ref{le:pitto} for the analysis of the asymptotic  properties of the conditional quantile estimators in Section \ref{S:quantasy}.
One noteworthy consequence of the Lemma is the fact that
\[
\Pro{\hat q^{IP}(.|x)\equiv\hat q(.|x)}\rightarrow 1,
\]
which follows because the mappings $\Psi$ and the right continuous inversion mapping coincide on the set of nondecreasing functions.
In particular, this indicates  that, from an asymptotic  point of view, it does not matter which of the estimators $\hat q, \hat q^{IP}$ is used.
The difference between both estimators will only be visible in finite samples - see Section \ref{secsim}. In fact, it can only occur if one of the estimators
$H_n,H_{k,n}$ is decreasing at some point.


\subsection{Weak convergence of the estimate of the conditional distribution}
We are  now ready to  describe the asymptotic  properties of  the   estimates defined in Section 2.
Our first   result deals with the weak uniform consistency of the estimate $F_{T,n}(.|x)$ under some rather weak conditions. In
particular, it does neither require the existence of densities of the conditional distribution functions [see \ref{D0}] nor integrability conditions like \ref{D1}.
\begin{theorem} \label{Th1:K}
If conditions \ref{D0000}, \ref{D0}, \ref{D8}, \ref{W1a}-\ref{W1b} and \ref{W3} are satisfied, then the following statements are correct.
\begin{enumerate}
\item The estimate $F_{T,n}(.|x)$ defined in (\ref{es3}) is weakly uniformly consistent on the interval $[0,\tau]$ for any $\tau$ such that $F_S(\tau|x)<1$.
\item If additionally $F_S(\tau_{T,1}(x)|x)=1$, where
\[
\tau_{T,1}(x) := \sup\{t:F_T(t|x)<1\},
\]
and $F_{T,n}(.|x)$ is increasing and takes values in the interval $[0,1]$,
the weak uniform consistency of the estimate $F_{T,n}(.|x)$ holds on the interval $[0,\infty)$.
\end{enumerate}
\end{theorem}
The next two results deal with the weak convergence of $F_{T,n}$ and require additional assumptions on the
censoring distribution.  We begin with a result for the
estimator $F_{L,n}$, which is computed in the first step of our procedure by formulas (\ref{E.1}) and (\ref{E.2}).
\begin{theorem} \label{Th1} \
\begin{enumerate}
\item Let the weights used for $H_{2,n}$ and $H_n$ in the definition of the estimate $M^-_{2,n}$ in (\ref{es2}) satisfy conditions \ref{W1} and \ref{W2}. Moreover, assume that conditions \ref{B1}, \ref{D0000} and \ref{D0}-\ref{D7} hold.
Then we have as $n \to \infty$
\[
\sqrt{nh^d}(H_n-H,H_{0,n}-H_0,M_{n,2}^- - M_2^-) \Rightarrow (G,G_0,G_M)
\]
in $D^3([t_{00},\infty])$, where $(G,G_0,G_M)$ denotes a centered Gaussian process with a.s.\ continuous sample paths and   $G_M(t) = A(t) - B(t)$ is
defined by
\begin{equation}
A(t) = \int_{t}^{\infty} \frac{dG_2(u)}{H(u|x)}, \quad \quad B(t) := \int_{t}^{\infty} \frac{G(u)}{H^2(u|x)}H_2(du|x) \label{dGM}.
\end{equation}
Here the process $(G_0, G_2, G)$ is specified in assumption \ref{W2} and the integral with respect to the process $G_2(t)$ is defined via integration-by-parts.
\item Under the conditions of the first part we have
\[
\sqrt{nh^d}(H_n-H,H_{0,n}-H_0,F_{L,n}-F_L) \Rightarrow (G,G_0,G_3)
\]
in  $D^3([t_{00},\infty])$,  where the process $(G_0, G_2, G)$ is specified in assumption \ref{W2} and $G_3$ is a centered Gaussian process with a.s.\ continuous sample paths which is defined by
\[
G_3(t) = F_L(t|x)G_M(t).
\]
\end{enumerate}
\end{theorem}
\begin{remark} \label{remGM}
{\rm The value of the process $G_M$ at the point $t_{00}$ is defined as its path-wise limit. The existence of this limit follows from assumption \ref{D1} and
the representation
\[
\E[G_M(s)G_M(t)] = b(x)\int_{s\vee t}^{\infty} \frac{1}{H(u|x)}M_2^-(du|x)
\]
for the covariance structure of $G_M$, which can be derived by computations similar to those in \cite{patiroli2001}.}
\end{remark}


\begin{theorem} \label{Th2}
Assume that the conditions of Theorem \ref{Th1} and condition \ref{D8} are satisfied. Moreover, let $t_{00}<\tau$ such that $F_S([0,\tau]|x)<1$.
Then we have the following weak convergence
\begin{enumerate}
\item
\[
\sqrt{nh^d} (\Lambda_{T,n}^- - \Lambda_T^-) \Rightarrow V
\]
in $D([0,\tau])$, where
\[
V(t) := \int_0^t \frac{G_0(du)}{(F_L-H)(u-|x)} - \int_0^t \frac{G_3(u-)-G(u-)}{(F_L-H)^2(u-|x)}H_0(du|x)
\]
is a centered Gaussian process with a.s. continuous sample paths and the integral with respect to $G_0$ is defined via integration-by-parts.
\item
\[
\sqrt{nh^d} (F_{T,n} - F_T) \Rightarrow W
\]
 in   $D([0,\tau])$, where
\[
W(t) := (1-F_T(t|x))V(t),
\]
is a centered Gaussian process with a.s. continuous sample paths.
\end{enumerate}
\end{theorem}

Note that the second part of Theorem \ref{Th2} follows from   the first part using the representation
(\ref{es4}) and the delta method.

\subsection{Weak convergence of conditional quantile estimators} \label{S:quantasy}
In this subsection we discuss the asymptotic properties of the two conditional quantile estimates $\hat q$ and $\hat q^{IP}$
defined in (\ref{qes1}) and (\ref{IP4}), respectively.
As an immediate consequence of Theorem \ref{Th1:K} and the continuity of the quantile mapping [see \cite{gill1989}, Proposition 1] we obtain the weak
consistency result.
\begin{theorem} \label{Th2:K}
If  the assumptions   of  the first part of Theorem \ref{Th1:K} are satisfied and  additionally the conditions
$F_S(F_T^{-1}(\tau|x)|x)<1$ and $\inf_{\eps\leq t \leq \tau} f_T(t|x)>0$ hold some some $\varepsilon > 0$, then the estimators $\hat q(.|x)$ and $q^{IP}(.|x)$
defined in  (\ref{qes1}) and (\ref{IP4}) are weakly uniformly
consistent on the interval $[\eps,\tau]$.
\end{theorem}
The compact differentiability of the quantile mapping and the delta method yield the following result.
\begin{theorem}\label{Th3}
If the assumptions of Theorem \ref{Th2} are satisfied, then we have for any $\eps>0$ and $\tau>0$ with
$F_S(F_T^{-1}(\tau|x)|x)<1$ and $\inf_{\eps \leq t \leq \tau} f_T(t|x)>0$
\bea
\sqrt{nh^d}(\hat q(.|x) - F_T^{-1}(.|x)) \Rightarrow Z(.) \quad &\mbox{on}& \quad D([\eps,\tau]),
\\
\sqrt{nh^d}(\hat q^{IP}(.|x) - F_T^{-1}(.|x)) \Rightarrow Z(.) \quad &\mbox{on}& \quad D([\eps,\tau]),
\eea
where $Z$ is a centered Gaussian process
defined by
\[
Z(.) = - \frac{W\circ F_T^{-1}(.|x)}{f_T(.|x)\circ F_T^{-1}(.|x)}
\]
and the centered Gaussian process $W$ is defined in part 2 of Theorem \ref{Th2}.
\end{theorem}

\medskip

The proof Theorem \ref{Th1:K} - \ref{Th3} is presented in the Appendix A and requires several
separate steps. A main step in the proof is a result regarding the weak convergence
of the Beran estimator on the maximal possible domain in the setting of conditional right censorship.
We were not able to find such a result in the literature. Because this question is of independent interest,
it is presented separately in the following Subsection.

\subsection{A new result for the Beran estimator} \label{s:beran}
We consider the common conditional right censorship model [see \cite{dabrowska1987} for details]. Assume that our observations consist of the triples $(X_i,Z_i,\Delta_i)$ where $Z_i = \min(B_i,D_i), \Delta_i = \Ind{Z_i=D_i}$, the random variables $B_i,D_i$ are independent conditionally on $X_i$ and nonnegative almost surely. The aim is to estimate the conditional distribution function $F_D$ of $D_i$. Following \cite{beran1981} this can be done by estimating $F_Z$, the conditional distribution function of $Z$, and $\pi_k(t|x) := \Pro{Z_i\leq t, \Delta_i=k|X=x}$ ($k=0,1$) through
\beq \label{ber1}
F_{Z,n}(t|x) := W_i(x)\Ind{Z_i\leq t}, \quad \quad \pi_{k,n}(t|x) := W_i(x)\Ind{Z_i\leq t, \Delta_i = k} \ \ (k =0,1)
\eeq
and then defining an estimator for $F_D$ as
\beq \label{be:deffdach}
F_{D,n}(t|x) := 1 - \prod_{[0,t]}(1-\Lambda_{D,n}^-(ds|x)),
\eeq
where the quantity $\Lambda_{D,n}^-(ds|x)$ is given by
\beq \label{be:deflambda}
\Lambda_{D,n}^-(ds|x) := \frac{\pi_{0,n} (ds|x)}{1 - F_{Z,n}(s-|x)},
\eeq
and the $W_i(x)$ denote local weights depending on $X_1,...,X_n$ [see also the discussion at the beginning of Section \ref{main}].  \\
The weak convergence of the process $\sqrt{nh^d}(F_{D,n}(t|x) - F_D(t|x))_t$ in $D([0,\tau])$ with $\pi_0(\tau|x)<1$ was first established by \cite{dabrowska1987}. An important problem is to establish conditions that ensure that the weak convergence can be extended to $D([0,t_0])$ where $t_0 := \sup \{s: \pi_0(s|x)<1 \}$.\\
In the unconditional case, such conditions were derived by \cite{gill1983} who used counting process techniques. A generalization of this method to the conditional case was first considered by \cite{mckeagueutikal1990} and later exploited by \cite{dabrowska1992b} and \cite{lidoss1995}. However, none of those authors considered weak convergence on the maximal possible interval $[0,t_0]$. The following Theorem provides sufficient conditions for the weak convergence on the maximal possible domain.
\begin{theorem}\label{satz:beran}

Assume that for some $\eps>0$
\begin{enumerate}[label=(R\arabic{*})]
\item The conditional distribution functions $F_D(.|x)$ and $F_B(.|x)$ have densities, say $f_D(.|x) $ and $f_B(.|x)$,  with respect to the Lebesque measure
\item \label{1'} $\int_0^{t_0} \frac{\lambda_D(t|x)}{1-F_Z(t-|x)} dt<\infty$,
\item \label{2'} $\sup_{k=1,...,d}\int_0^{t_0} \frac{|\partial_{x_k}\lambda_D(t|x)|}{1-F_Z(t-|x)} dt<\infty$,
\item \label{3'} $\sup_{j,k=1,...,d}\sup_{(t,y)\in (0,t_0)\times U_{\epsilon}(x)}\left| \partial_{y_k}\partial_{y_j} \lambda_D(t|y)\right| < \infty$,
\item \label{4'} $1-F_Z(t|y) \geq C(1-F_Z(t|x))$ for all $(t,y)\in (t_0-\eps,t_0]\times I$ where $I$ is a set with the property $\int_{I \cap U_\delta(x)} f_X(s) ds \geq c\delta^d$ for some $c>0$ and all $0<\delta\leq\eps$.
\item \label{6'} $\lambda_D(t|y) = \lambda_D(t|x)(1+o(1))$ uniformly in $t \in (t_{0}-\eps,t_{0}]$ as $y \rightarrow x$.
\end{enumerate}
Moreover, let the weights in (\ref{ber1}) satisfy condition \ref{W1} and let the weak convergence
\[
\sqrt{nh^d}(F_{Z,n}(.|x) - F_Z(.|x) ,\pi_{0,n}(.|x)-\pi_0(.|x)) \Rightarrow (G,G_0) \quad \mbox{on} \quad D([0,\infty))
\]
to a centered Gaussian process $(G,G_0)$ with covariance structure given by
\bea
\mbox{\rm{Cov}}(G_0(s|x),G_0(t|x)) &=& b(x)(\pi_0(s\wedge t|x) - \pi_0(s|x)\pi_0(t|x))\\
\mbox{\rm{Cov}}(G(s|x),G(t|x))     &=& b(x)(F_Z(s\wedge t|x) - F_Z(s|x)F_Z(t|x))\\
\mbox{\rm{Cov}}(G_0(s|x),G(t|x))   &=& b(x)(\pi_0(s\wedge t|x) - \pi_0(s|x)F_Z(t|x))
\eea
for some function $b(x)$ hold [this is the case for Nadaraya-Watson or local linear weights, see Lemma \ref{le:g1}]. Then under assumption \ref{B1}
\beq
\sqrt{nh^d}(F_{D,n}(.|x) - F_D(.|x))_t \Rightarrow G_D(.) \quad \mbox{in} \quad  D([0,t_0]),
\eeq
where $G_D$ denotes a centered Gaussian process with covariance structure taking the form
\[
\mbox{\rm{Cov}}(G_D(t),G_D(s)) = b(x)(1-F_D(s|x))(1-F_D(t|x))\int_0^{s\wedge t} \frac{\Lambda_D(du|x)}{1 - F_Z(u|x)}.
\]
\end{theorem}
\newpage
\section{Finite sample properties} \label{secsim}
We have performed a small simulation study in order to investigate the finite sample properties of the
proposed estimates. An important but difficult question in the estimation of the conditional
distribution function from censored data is the choice of the
smoothing parameter.  For  conditional right censored data some proposals regarding the choice of the
bandwidth have been made by \cite{dabrowska1992b} and \cite{lidatta2001}.
In order to obtain a reasonable bandwidth parameter for our simulations, we used a modification of the cross validation procedure proposed by \cite{abberger1998} in the context of nonparametric quantile regression. To address
 the presence of censoring in the cross validation procedure, we proceeded as follows:
\begin{enumerate}
\item Divide the data in blocks of size $K$ with respect to the (ordered) $X$-components.
 Let $\{ (Y_{jk},X_{jk}, \delta_{jk} ) |~j=1, \ldots ,J_k \}$
 denote the points among $\{ (Y_i,X_i, \delta_i) |~i=1, \ldots ,n \}$ which
 fall in block $k$ ($k=1,\ldots ,K$). For our simulations we used $K=25$ blocks.
\item In each block, estimate the distribution function $F_T$ as described in Section \ref{subsecestcdf}.
Denote the sizes of the jumps at the $j$th uncensored observation in the $k$th block by $w_{jk}$
\item Define
\[
h := \mbox{argmin}_{\alpha} \sum_{k=1}^K \sum_{j=1}^{J_k} w_{jk}\rho_{\tau}(Y_{jk}- \tilde q_{\alpha}^{j,k}(\tau|X_{jk}))
\]
where $\rho_{\tau}$ denotes the check function and $\tilde q_{\alpha}^{j,k}$ is either the estimator
$\hat q^{IP}$ or $\hat q $ with  bandwidth $\alpha$  based
on the sample $\{ (Y_i,X_i, \delta_i) |~i=1, \ldots ,n \}$
 without the observation
$(Y_{jk},X_{jk}, \delta_{jk} )$.
\end{enumerate}
 For a motivation of  the proposed procedure, observe that the classical cross validation
 is based on the fact that each observation is an unbiased 'estimator' for the
 regression function at the corresponding covariate.
 In the presence of censoring, such an estimator is not available. Therefore,
 the cross validation criterion discussed above tries to mimic this property by
 introducing the weights $w_{jk}$.
 A deeper investigation of the  theoretical  properties of the procedure is  beyond the scope of the present
 paper and postponed to future research.  In order to save computing time
 the bandwidth that we used for our simulations is an average of
$100$ cross validation runs in each scenario.

\begin{figure}[!hc]
\includegraphics[width = 5.5cm]{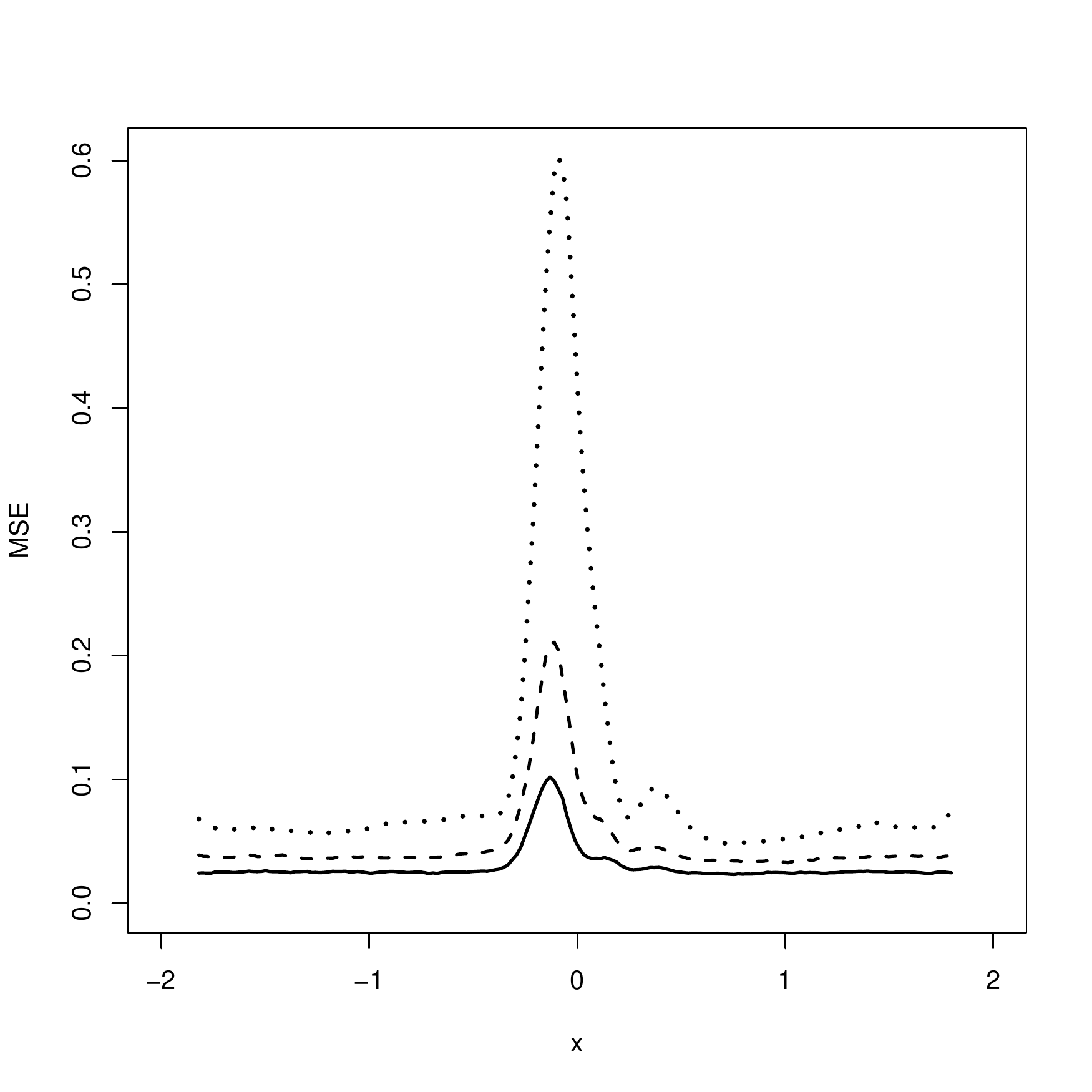}
\includegraphics[width = 5.5cm]{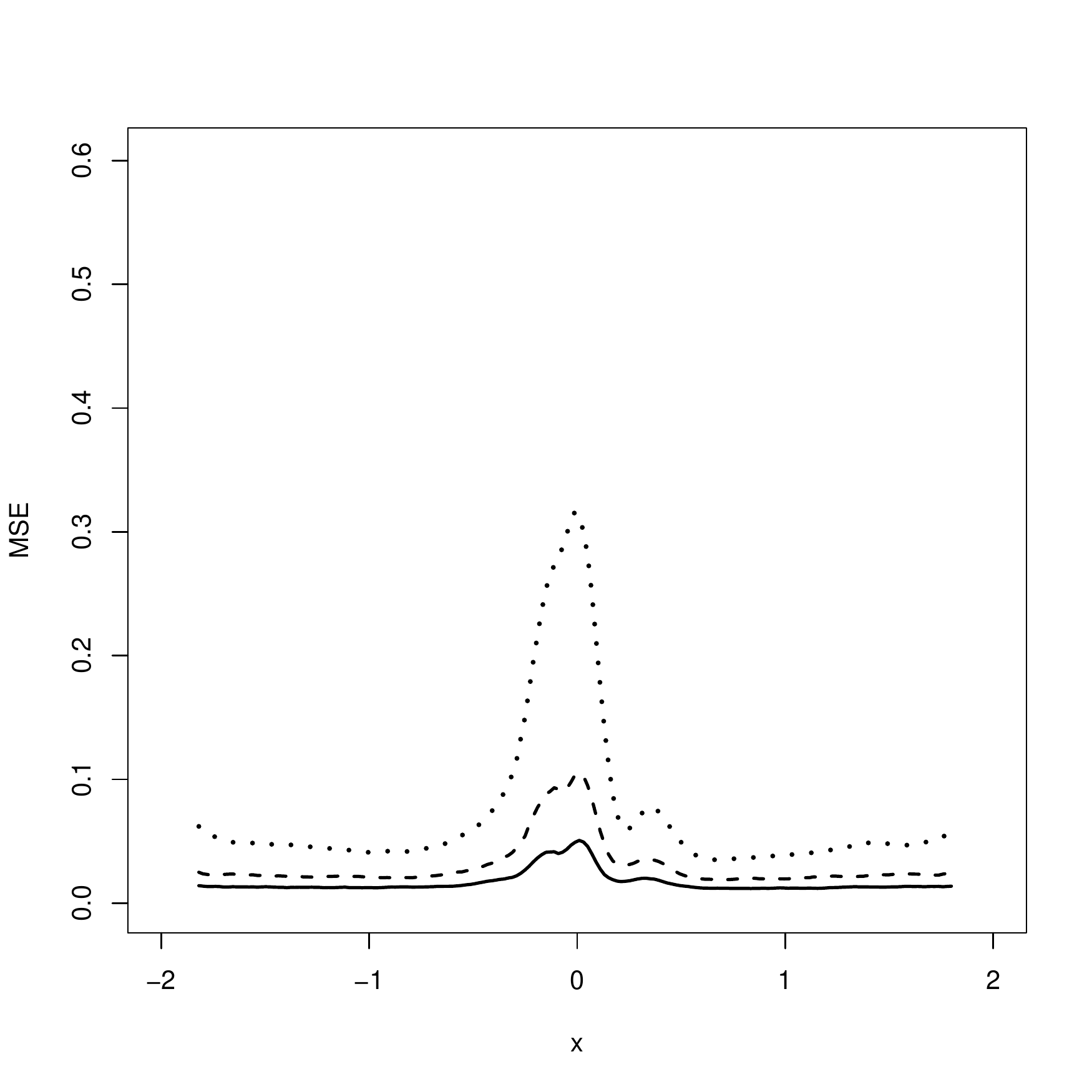}
\includegraphics[width = 5.5cm]{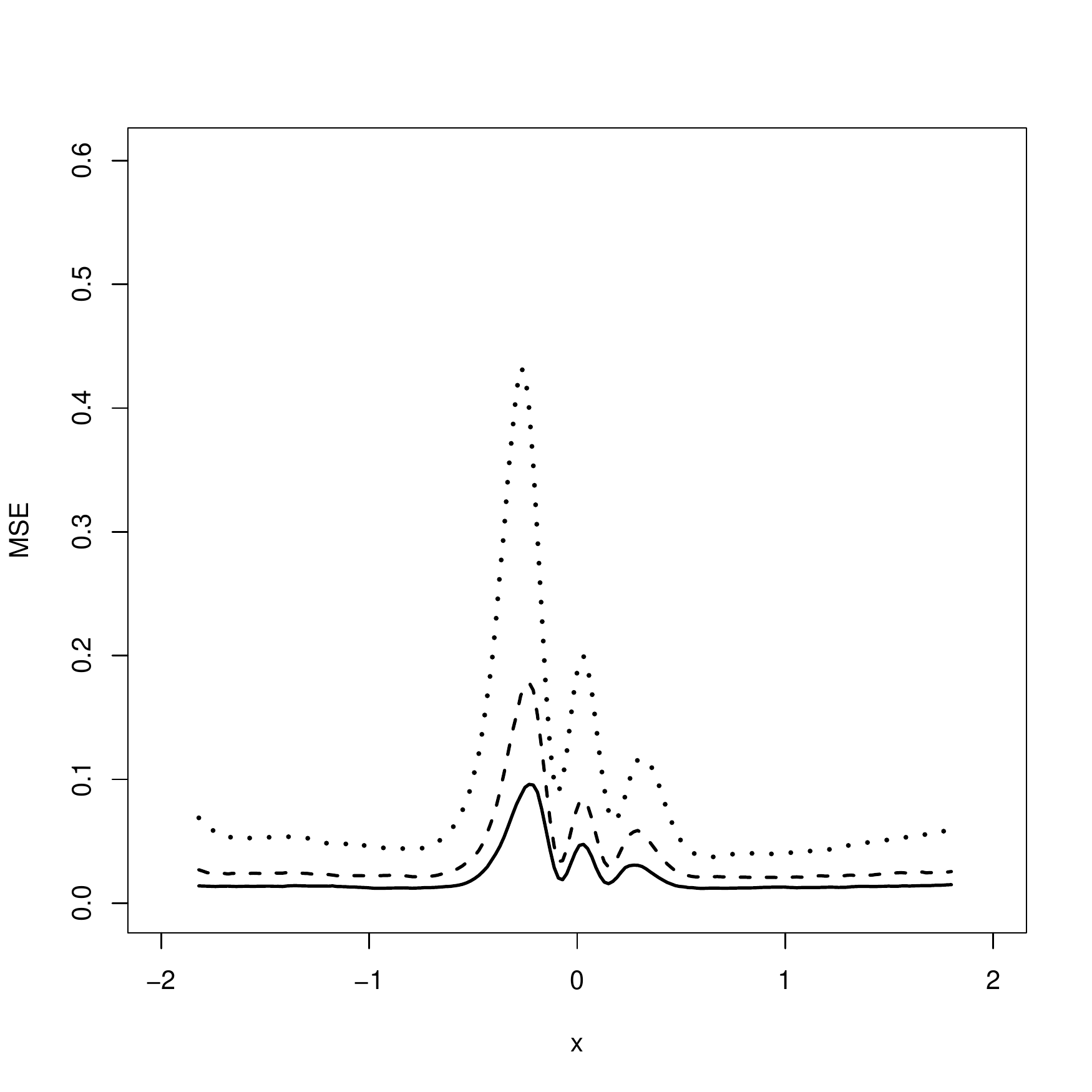}
\caption{\label{figm22mse}
\it Mean squared error curves of the estimates of the  quantile curves in  model 1 for
different sample sizes:   $n=100$ (dotted line);  $n=250$ (dashed line);  $n=500$ (solid line).
Left panel: estimates of the $25\%$-quantile curves;  middle panel: estimates of the $50\%$-quantile curves;
right panel: estimates of the $75\%$-quantile curves. $10\%$ of the observations are censored by
type $\delta=1$ and  $\delta=2$, respectively.}
\end{figure}

\begin{figure}[!hb]
\includegraphics[width = 5.6cm]{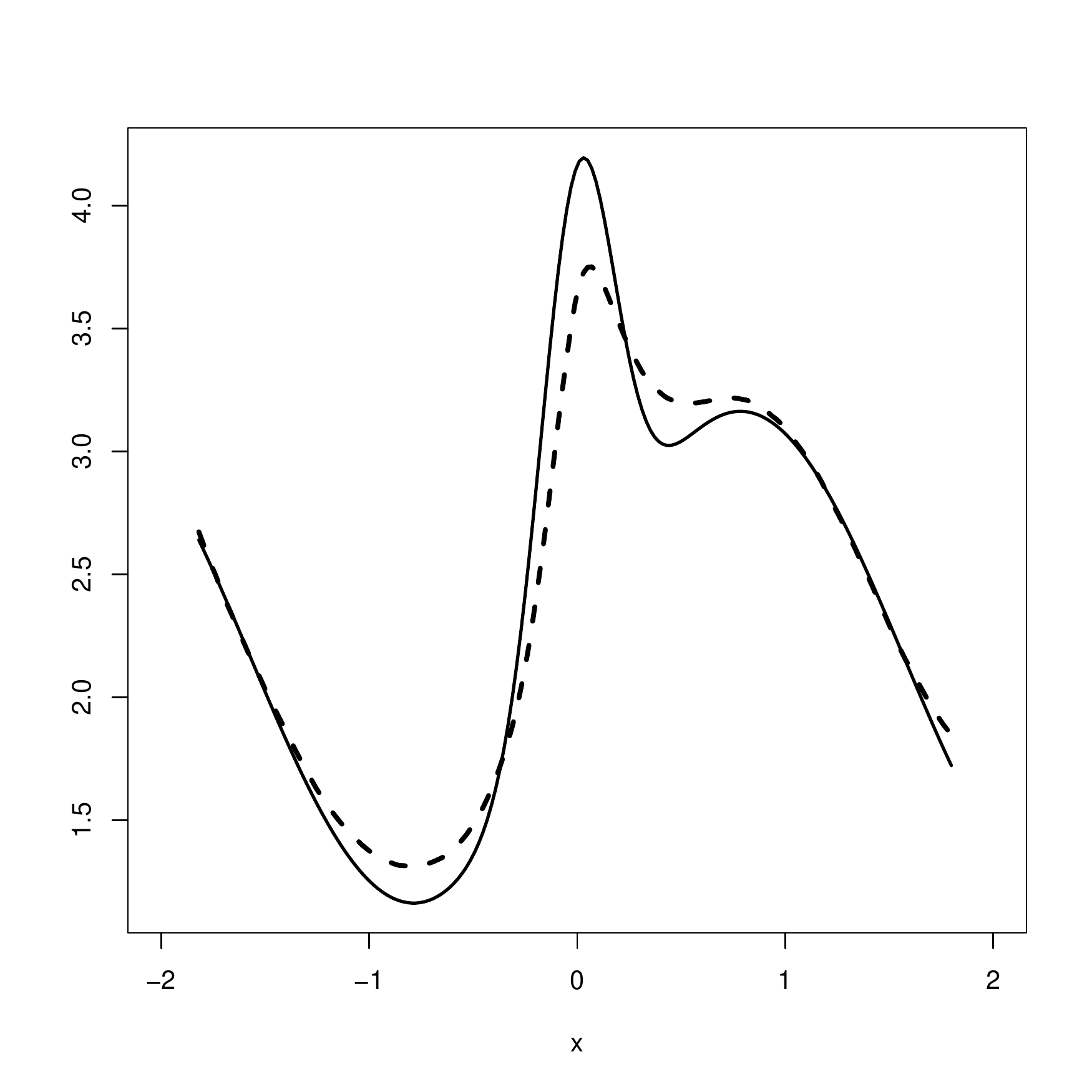}
\includegraphics[width = 5.6cm]{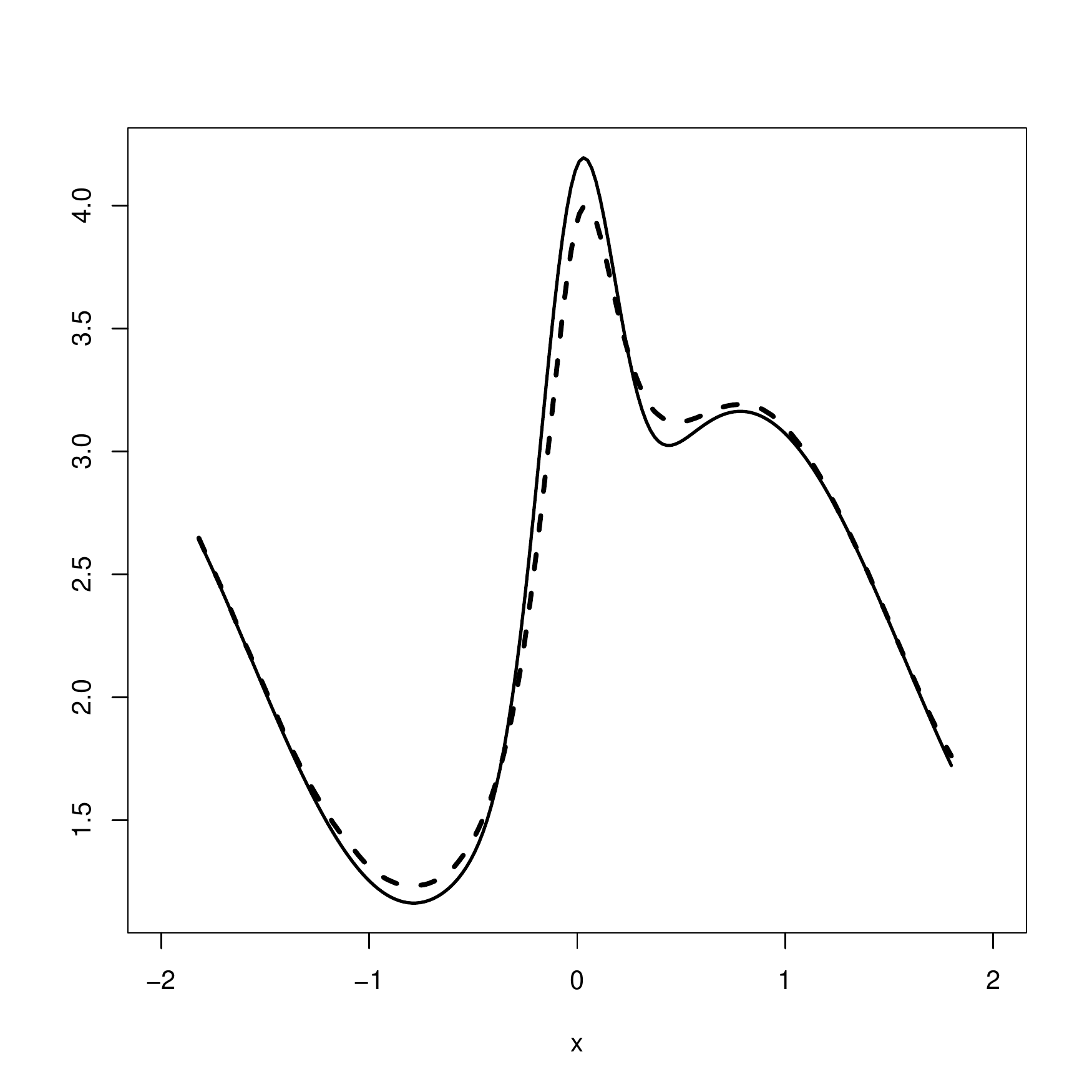}
\includegraphics[width = 5.6cm]{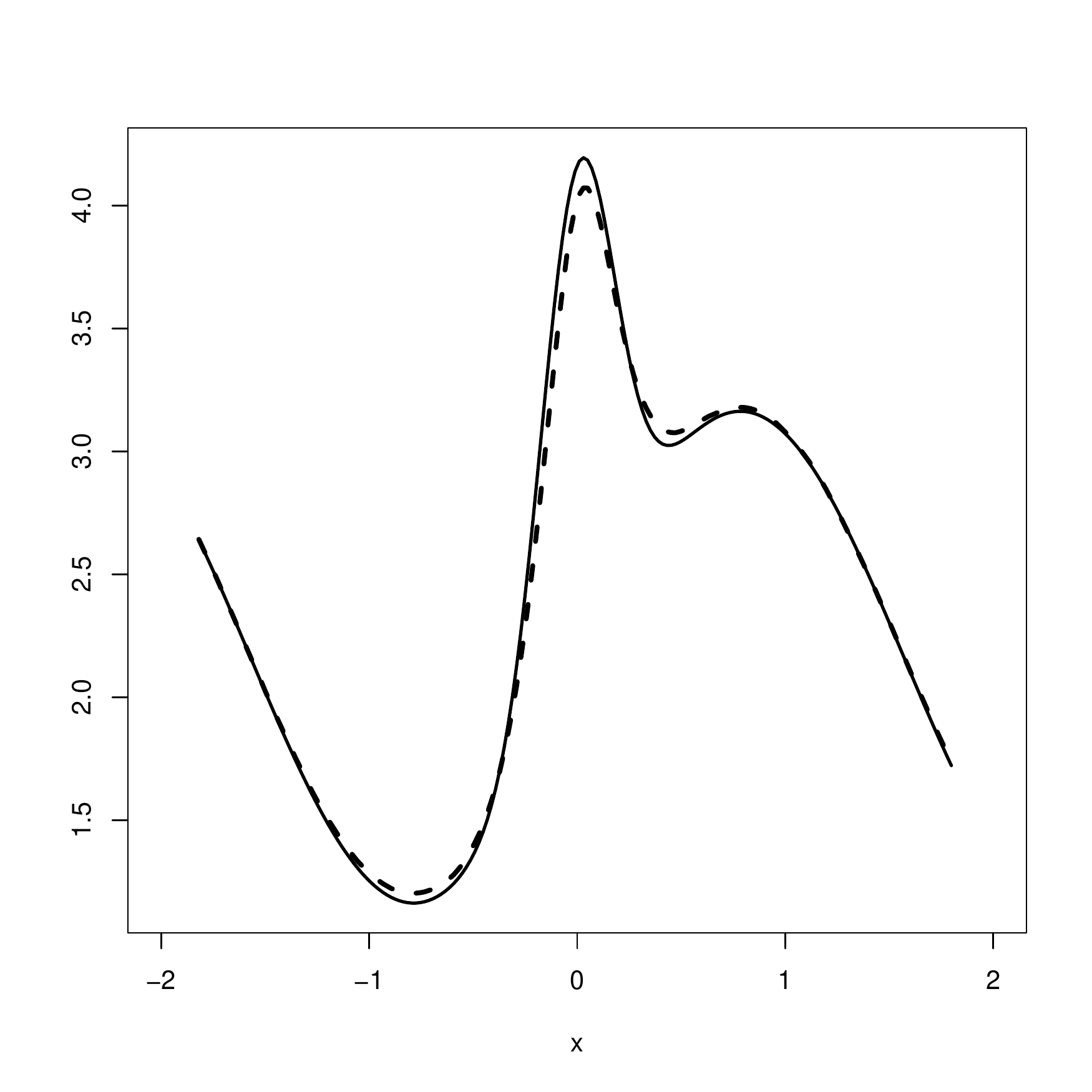}
\includegraphics[width = 5.6cm]{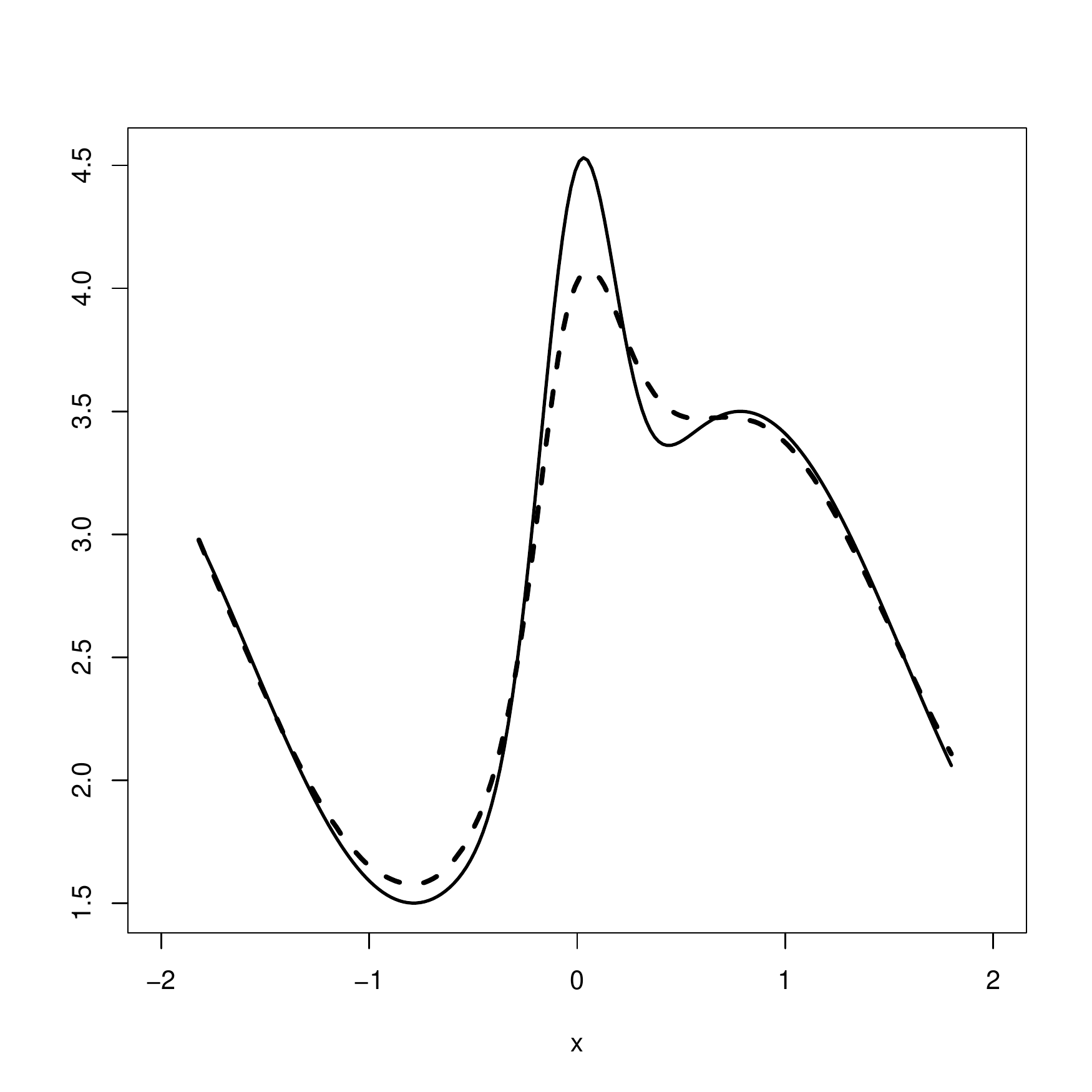}
\includegraphics[width = 5.6cm]{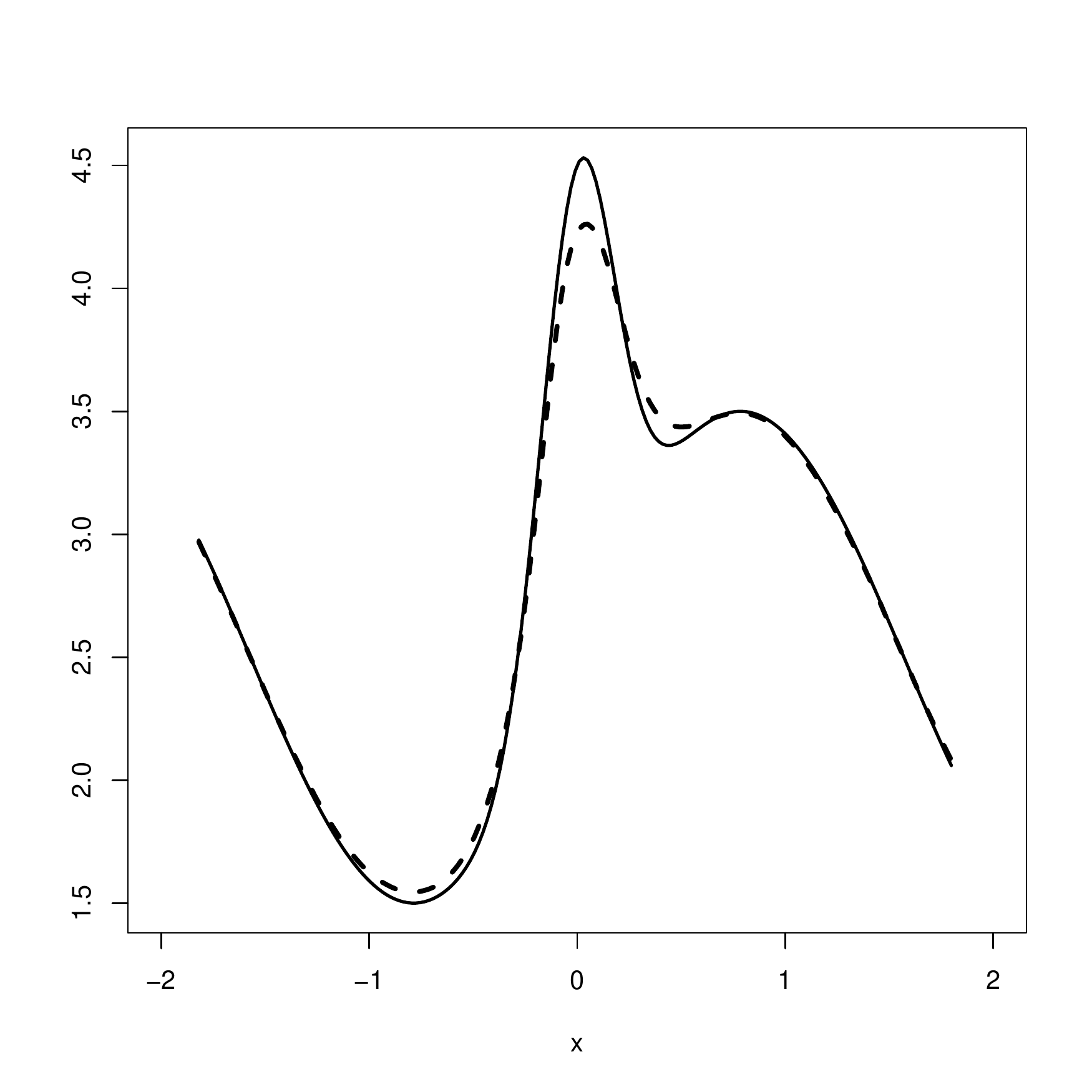}
\includegraphics[width = 5.6cm]{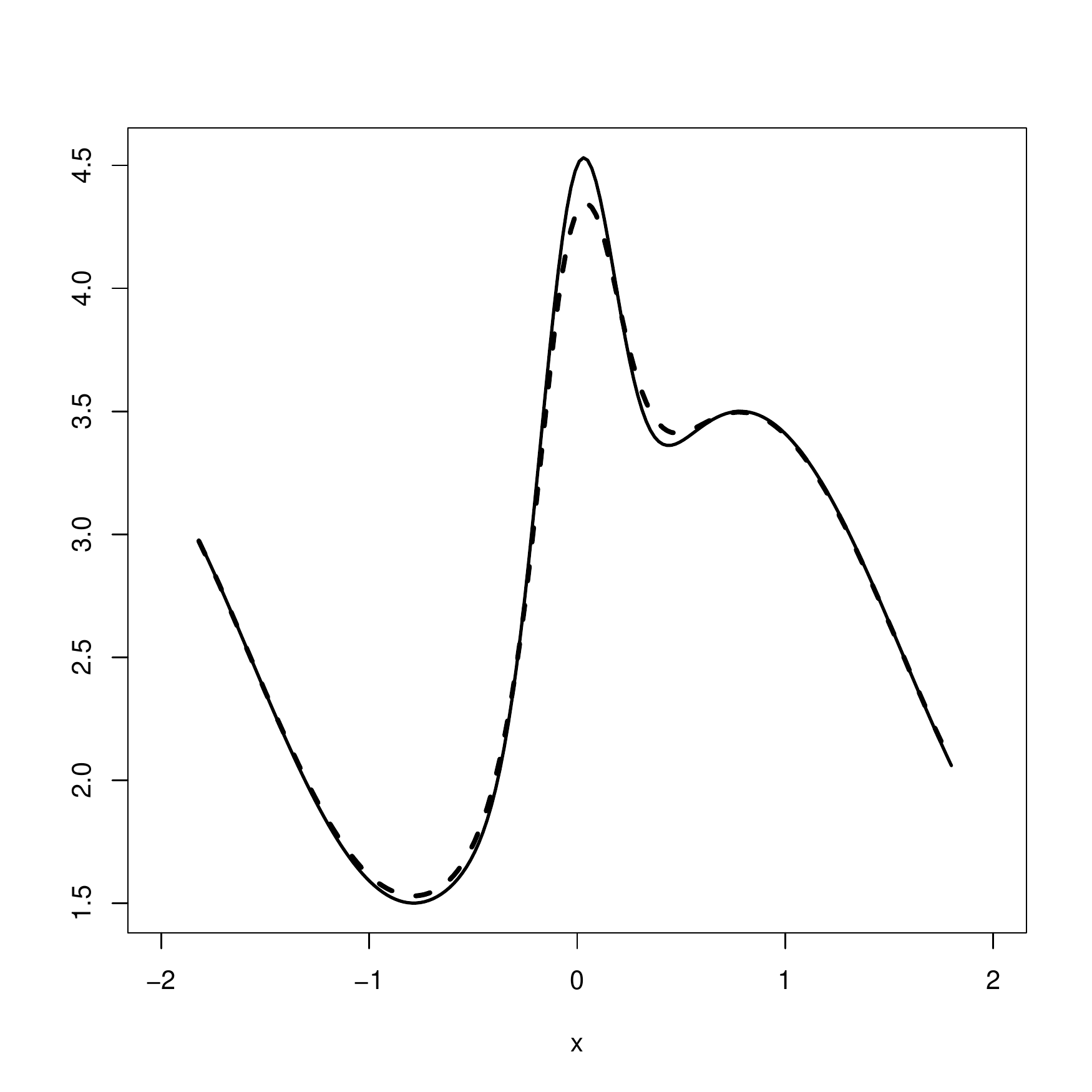}
\includegraphics[width = 5.6cm]{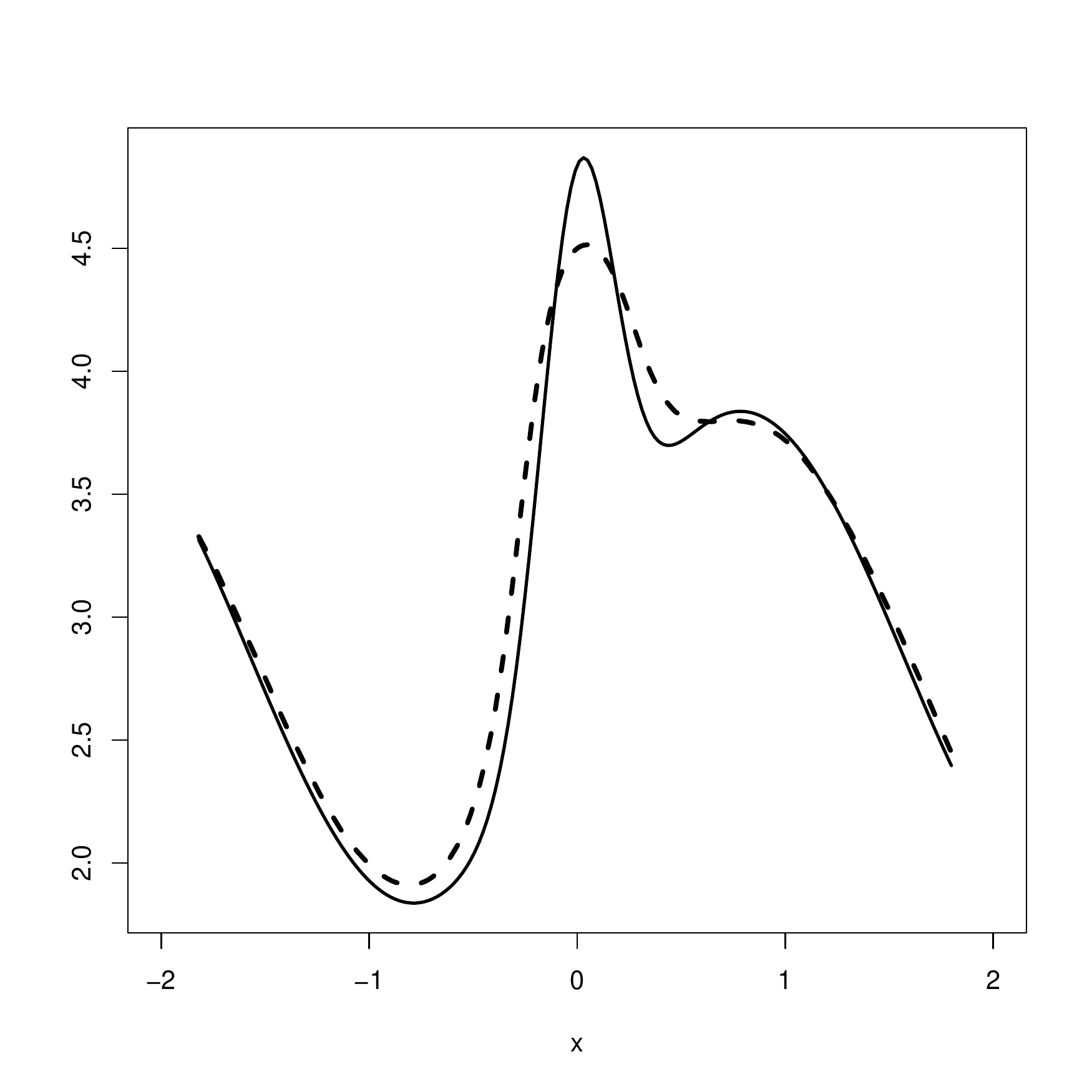}
\includegraphics[width = 5.6cm]{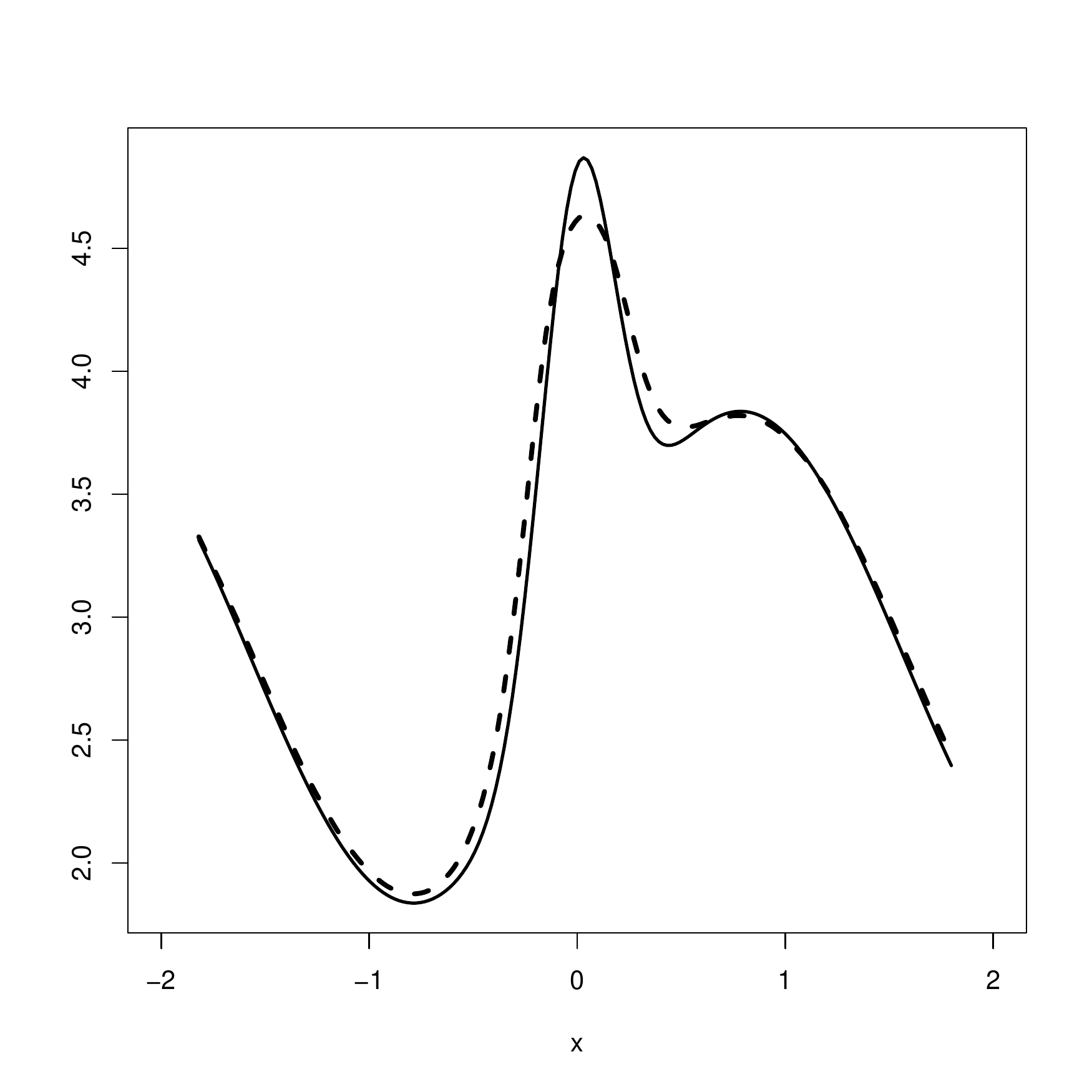}
\includegraphics[width = 5.6cm]{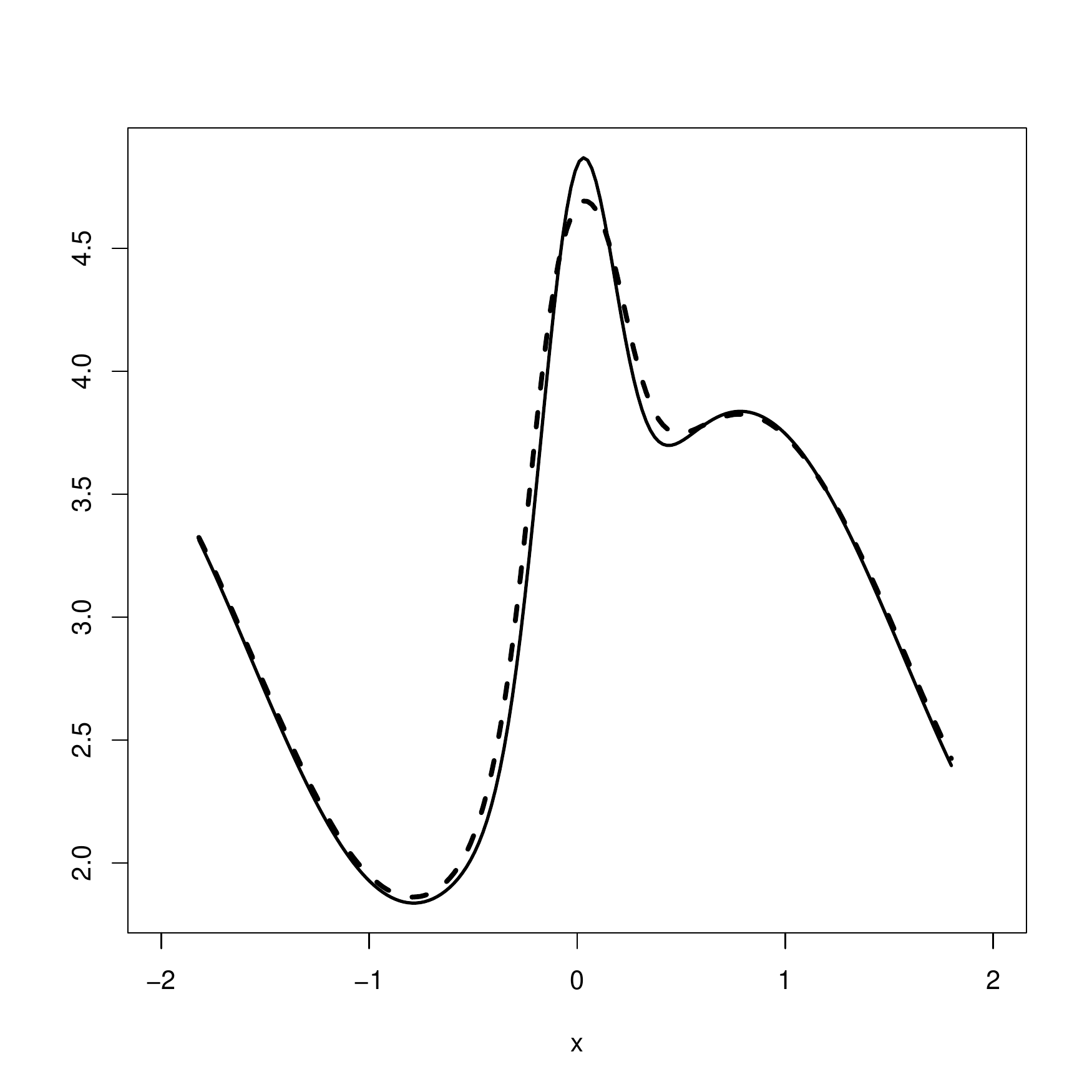}
\caption{ \label{figm22mc}
\it Mean (dashed lines) and true (solid lines) quantile curves for model 1
 for different sample sizes: $n=100$ (left column), $n=250$ (middle column)
and $n=500$ (right column). Upper row: estimates of the $25\%$ quantile curves; middle row:
estimates of the $50\%$ quantile curves;
lower row: estimates of the $75\%$ quantile curves. $10\%$ of the observations are censored by
type $\delta=1$ and  $\delta=2$, respectively.}
\end{figure}

For the calculation of the estimators of the conditional sub-distribution functions, we chose local linear weights [see Remark \ref{nwll}] with a truncated version of the Gaussian Kernel,
i.e. $$K(x) = \phi (x)I_{\{\phi(x)>0.001\}},$$
where $\phi $ denotes the density of the  standard normal distribution.
\\
We investigate the finite sample properties of the new  estimators in
a similar scenario as models 2 and 3 in \cite{yujon1997} [note that we additionally introduce
 a censoring mechanism]. The first model is given by
\[
\mbox{(model 1)} ~~~~~~~~~~~
\left\{
\begin{array}{ll}
& T_i = 2.5 + \sin(2X_i) + 2\exp(-16X_i^2) + 0.5\mathcal{N}(0,1) \\
& L_i = 2.6 + \sin(2X_i) + 2\exp(-16X_i^2) + 0.5(\mathcal{N}(0,1)+q_{0.1})\\
& R_i = 3.4 + \sin(2X_i) + 2\exp(-16X_i^2) + 0.5(\mathcal{N}(0,1)+q_{0.9})
\end{array}
\right.
\]
where the covariates $X_i$ are uniformly distributed on the interval $[-2,2]$
and $q_{p}$ denotes the $p$-quantile of a standard normal distribution.
This means that about $10\%$ of the observations are censored by
type $\delta=1$ and  $\delta=2$, respectively. For the sample size we use  $n=100,250,500$.
In  Figures \ref{figm22mc} and \ref{figm22mse}   we show the mean conditional quantile curves and corresponding mean squared error
curves for the  $25\%$, $50\%$ and $75\%$ quantile based on $5000$ simulation runs. The cases where
the $\hat q^{IP}(\tau|x)$ is not defined are omitted in the estimation of the mean squared error and mean
curves [this phenomenon occurred in less than $3\%$ of the simulation runs].
Only results for the  the estimator $\hat q^{IP}$ are presented because
it shows a  slightly better performance than the estimator $\hat q$. We observe no
substantial differences in the performance of the estimates for the
 $25\%$, $50\%$ and $75\%$ quantile curves with respect to bias.
On the other hand it can be seen from Figure \ref{figm22mse} that
 the estimates of the quantile curves corresponding to the $25\%$ and $75\%$ quantile have larger variability. In particular the mse is large at the point $0$, where the quantile
 curves attain their maximum.

As a second example  we investigate the effect of different censoring types. To this end, we consider
a similar example as in model 3 of \cite{yujon1997}, that is
$$
\mbox{(model 2)}~~~~~~~~~~~~~~~~~~~~
\left\{
\begin{array}{ll}
& T_i = 2 + 2\cos(X_i) + \exp(-4X_i^2) + \mathcal{E}(1)\\
& L_i = 2 + 2\cos(X_i) + \exp(-4X_i^2) + (c_L + \mathcal{U}[0,1])\\
& R_i = 2 + 2\cos(X_i) + \exp(-4X_i^2) + (c_R + \mathcal{E}(1))
\end{array}
\right.
$$
where the covariates $X_i$ are uniformly distributed on the interval $[-2,2]$, $\mathcal{E}(1)$ denotes an exponentially distributed random variable with parameter 1, $\mathcal{U}[0,1]$ is a uniformly distributed random variable
 on $[0,1]$ and the parameters $(c_L,c_R)$
are used to control the amount of censoring. For this purpose we investigate
 three different cases for the parameters $(c_L,c_R)$, namely $(-0.5,1.5)$, $(-0.5,0.5)$ and $(-0.2,1.5)$,
 which corresponds to approximately $(10\%,11\%)$, $(30\%,11\%)$ and $(11\%,25\%)$ of
type $\delta=1$ and $\delta=2$ censoring, respectively.
The corresponding results for the estimators of the $25\%$, $50\%$ and $75\%$ quantile on the basis
of a sample of $n = 250$ observations are presented in Figures 3 and 4.
\begin{figure}[!hc]
\includegraphics[width = 5.5cm]{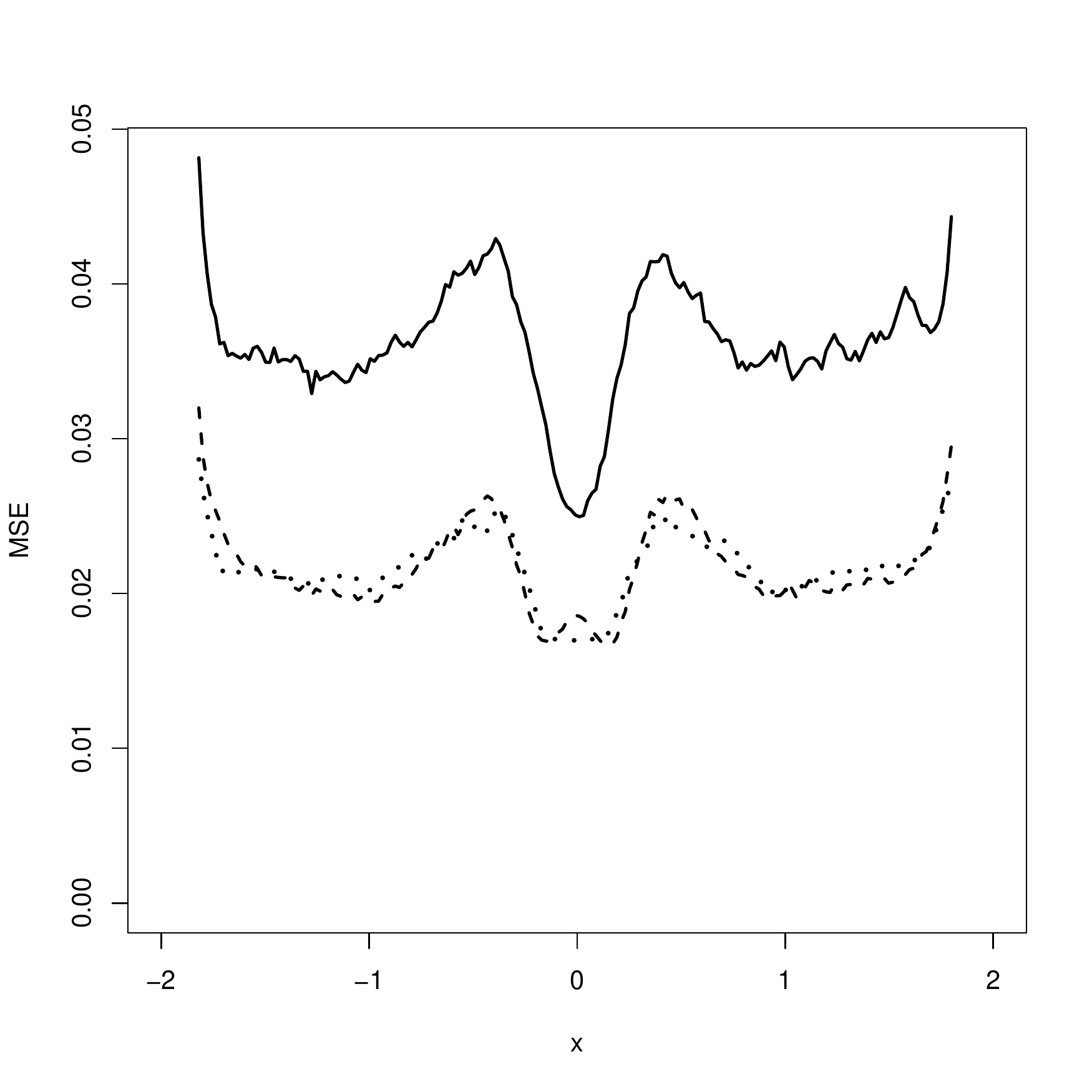}
\includegraphics[width = 5.5cm]{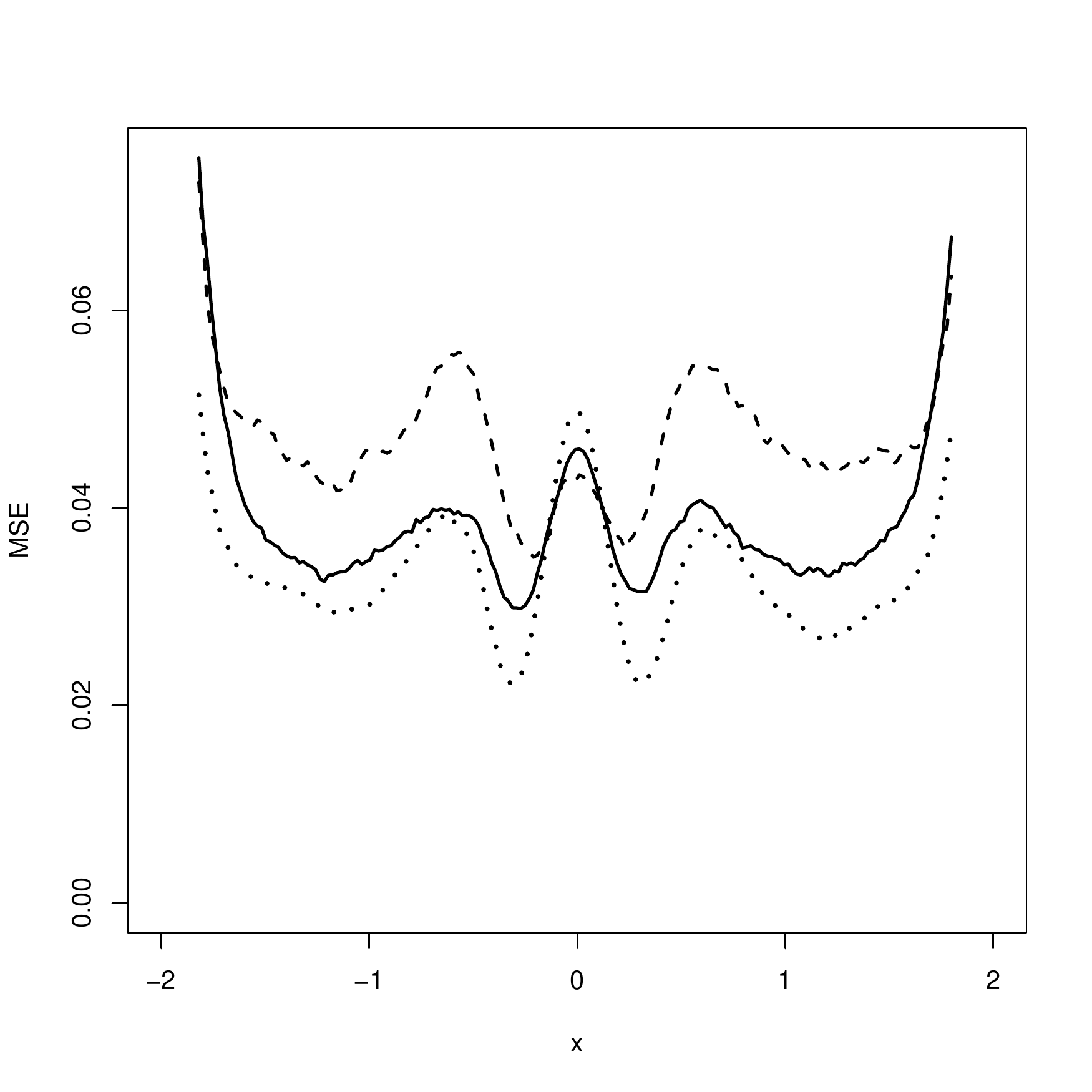}
\includegraphics[width = 5.5cm]{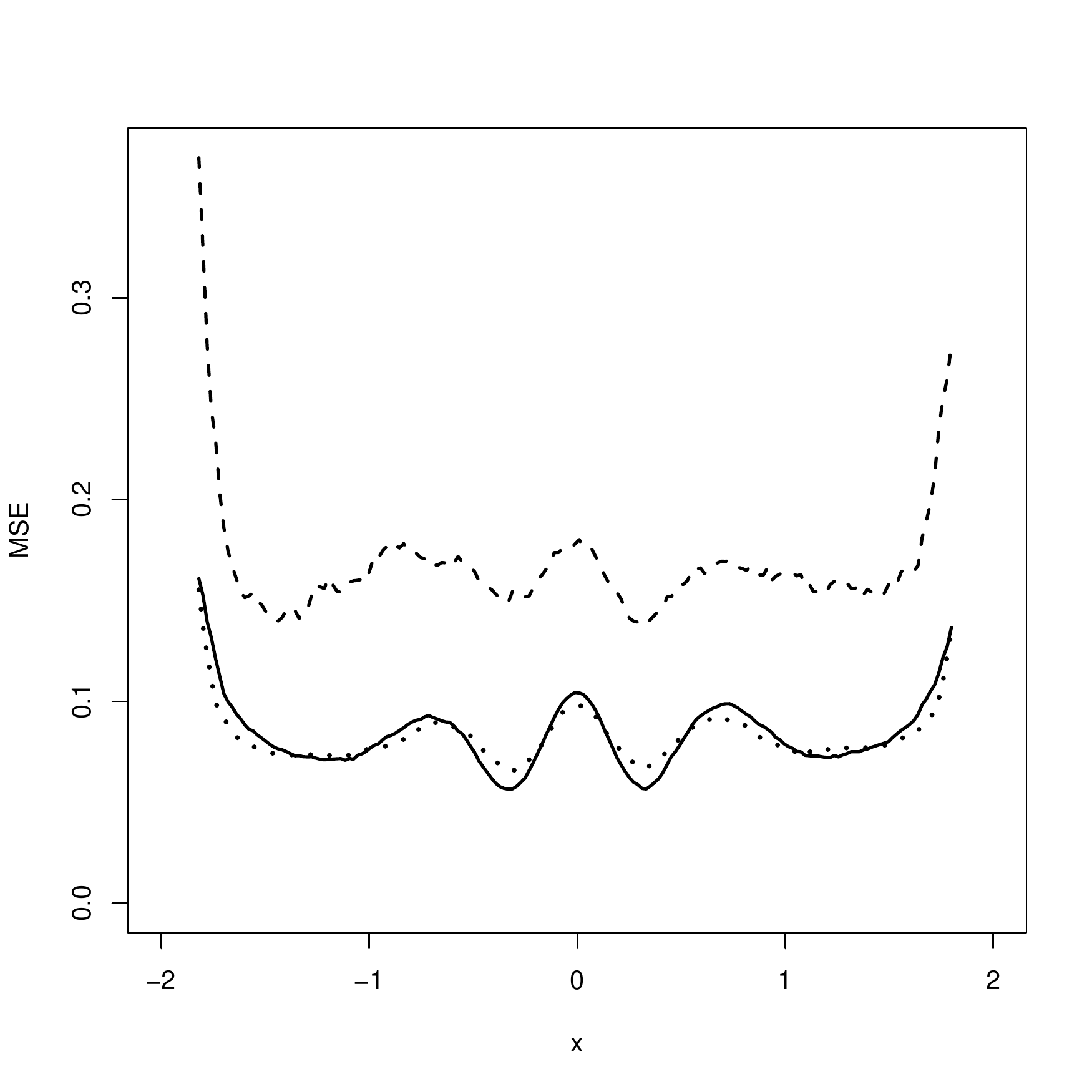}
\caption{\label{figm33mse}
\it Mean squared error curves of the estimates of the  quantile curves in  model 2 for
different censoring: $(10\%,11\%)$ censoring (dotted line);  $(30\%,11\%)$  censoring (dashed line);
$(11\%,25\%)$ censoring (solid line). Left panel: estimates of the $25\%$-quantile curves;  middle panel: estimates of the $50\%$-quantile curves;
right panel: estimates of the $75\%$-quantile curves. The sample size is $n=250$.  }
\end{figure}

{We observe a slight increase in bias when estimating upper quantile curves}. An additional amount of censoring  results in a slightly worse average behavior of the estimates. More censoring of type $\delta = 2$ has an impact on the accuracy of the estimates of the lower quantiles, while more censoring of type $\delta = 1$ has a stronger effect for the upper quantile curves.
Upper quantile curves are always estimated with more variability which is in accordance with the factor $1/f_T(F_T^{-1}(p|x)|x)$ in their limiting process.

\begin{figure}[!hc]
\includegraphics[width = 5.8cm]{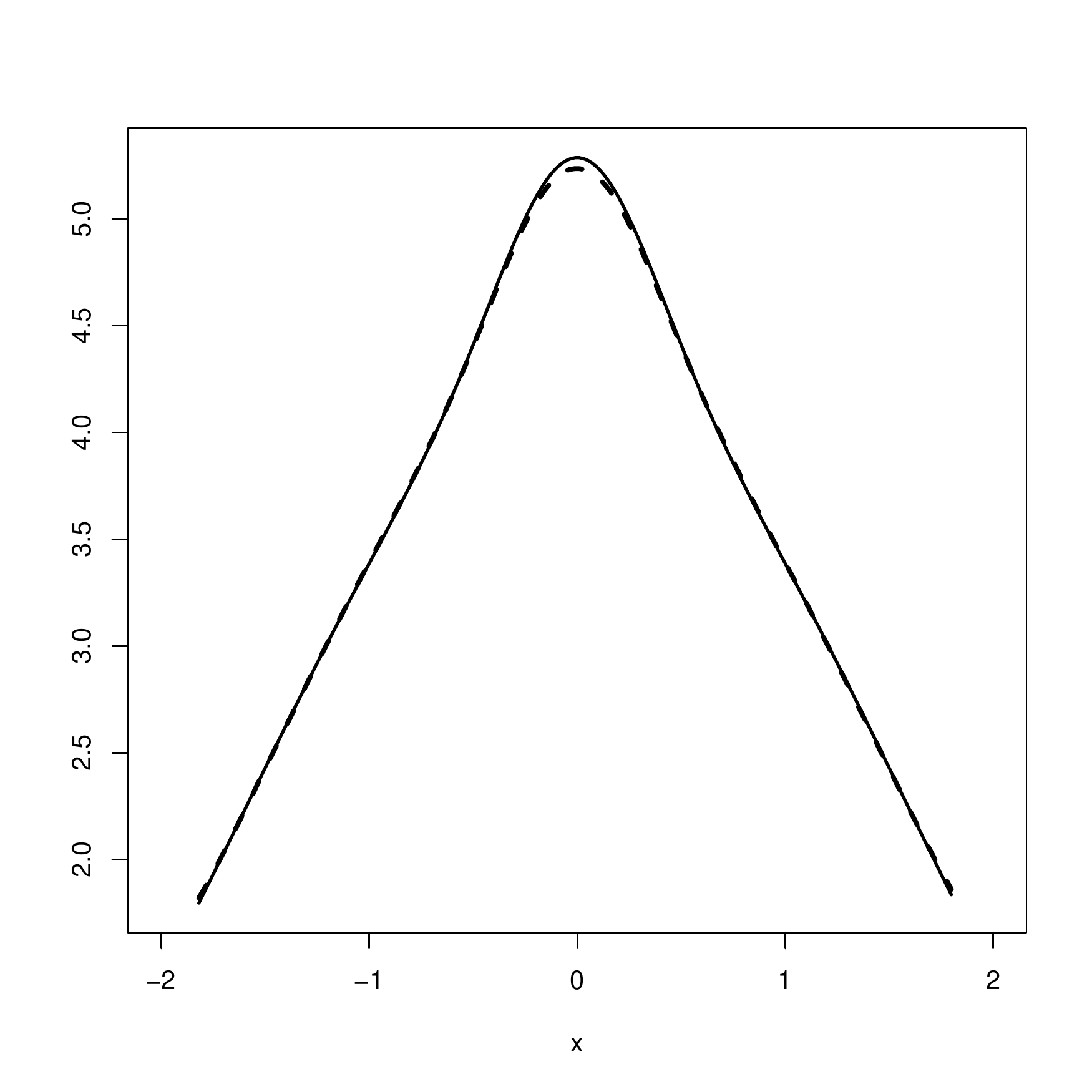}
\includegraphics[width = 5.8cm]{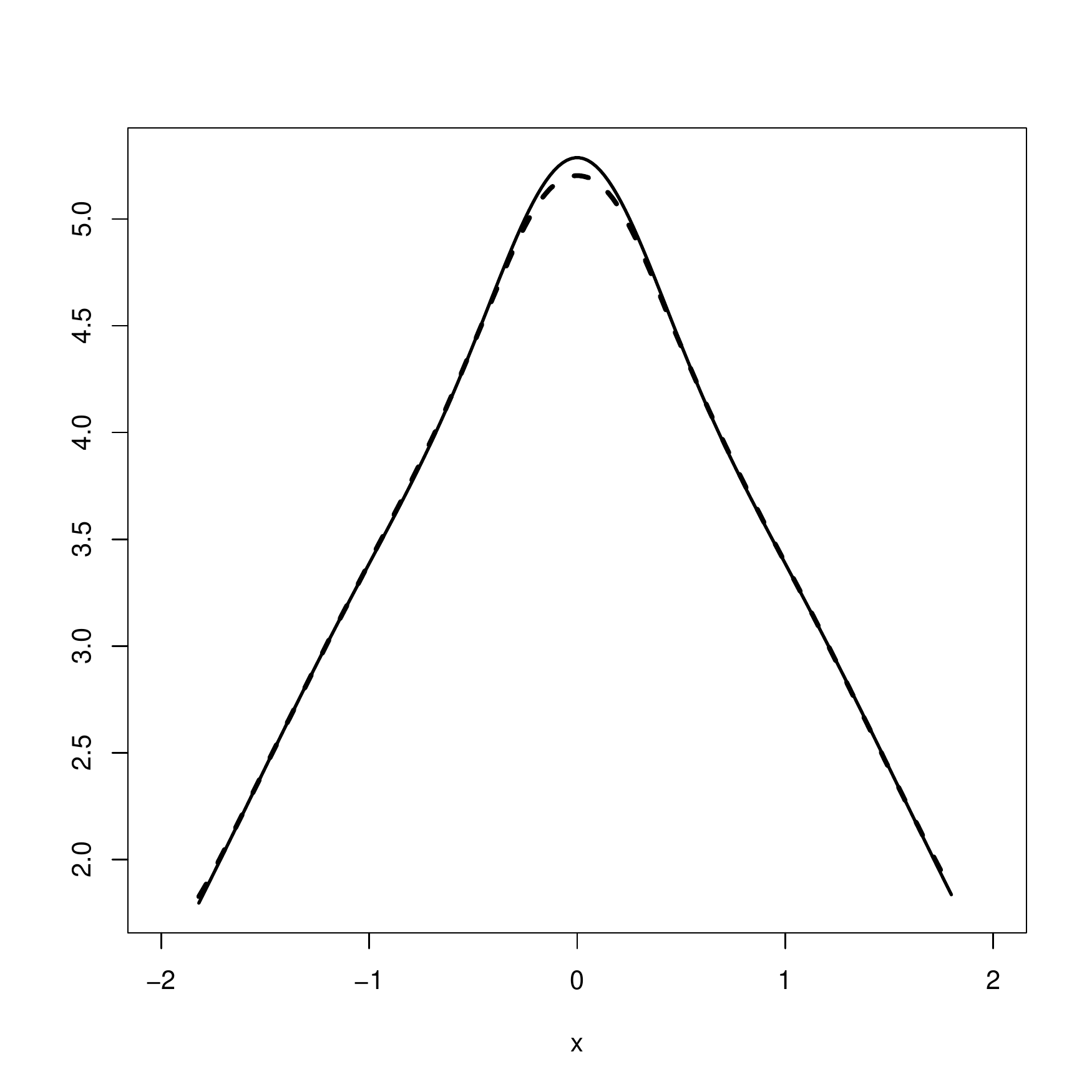}
\includegraphics[width = 5.8cm]{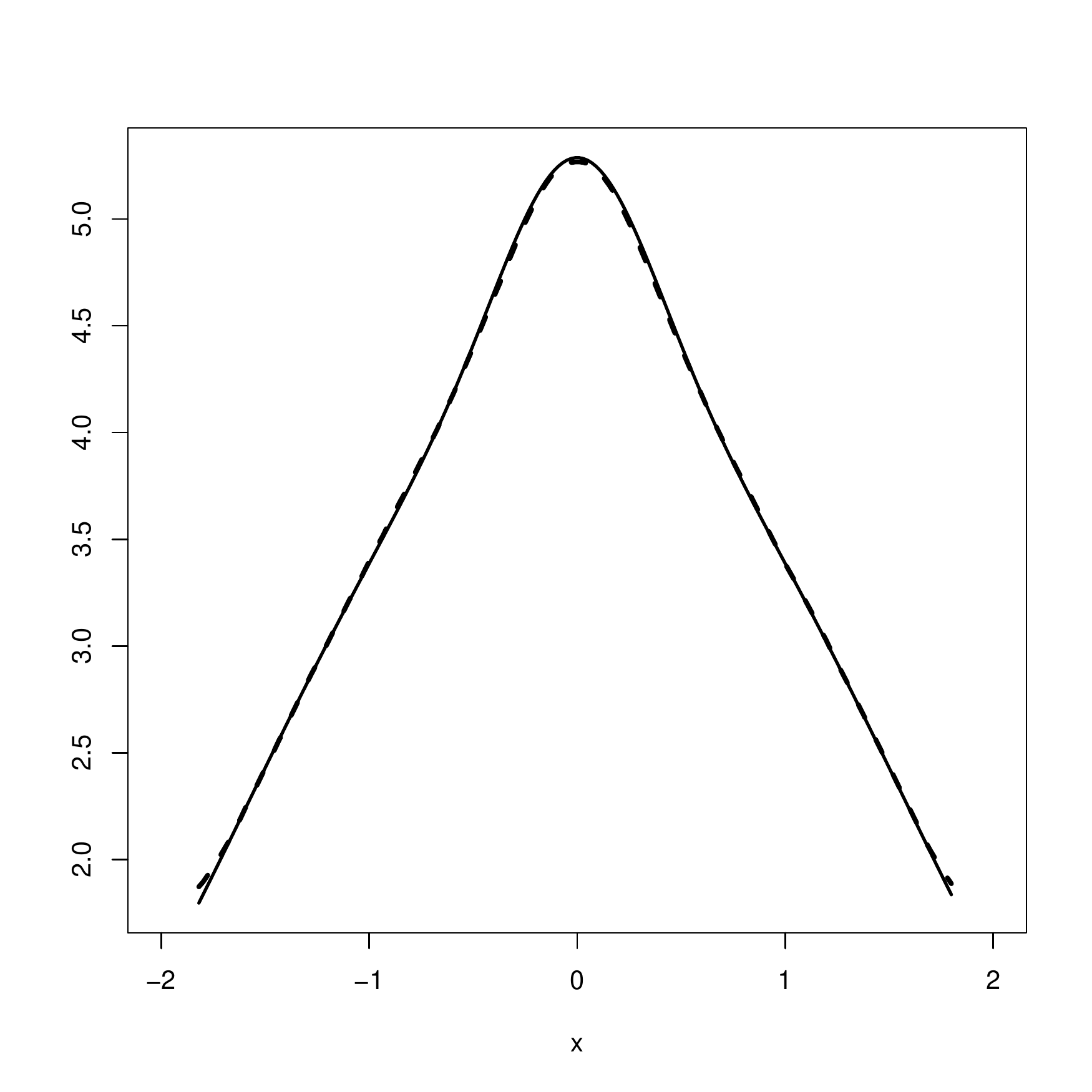}
\includegraphics[width = 5.8cm]{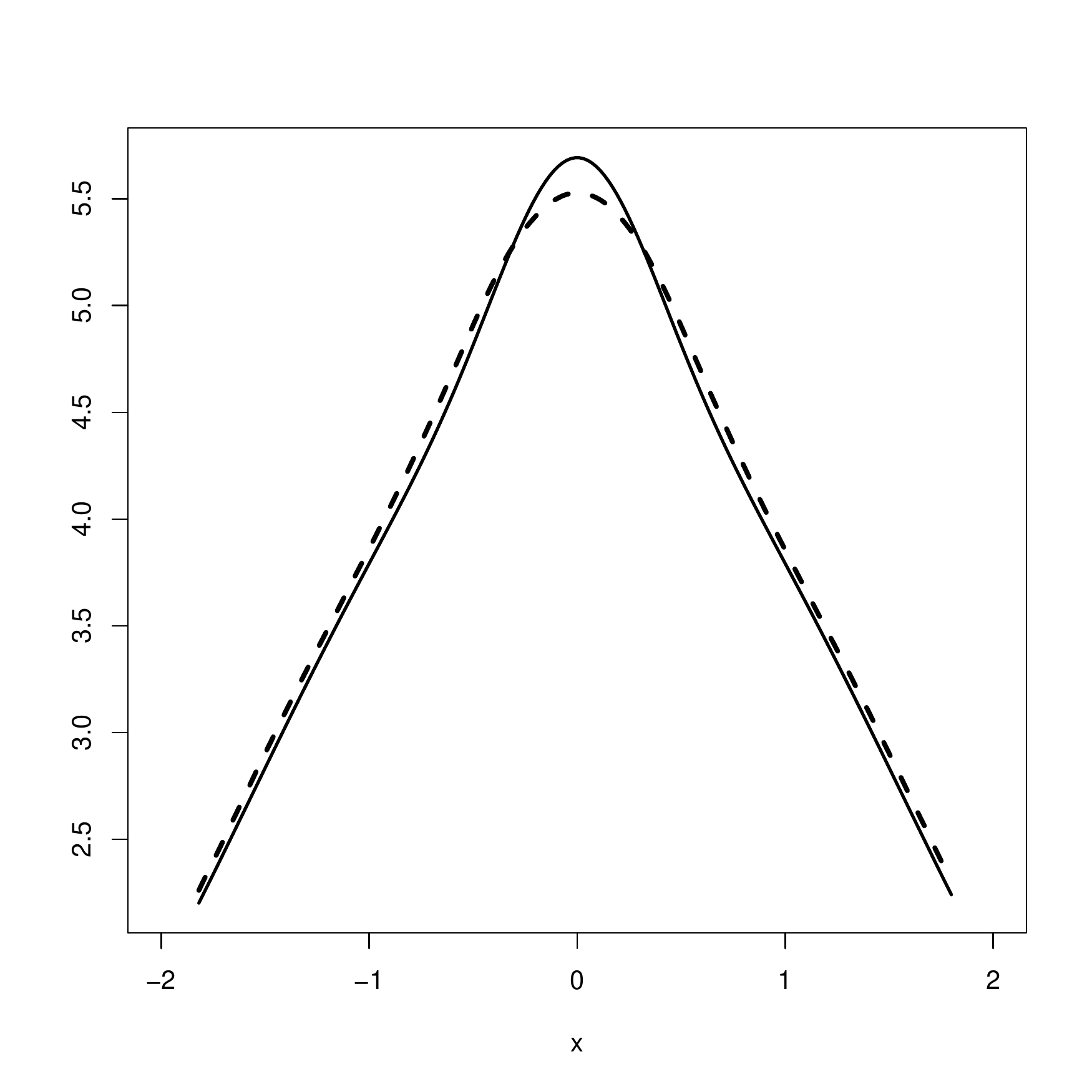}
\includegraphics[width = 5.8cm]{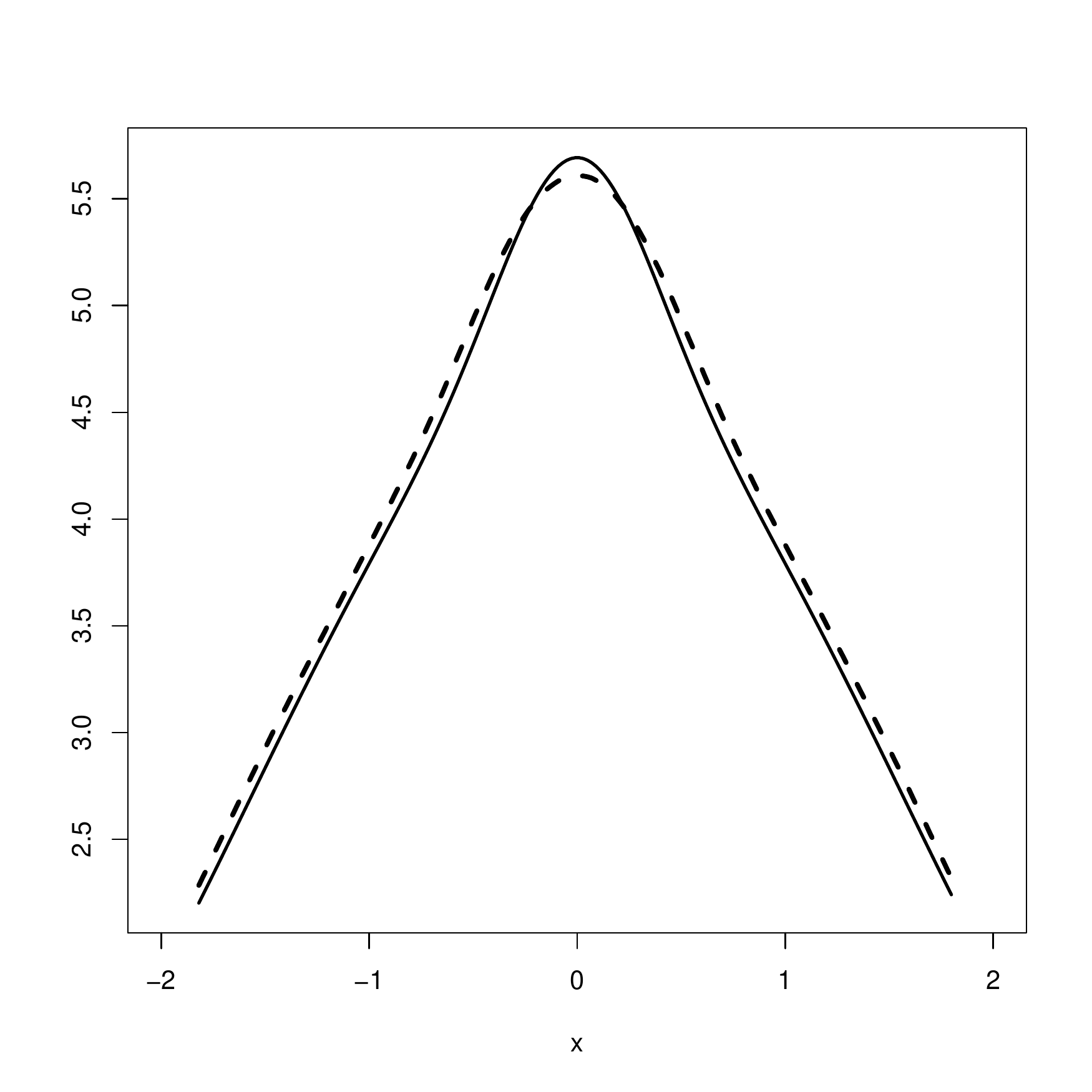}
\includegraphics[width = 5.8cm]{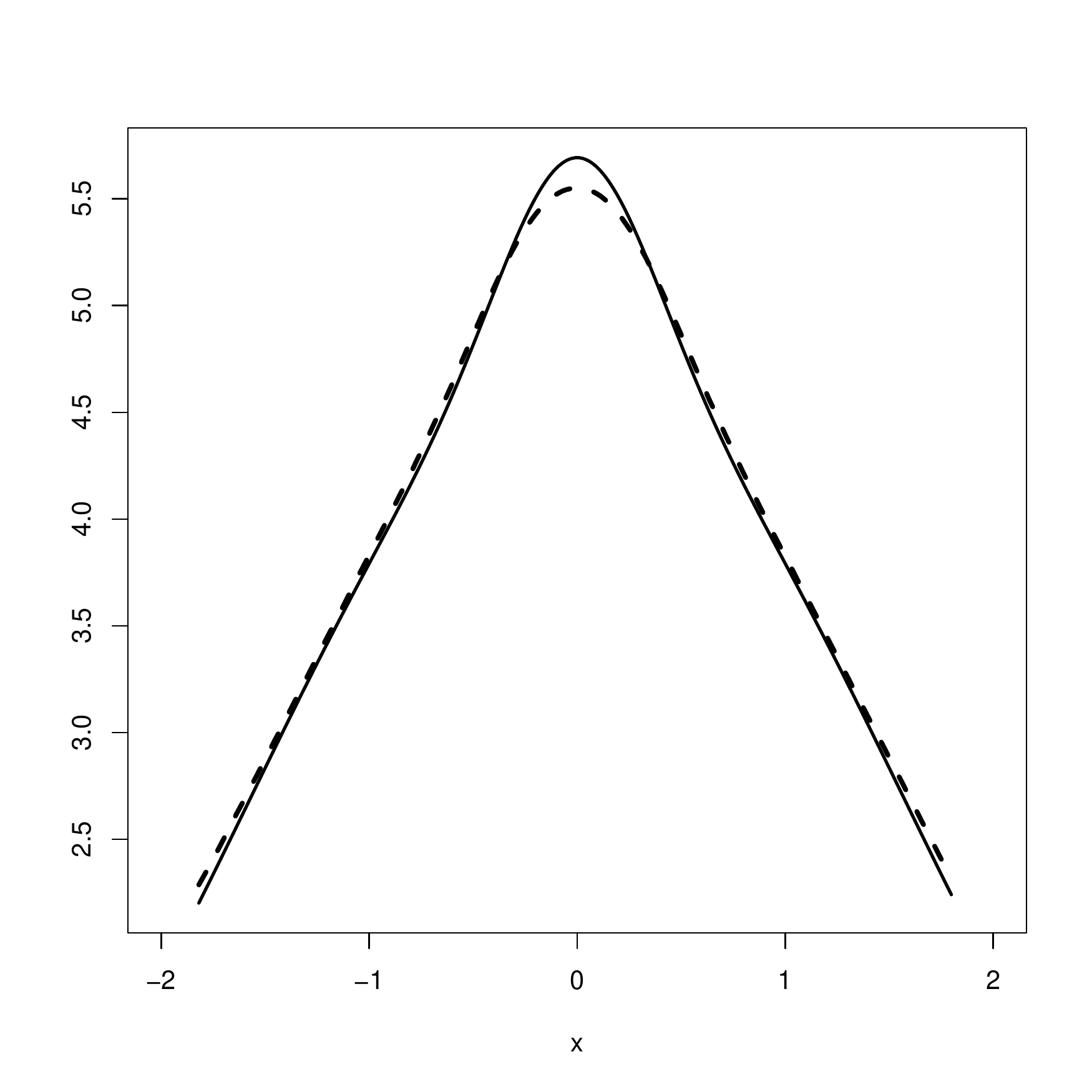}
\includegraphics[width = 5.8cm]{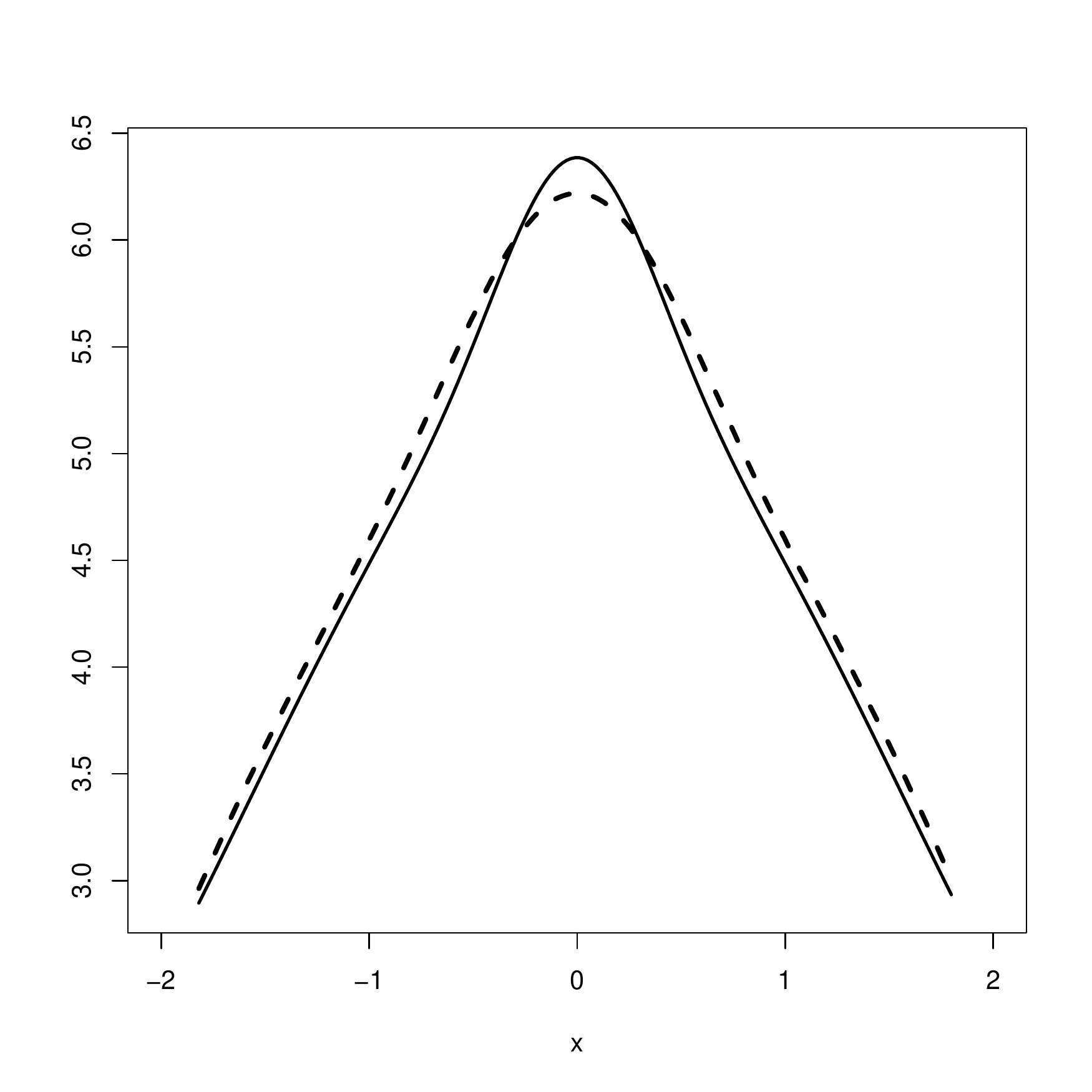}
\includegraphics[width = 5.8cm]{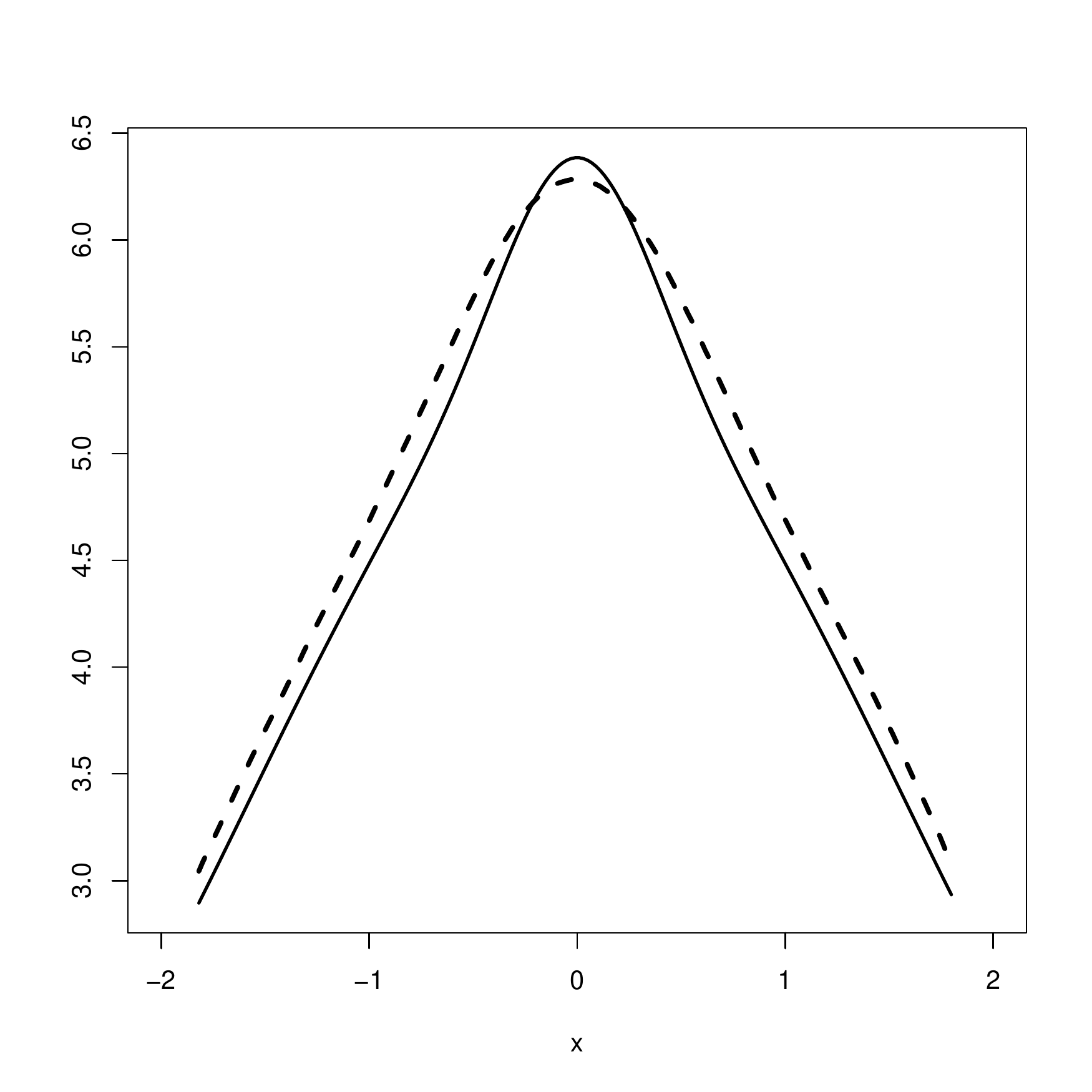}
\includegraphics[width = 5.8cm]{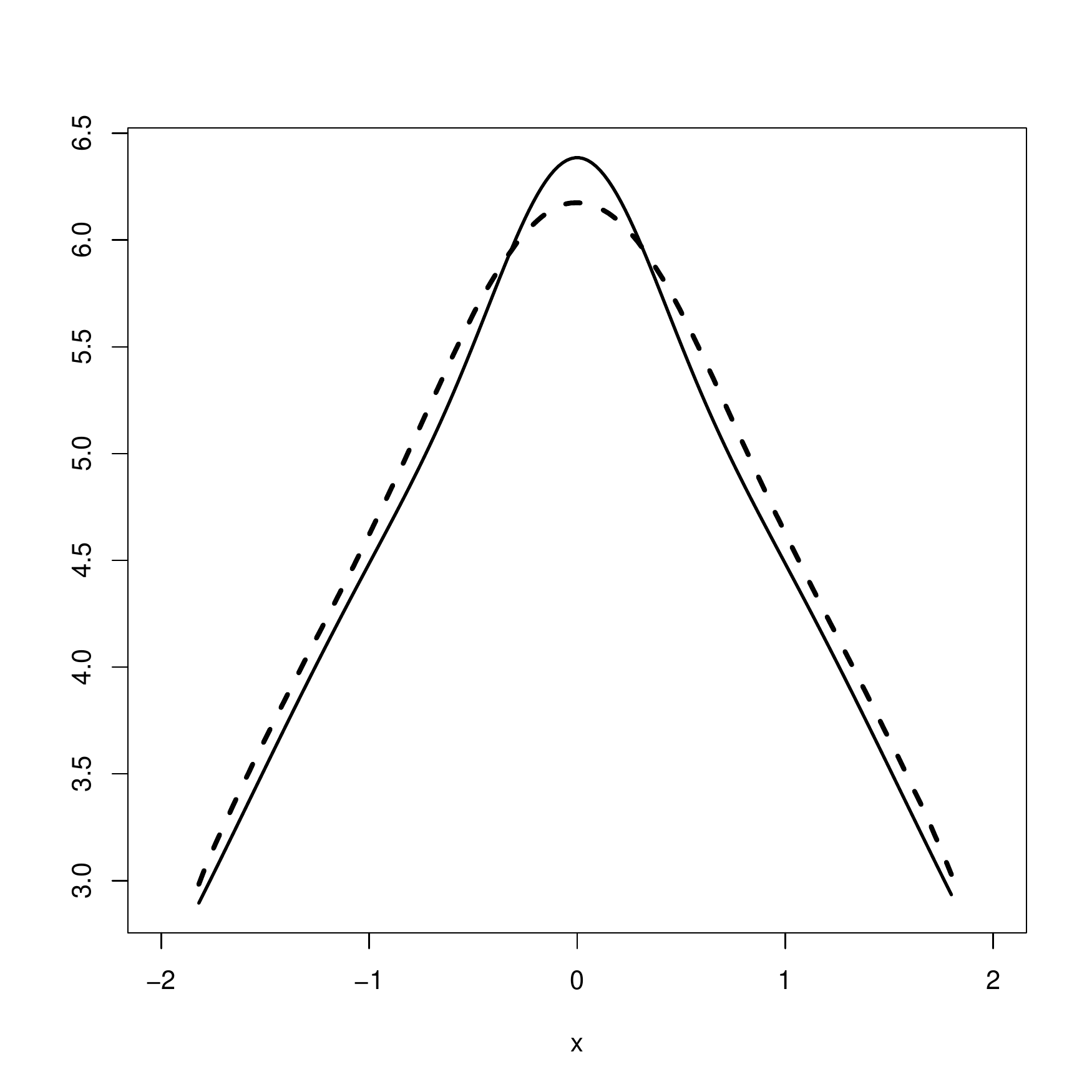}
\caption{\label{figm33mc}
\it Mean (dashed lines) and true (solid lines) quantile curves for model 2 and different
censoring: left column:  $(10\%,11\%)$
 censoring; middle column: $(30\%,11\%)$ censoring; right column: $(11\%,25\%)$ censoring.
 Upper row: $25\%$ quantile curves; middle row: $50\%$ quantile curves;
lower row: $75\%$ quantile curves. The sample sizes is $250$.}
 \end{figure}

\newpage

{\bf Acknowledgements.} The authors are grateful to Martina Stein  who typed parts of this paper with
considerable technical expertise.
This work has been supported in part by the Collaborative
Research Center ``Statistical modeling of nonlinear dynamic processes'' (SFB 823) of the German Research Foundation (DFG) and
in part by an NIH grant award
IR01GM072876:01A1.

\begin{appendix}
\section{Appendix: Proofs}
\def\theequation{A.\arabic{equation}}
\setcounter{equation}{0}
\textbf{\underline{Proof of Lemma \ref{le:g1}}}
We begin with the proof of the first part. Recalling the definition of the  Nadaraya-Watson weights in  (\ref{nw}), we see that \ref{W1a} follows easily from the inequality $c_1\leq K(x)\leq c_2$ for all $x$ in the support of $K$. Conditions \ref{W1b}   and \ref{W1c} hold  with $C(x)=f_X(x)$, which is a standard result from density estimation [see e.g. \cite{parzen1962}].\\
Finally, for assumption \ref{W1d}  we note that, as soon as the function $f_X(.)F_Y(t|.)$ is continuously differentiable in a neighborhood of $x$ with
uniformly (in $t$) bounded derivative, we have
\[
\sup_t \Bigl\|\frac{1}{nh^d}\E\Bigl[ \sum_i K_h(x-X_i)(x-X_i)\Ind{Y_i\leq t}\Bigr]\Bigr\| = O(h^2).
\]
From standard empirical process  arguments [see for example \cite{pollard1984}] we therefore
obtain
\[
\sup_t \frac{1}{nh^d}\Bigl\| \sum_i K_h(x-X_i)(x-X_i)\Ind{Y_i\leq t} - \E\Bigl[ \sum_i K_h(x-X_i)(x-X_i)\Ind{Y_i\leq t}\Bigr]\Bigr\|
= O\Bigl(\sqrt{\frac{h^2\log n}{nh^{d}}}\Bigr)
\]
a.s.\ and the  assertion now follows from condition \ref{B1}.
\\
To see that we can also use the local linear weights defined in (\ref{ll}), we note that
 \begin{eqnarray} \label{s0}
 S_{n,0} &=& f_X(x)(1+o_P(1))  \\
 \label{s1} S_{n,1} &=&  h^2\mu_2(K)f_X'(x) + o_P(h^2),\\
 \label{s2}
 S_{n,2} &=& h^2\mu_2(K)f_X(x) + o_P(h^2)
   \end{eqnarray}
and from the compactness of the support of $K$, which implies: $|x-X_j|=O(h)$ uniformly in $j$, we obtain the
representation $V_i^{LL} = V_i^{NW}(1+o_P(1))$ uniformly in $i$. Conditions \ref{W1a} and \ref{W1d} for the local linear follow from the
corresponding properties of
 the Nadaraya-Watson weights  (possibly with slightly smaller and larger constants
$\underline{c}$ and $\overline{c}$, respectively).

Finally, from the fact that, with probability tending to one, the local linear weights are positive, it follows that the corresponding estimators
$H_n,H_{ni}$ are increasing and hence unchanged by the rearrangement. This implies
$
\Pro{\exists i\in 1,...,n: W_i^{LL} \neq W_i^{LLI}} \stackrel{\nti}{\longrightarrow} 0,
$
where $W_i^{LLI}$ denote the weights of the rearranged local linear estimator. Thus condition (W1) also holds for the weights $W_i^{LLI}$ and the proof of the first part is complete.

For a proof of the second part of the Lemma we note that
  the same arguments as given in \cite{dabrowska1987}, Section 3.2,  yield   condition \ref{W2}  for the Nadaraya-Watson weights  [here we used assumptions \ref{D4}, \ref{D5} and \ref{B1}].

The corresponding result for the local linear weights can be derived by a closer examination of the weights $W_i^{LL}$. For the sake of
brevity, we will only consider the estimate  $H_n$ defined in (\ref{DS1}), the results for $H_{k,n}$ $(k=0,1,2)$ follow analogously. From the definition of the weights $W_i^{LL}$
we obtain the representation \bea H_n^{LL}(t|x) &=& \frac{1}{nh}\sum_{i=1}^n \frac{\Kern{}{x}{X_i}{h}\left(S_{n,2} - (x-X_i)S_{n,1}
\right)}{S_{n,2}S_{n,0}-S_{n,1}^2}\Ind{Y_i\leq t}
\\
&=& \frac{1}{nh}\sum_{i=1}^n \frac{\Kern{}{x}{X_i}{h}}{S_{n,0}}\frac{1}{1-S_{n,1}^2/(S_{n,0}S_{n,2})}\Ind{Y_i\leq t} - \frac{1}{nh}\sum_{i=1}^n
\frac{\Kern{}{x}{X_i}{h} (x-X_i)S_{n,1} }{S_{n,2}S_{n,0}-S_{n,1}^2}\Ind{Y_i\leq t}
\\
&=& H_n^{NW}(t|x) + O_P(h^2) \eea uniformly in $t$ where the last equality follows from the estimates $H_n^{NW}(t|x) = O_P(1)$ and (\ref{s0}) - (\ref{s2}).
Now condition \ref{B1} ensures $h^2 = o(1/\sqrt{nh})$ and thus the difference $H_n^{NW}-H_n^{LL}$ is
asymptotically negligible.
From Lemma \ref{le:pitto} we immediately obtain that, with probability tending to one, the rearranged estimators $H_n^{LLI}$ and $H_{i,n}^{LLI}$ defined in (\ref{ra1}) and (\ref{ra2}) coincide with the estimates $H_n^{LL}$ and $H_{i,n}^{LL}$ respectively.
Thus condition $(W2)$ also holds for $(H_n^{LLI},H_{0,n}^{LLI},H_{2,n}^{LLI})$ and the second part of Lemma \ref{le:g1} has been established.

We now turn to the proof of the last part. Again we only consider the process $H_n(.|x)$, and note that the uniform consistency of $H_{k,n}(.|x)$ follows analogously.
First, observe the estimate
\[
\E\Bigl[ \frac{1}{nh^d}\sum_i K_h(x-X_i)\Ind{Y_i\leq t}\Bigr] = \frac{1}{h^d}\int K_h(x-u)F_Y(t|u)f_X(u)du = f_X(x)F_Y(t|u)(1+o(1))
\]
uniformly in $t$, which is a consequence of condition \ref{D00}. From standard empirical process arguments [see \cite{pollard1984}] it
follows that almost surely
\[
\sup_t \Bigl|\frac{1}{nh^d}\sum_i K_h(x-X_i)\Ind{Y_i\leq t}-\E\Bigl[ \frac{1}{nh^d}\sum_i K_h(x-X_i)\Ind{Y_i\leq t}\Bigr] \Bigr| =
O\Bigl(\sqrt{\frac{\log n}{nh^d}}\Bigr),
\]
and with condition \ref{B2} the assertion for the Nadaraya-Watson weights follows. The extension of the result to local linear and rearranged
local linear weights can be established by the same arguments as presented in the second part of the proof.
\hfill$\Box$

\begin{remark} \rm
Before we begin with the proof of Theorem \ref{Th1:K}, we observe that condition \ref{W1} implies that we can write the weights $W_i(x)$ in the estimates (\ref{DS1}) in the form
\[
W_i(x) = W_i^{(1)}(x)I_{A_n} + W_i^{(2)}(x)I_{A_n^C},
\]
where $A_n$ is some event with $\Pro{A_n} \rightarrow 1$,  $W_i^{(1)}(x) = V_i(x)/\sum_j V_j(x)$ and
$W_i^{(2)}(x)$ denote some other weights. If we now define modified weights
\[
\tilde{W}_i(x) := W_i^{(1)}(x)I_{A_n} + W_i^{NW}(x)I_{A_n^C},
\]
where $W_i^{NW}(x)$ denote Nadaraya-Watson weights, we obtain: $\mbox{P}(\exists i\in 1,...,n: \tilde W_i \neq W_i)\rightarrow 0$, i.e. any estimator constructed with the weights $\tilde W_i(x)$ will have the same asymptotic properties as an estimator based on the original weights $W_i(x)$. Thus we may confine ourselves to the investigation of the asymptotic distribution of estimators constructed from the statistics in (\ref{DS1}) that are based on the weights $\tilde{W}_i(x)$. In order to keep the notation simple, the modified estimates are also denoted by $H_n, H_{k,n}, $ etc. Finally, observe that we have the representation $\tilde W_i(x) = \frac{\tilde V_i(x)}{\sum_j \tilde V_j(x)}$ with $\tilde V_i := V_iI_{A_n} + V_i^{NW}(x)I_{A_n^C}$. Note that by construction, the random variables $\tilde V_i$ satisfy conditions \ref{W1a}-\ref{W1d} if the kernel in the definition of $W_i^{NW}(x)$ satisfies assumption (K1).

\end{remark}

\textbf{\underline{Proof of Theorem \ref{Th1:K}}:}
Let $S$ denote the set of pairs of functions $( H_2(.|x),H(.|x))$ of bounded variation such that $ H(.|x)\geq\beta>0$. Since the map $(H_2(.|x),H(.|x))\mapsto M_2^-(.|x)$ is continuous on $S$ with respect to the supremum norm [see the discussion in \cite{andborgilkei1993}  following Proposition II.8.6], and $H_n$ is uniformly consistent
[which implies $\mbox{P}((H_{2,n},H_n) \in S ] \rightarrow 1$], the weak uniform consistency of $M_{2n}^-$ on $[t_{00}+\eps,\infty)$ [$\eps>0$
is arbitrary]
follows from the uniform consistency of $H_{2,n}$ and $H_n$.  This can be seen by similar arguments as given in \cite{dabrowska1987}, p.\ 184.

Moreover, the map $M_2^-(.|x) \mapsto F_L(.|x)$ is continuous on the set of functions of bounded variation [reverse time and use the discussion
in Andersen et.al. (1993)
following Proposition II.8.7], and thus the uniform consistency of $F_{L,n}(.|x)$ on $[t_{00}+\eps,\infty)$ follows for any positive $\eps > 0$.

In the next step, we consider the map
\[
(H_{0,n}(.|x),H_n(.|x),F_{L,n}(.|x)) \mapsto \Lambda_{T,n}(.|x) = \int_0^. \frac{H_{0,n}(dt|x)}{F_{L,n}(t-|x)-H_n(t-|x)}
\]
and split the range of integration into the intervals $[0,t_{00}+\eps)$ and $[t_{00}+\eps,t)$. The continuity of the integration and fraction
mappings yields the uniform convergence
\beq \label{B:Th1:K:1}
\sup_{t\in [t_{00}+\eps,\tau)} \left| \int_{[t_{00}+\eps,t)}\frac{H_{0,n}(dt|x)}{F_{L,n}(t-|x)-H_n(t-|x)} - \int_{[t_{00}+\eps,t)}\frac{H_{0}(dt|x)}{F_L(t-|x)-H(t-|x)}\right| \Pkonv 0
\eeq
for any $\tau$ with $F_S(\tau|x)<1$ [note that $\inf_{t\in[t_{00}+\eps,\tau)}F_L(t-|x)-H(t-|x)>0$ since $F_L(t-|x)-H(t-|x) = F_L(t-|x)(1 - F_S(t-|x))$ and $F_L(t_{00}-|x)>0$ by assumption \ref{D8} and continuity of the conditional distribution function $F_L(.|x)$]. We now will show that the integral over the interval $[0,t_{00}+\eps)$ can be made arbitrarily small by an appropriate choice of $\eps$. To this end, denote by $W_1(x,n),...,W_k(x,n)$ those values of $Y_1,...,Y_n$, whose weights fulfill $W_i(x)\neq 0$ and by $W_{(1)}(x,n),...,W_{(k)}(x,n)$ the corresponding increasingly ordered values.
By Lemma \ref{le:PFL:1} in Appendix B we can find an $\eps>0$ such that:
\[
\sup_{t_{00} +\eps \geq t\geq W_{(2)}(x,n)} \frac{1}{F_{L,n}(s-|x)- H_n(s-|x)} = O_P(1),
\]
and it follows
\[
\int_{[W_{(2)}(x,n),t_{00}+\eps)}\frac{H_{0,n}(ds|x)}{F_{L,n}(s-|x)- H_n(s-|x)} \leq H_{0,n}(t_{00}+\eps|x)O_P(1).
\]
Therefore it remains to find a bound for the integral $\int_{[0,W_{(2)}(x,n))}\frac{H_{0,n}(ds|x)}{F_{L,n}(s-|x)- H_n(s-|x)}$. For this purpose we consider two cases. The first one appears if the $\delta_i$ corresponding to $W_{(1)}(x,n)$ equals $0$. In this case there is positive mass at the point $W_{(1)}(x,n)$ but at the same time $F_{L,n}(s|x) = F_{L,n}(W_{(2)}(x,n)|x)$ for all $s \in [0,W_{(2)}(x,n))$ and hence
$\int_{[0,t_{00}+\eps)}\frac{H_{0,n}(ds|x)}{F_{L,n}(s-|x)- H_n(s-|x)} \leq H_{0,n}(t_{00}+\eps|x)O_P(1)$. For all other values of the corresponding
$\delta_i$ the mass of $H_{0,n}(ds|x)$ at the point $W_{(1)}(x,n)$ equals zero and thus the integral vanishes. Summarizing, we have obtained the estimate
\[
\int_{[0,t_{00}+\eps)}\frac{H_{0,n}(ds|x)}{F_{L,n}(s-|x)- H_n(s-|x)} \leq H_{0,n}(t_{00}+\eps|x)O_P(1) = H_{0}(t_{00}+\eps|x)O_P(1),
\]
where the last equality follows from the uniform consistency of $H_{0,n}$ and the remainder  $O_P(1)$ does not depend on $\eps$. Moreover, since the function $\Lambda_{T,n}(.|x)$ is increasing [see Lemma \ref{le:LambdaT}], the inequality
\beq \label{l1}
\sup_{t\leq t_{00}+\eps} |\Lambda_{T,n}(t|x)| = \int_{[0,t_{00}+\eps)}\frac{H_{0,n}(ds|x)}{F_{L,n}(s-|x)- H_n(s-|x)} \leq H_{0}(t_{00}+\eps|x)O_P(1)
\eeq
follows. Now for any $\delta>0$ we can choose an $\eps_\delta>0$ such that $H_{0}(t_{00}+\eps_\delta|x)<\delta$ [recall the definition of $t_{00}$ in (\ref{t0})] and we have
\bea
\Pro{ \sup_{t\in [0,t_{00}+\eps_\delta)} |\Lambda_{T,n}(t|x)-\Lambda_T(t|x)| > 2\alpha} \leq
\Pro{ \sup_{t\in [0,t_{00}+\eps_\delta)} |\Lambda_{T,n}(t|x)| > \alpha} \leq \Pro{O_P(1) > \alpha/\delta },
\eea
whenever $\Lambda_T(t_{00}+\eps|x)<\alpha$, where the last inequality follows from (\ref{l1}) and the remainder $O_P(1)$ does not depend on $\alpha$ and $\delta$. From this estimate we obtain for any $\tau$ with $F_S(\tau|x)<1$
\[
\Pro{ \sup_{t\in [0,\tau)} |\Lambda_{T,n}(t|x)-\Lambda_T(t|x)| > 4\alpha} \leq  \Pro{ \sup_{t\in [t_{00}+\eps_\delta,\tau)}
|\Lambda_{T,n}(t|x)-\Lambda_T(t|x)| > 2\alpha} + \Pro{O_P(1) > \alpha/\delta }.
\]
By (\ref{B:Th1:K:1}) The first probability on the right hand side of the inequality converges to zero as $n$ tends to infinity for any $\alpha,\eps_\delta>0$, and
the limit of the second one can be made arbitrarily small by choosing $\delta$ appropriately. Thus we   obtain   $\lim_{\nti}\Pro{ \sup_{t\in
[0,\tau)} |\Lambda_{T,n}(t|x)-\Lambda_T(t|x)| > 4\alpha}=0$, which implies the weak
uniform consistency of
$\Lambda_{T,n}(.|x)$ on the interval $[0,\tau)$.

Finally, the continuity of the mapping $\Lambda_T\mapsto F_T$ [see the discussion in \cite{andborgilkei1993}  following Proposition II.8.7] yields the weak
uniform consistency of the estimate $F_{T,n}$ and the first part of the theorem is established.

For a proof of the second part, we use an idea from \cite{wang1987}. Note that, as soon as $F_{T,n}(.|x)$ is increasing and bounded
by $1$ from above, we have the inequality $\sup_{t\geq a} |F_{T,n}(t|x)-F_T(t|x)| \leq |F_{T,n}(a|x)-F_T(a|x)| + (1-F_T(a|x))$. Thus
\[
\sup_{t\geq 0}\abs{F_{T,n}(t|x)-F_T(t|x)} \leq 2\sup_{0\leq t \leq a}\abs{F_{T,n}(t|x)-F_T(t|x)} + 2(1-F_T(a|x)),
\]
and   by assumption and part one of the theorem we can make $1-F_T(a|x)$ arbitrarily small with uniform consistency on the interval $[0,a]$ still holding.
Consequently, we obtain the uniform consistency on $[0,\infty)$, which completes the proof of Theorem \ref{Th1:K}. \hfill{$\Box$}\\
\\
\underline{\textbf{Proof of Theorem \ref{Th1}:}}
The second part follows from the first one by the Hadamard differentiability of the map $A \mapsto \prod_{(t,\infty]}(1-A(ds))$ in definition (\ref{es1}) [see \cite{patiroli2001}, Lemma A.1]  and the delta method [\cite{gill1989}].  Note that these results require a.s.\ continuity of the sample paths which follows from the fact that the process $G_M$ defined in the first part of the Theorem has a.s. continuous sample paths together with the continuity of $F_L(.|x)$.\\
The  proof will now proceed in two steps: first we will show that weak convergence holds in $D^3([\sigma,\infty])$ for any $\sigma>t_{00}$
and secondly we will extend this convergence to $D^3([t_{00},\infty]).$ Note that from condition \ref{D1} we obtain $F_L(t_{00}|x)>0$, and the continuity of $F_L(.|x)$ yields $t_{00}>0.$
\\
Set $\epsilon>0$ and choose $\sigma > t_{00}$ such that $H(\sigma|x)>\epsilon$. Recall that the map
\[
(H,H_0,H_2) \mapsto (H,H_0,M_2^-)
\]
is Hadamard differentiable on the domain $\tilde D :=\left\{ (A_1,A_2,A_3)\in BV_1^3([\sigma,\infty]): A_1\geq 0, A_3\geq\epsilon/2 \right\}$
[see \cite{patiroli2001}] and takes values in $BV_C^3([\sigma,\infty])$.
Here $BV_C$ denotes the space of functions of bounded variation with elements uniformly bounded by the constant $C$.
Moreover, assumption \ref{W2} implies weak convergence and weak uniform consistency of the estimator $H_n$ on $D([\sigma,\infty])$. Therefore $(H_{0,n},H_{2,n},H_n)$ will belong to the domain $\tilde D$ with probability tending to one if $n \to \infty$.
Hence, we can define the random variable $\bar H_n := I_{A_n}H_n + I_{A_n^C}$  where $A_n := \left\{ \inf_{t\in [\sigma,\infty]} H_n(t) \geq
\epsilon/2\right \}$, which certainly has the property $\bar H_n \geq \epsilon/2$ on $[\sigma,\infty]$ almost surely. Now, since $\mbox{P}(\bar
H_n\neq H_n] = 1 - \mbox{P}(A_n) \rightarrow 0$, the weak convergence result in \ref{W2} continues to hold on $D^3([\sigma,\infty])$ with $H_n$ replaced by $\bar
H_n$. By the same argument, we may replace the $H_n$ in the definition of $M_{2,n}^-$ by $\bar H_n$ without changing the asymptotics. Thus we can apply the delta method [see \cite{gill1989}, Theorem 3] to $(H_{0,n},H_{2,n},\bar H_n)$ and deduce the weak convergence
\[
\sqrt{nh^d}(H_n-H,H_{0,n}-H_0,M_{2,n}^- - M_2^-) \Rightarrow (G,G_0,G_{M_\sigma})\quad \rm{in} \quad D^3([\sigma,\infty]).
\]
To obtain the weak convergence in $D^3([t_{00},\infty])$, we apply a Lemma from Pollard (1984, page 70, Example 11). First define $G_M$ as the
pathwise limit of $G_{M_{\sigma}}(\sigma)$ for $\sigma\downarrow t_{00}$, the existence of this limit is discussed in Remark \ref{remGM}.
Note that there exist versions of $G_M, G, G_0$ with a.s.\ continuous paths (this holds for $G$ and $G_0$ by assumption, whereas the paths of
$G_M$ are obtained from those of $G_2, G$ by a transformation that preserves continuity [see equation ($\ref{dGM}$)]), and hence the condition on the
limit process in the Lemma is fulfilled.


Hereby we have obtained a Gaussian process $G_M$ on the interval $[t_{00},\infty]$ and have taken care of condition $(iii)$ in the Lemma in \cite{pollard1984}. For
arbitrary positive $\epsilon$ and $\delta$ we now have to find a $\sigma = \sigma(\delta,\epsilon)>t_{00}$ such that
\begin{eqnarray}
 && P\left(\sup_{t_{00}<t\leq\sigma} \left| G_M(t) \right| \geq \delta \right)<\epsilon \label{BTh1.1}  \\
 && \limsup_{\nti}P\left( \sup_{t_{00}<t\leq\sigma}\sqrt{nh^d}\left| (M_{2,n}^- - M_{2}^- )(\sigma-|x) - (M_{2,n}^- - M_{2}^- )(t-|x)
 \right| \geq \delta\right)<\epsilon. \label{BTh1.2}
\end{eqnarray}
Note that once we have found a $\sigma$ such that $(\ref{BTh1.2})$ holds, we can make $\sigma$ smaller until $(\ref{BTh1.1})$ is fulfilled
with $(\ref{BTh1.2})$ still holding. This is possible because for every $\delta>0$, we have \\
$\lim_{\sigma \downarrow t_{00}} P\left(\sup_{t_{00}<t\leq\sigma} \left| G_M(t) \right| \geq \delta \right) = 0$, which can be established as follows. Define the function $\kappa(t) := \int_t^\infty \frac{M_2^-(ds|x)}{H(s|x)}$ and denote by $W_t$ a Brownian motion on $[0,\infty]$. Then we have
\[
\mbox{Cov}(\sqrt{b(x)}W_{\kappa(s)},\sqrt{b(x)}W_{\kappa(t)}) = b(x)(\kappa(s)\wedge\kappa(t)) = b(x)\int_{s\vee t}^\infty \frac{M_2^-(ds|x)}{H(s|x)} = \mbox{Cov}(G_M(s),G_M(t)),
\]
where the last equality follows from Remark \ref{remGM}. Thus we have represented the process $G_M$ in terms of a Brownian motion and the assertion follows from the finiteness of $\kappa(t_{00})$ [by assumption \ref{D1}] and the properties of the Brownian motion.\\
In order to prove the existence of a constant $\sigma$ that ensures (\ref{BTh1.2}), we reverse time and transform our problem into the setting
of conditional right censorship [see Section \ref{s:beran}].
To be more precise, define the function $a(t) := \frac{1}{t}$ which is strictly decreasing and maps the interval $[0,\infty]$ onto itself. Consider the
random variables $B_i := a(S_i)$, $D_i := a(L_i)$, $Z_i := B_i\wedge D_i$ and $\Delta_i:= \Ind{D_i\leq B_i} = \Ind{S_i\leq L_i}$. This is a conditional
right censorship model with the useful property that $\Lambda_D^-(.|X_i)$, the predictable hazard function of $D_i$, is closely connected to the reverse hazard function $M_{2}^-(.|X_i)$ by the   identity
$$
\Lambda_D^-(a(t)|x) = M_{2}^-(\infty|x) - M_{2}^-(t-|x)
$$
It is easy to verify that the conditional Nelson-Aalen estimator $\Lambda_{D,n}^-(dt|x)$ in the new model is related to
the estimator $M_{2,n}^-$ in a similar way, i.e.\ $\Lambda_{D,n}^-(a(t)|x) = M_{2,n}^-(\infty|x) - M_{2,n}^-(t|x)$. Thus to prove $(\ref{BTh1.2})$ it suffices to
find a $\sigma$ such that in the new model the following inequality is fulfilled
\begin{equation}\label{help1}
\limsup_{\nti} P\left(\sup_{\sigma\leq t< t_0} \sqrt{nh^d}\left| (\Lambda_{D,n}^- - \Lambda_D^-)(t|x) - (\Lambda_{D,n}^- - \Lambda_D^-)(\sigma-|x) \right| > \delta \right) <
\epsilon,
\end{equation}
where we define $t_0 = a(t_{00}) < \infty.$ This assertion is established in the proof of Theorem \ref{satz:beran} [note that the assumptions \ref{1'}-\ref{6'} can be directly identified with the assumptions of Theorem \ref{Th1}]. \hfill $\Box$\\
\\



\underline{\textbf{Proof of Theorem \ref{Th2}:}}
First of all note that the a.s.\ continuity of the sample paths of the processes $V(.)$ and $W(.)$ follows because these processes are constructed from
processes which already have a.s. continuous sample paths in a way that preserves continuity. Thus it remains to verify the weak convergence.
From  Theorem \ref{Th1} we obtain
\beq
\sqrt{nh^d}(H_n-H,H_{0,n}-H_0,F_{L,n} - F_L) \Rightarrow (G,G_0,G_3) \label{Th2.1}
\eeq
in  $D^3([t_{00},\infty])$.
Now from $F_L(s-|x)-H(s-|x)=F_L(s-|x)(1-F_S(s-|x))$ and the definition of $\tau$ it follows that
\[
F_L(s-|x)-H(s-|x)\geq\epsilon>0  \quad \forall s\in[t_{00},\tau]
\]
[note that the inequality $F_L(t_{00}-|x)>0$ was derived at the beginning of the proof of Theorem $\ref{Th1}$]. For positive numbers $\delta$ define the event
\[
A_n(\delta) := \left\{ \inf_{t\in [t_{00},\tau)} (F_{L,n}(t|x) - H_n(t|x)) > \delta \right\}.
\]
Because of (\ref{Th2.1}) [which implies the uniform consistency of $F_{L,n}(.|x)$ and $H_n(.|x)$], we have that for
$\delta<\epsilon$ $P(I_{A_n(\delta)}\neq 1)\knti 0$. Define $\tilde H_n := H_nI_{A_n(\delta)}$, $\tilde H_{0,n} := H_{0,n}I_{A_n(\delta)}$ and
$\tilde F_{L,n} := F_{L,n}I_{A_n(\delta)}+I_{A^C_n(\delta)}$, then it follows from $(\ref{Th2.1})$
\[
\sqrt{nh^d}(\tilde F_{L,n} - F_L - (\tilde H_n-H),\tilde H_{0,n}-H_0) \Rightarrow (G_3 - G,G_0) \quad \mbox{in} \quad D^3([t_{00},\tau])
\]
Moreover, the pair $(\tilde H_{0,n},\tilde F_{L,n} - \tilde H_n)$ is an element of $\left\{(A,B)\in BV_1^2([t_{00},\tau]): A\geq 0, B\geq\delta>0
\right\}$. Since the map $(A,B) \mapsto \int_{t_{00}}^t \frac{dA(s)}{B(s)}$ is Hadamard differentiable on this set [see \cite{andborgilkei1993},page 113], the
delta method [see \cite{gill1989}] yields
\[
\sqrt{nh^d} \left(\int_{t_{00}}^.\frac{H_{0,n}(ds|x)}{F_{L,n}(s-|x) - H_n(s-|x)} - \Lambda_T^-(.|x) \right) \Rightarrow V(.)
\]
in $D([t_{00},\tau]]$. Finally, observe that for $t\geq t_{00}$ we have
\[
\Lambda_{T,n}^- (t|x) = \int_{t_{00}}^t\frac{H_{0,n}(ds|x)}{F_{L,n}(s-|x)- H_n(s-|x)} + \int_{[0,t_{00})}\frac{H_{0,n}(ds|x)}{F_{L,n}(s-|x)-
H_n(s-|x)},
\]
and thus it remains to prove that the second term in this sum is of order $o_P(1/\sqrt{nh^d})$. From Lemma \ref{le:PFL:1} in the Appendix B we obtain the bound: $\sup_{t_{00} \geq t\geq W_{(2)}(x,n)} \frac{1}{F_{L,n}(s-|x)- H_n(s-|x)} = O_P(1)$, where $W_{(2)}(x,n)$ is defined in the proof of theorem \ref{Th1:K}, and it follows
\[
\int_{[W_{(2)}(x,n),t_{00})}\frac{H_{0,n}(ds|x)}{F_{L,n}(s-|x)- H_n(s-|x)} \leq H_{0,n}(t_{00}|x)O_P(1).
\]
Standard arguments yield  the estimate $H_{0,n}(t_{00}|x)=o_P(1/\sqrt{nh^d})$ and thus it remains to derive an estimate for  the integral
$\int_{[0,W_{(2)}(x,n))}\frac{H_{0,n}(ds|x)}{F_{L,n}(s-|x)- H_n(s-|x)}$. For this purpose we consider two cases. The first one
appears if the $\delta_i$ corresponding to $W_{(1)}(x,n)$ equals $0$. In this case there is positive mass at the point $W_{(1)}(x,n)$ but at the
same time $F_{L,n}(s|x) = F_{L,n}(W_{(2)}(x,n)|x)$ for all $s \in [0,W_{(2)}(x,n))$ and hence
$\int_{[0,t_{00})}\frac{H_{0,n}(ds|x)}{F_{L,n}(s-|x)- H_n(s-|x)} \leq H_{0,n}(t_{00}|x)O_P(1)$. For all other values of the corresponding
$\delta_i$ the mass of $H_{0,n}(ds|x)$ at the point $W_{(1)}(x,n)$ equals zero and thus the integral vanishes. Now the proof of the theorem is complete.
\hfill$\Box$
\\
\\
\textbf{\underline{Proof of Theorem \ref{Th2:K}}:}
Note that the estimator $F_{T,n}^{IP}(.|x)$ is nondecreasing by construction. The assertion for $\hat q^{IP}(.|x)$ now follows from the Hadamard differetiability
of the inversion mapping tangentially to the space of continuous functions [see Proposition 1 in \cite{gill1989}], the continuity of $F_T(.|x)$ and the weak uniform
consistency of $F_{T,n}^{IP}(.|x)$ on the interval $[0,\tau]$.
The corresponding result for the estimator $\hat q(.|x)$ follows from the convergence $\Pro{\hat q^{IP}(.|x)\equiv\hat q(.|x)}\rightarrow 1$ [see the
discussion after Lemma \ref{le:pitto}].
\hfill$\Box$ \\
\\
\underline{\textbf{Proof of Theorem \ref{Th3}}}:
Observe that the estimator $F_{T,n}^{IP}(.|x)$ is nondecreasing by construction and that  Theorem \ref{Th2} yields
 $\sqrt{nh^d}(F_{T,n}^{IP}(.|x) - F^T(.|x)) \Rightarrow W(.)$  on $D([0,\tau+\alpha])$ for some $\alpha>0$ where the process $W$ has a.s.\ continuous sample
paths. Note    that the convergence holds on $D([0,\tau+\alpha])$. This follows from the continuity of
$F_S(.|x)$ and $F_T^{-1}(.|x)$ at $\tau$ which implies $F_S(F_T^{-1}(\tau+\alpha|x)|x)<1$ for some $\alpha>0$. By the same arguments $f_T(.|x) \geq \delta > 0$ on the interval $[\eps-\alpha,\tau+\alpha]$ if we choose $\alpha$ sufficiently small. Thus Proposition 1 from \cite{gill1989}
 together with the
delta method   yield  the weak convergence
of the process for $\hat q^{IP}(.|x) $.
The corresponding result for $\hat q(.|x)$ follows from the fact that $\Pro{\hat q^{IP}(.|x) \equiv \hat q(.|x)}\rightarrow 1$. \hfill$\Box$ \\
\\

\underline{\textbf{Proof of Theorem \ref{satz:beran}}}:
By the delta method [\cite{gill1989}], formula (\ref{be:deffdach}), and the Hadamard differentiability of the product-limit mapping [\cite{andborgilkei1993}] it suffices to verify the weak convergence of $\sqrt{nh^d}(\Lambda_{D,n}^-(t|x) - \Lambda_D^-(t|x))_t$ on $D([0,t_0])$. The corresponding result on $D([0,\tau])$ with $\tau<t_0$ follows from the delta method and the Hadamard differentiability of the mapping $(\pi_{0,n},F_{Z,n})\mapsto \Lambda_{D,n}^-$. For the extension of the converegnce to $D([0,t_0])$ it suffices to establish condition (\ref{help1}) [this follows by arguments similar to those in the proof of Theorem \ref{Th1}].
Define the random variable $U$ as the largest $Z_i$ corresponding nonvanishing weight $\tilde W_i(x)$
i.e.
\[
U = U(x) := \max\left\{Z_i : \tilde W_i(x)\neq 0 \right\}.
\]
Note that for $t\geq U$ we have $F_{Z,n}(t|x) = 1$ for the corresponding estimate of $F_Z(.|x)$. We write
\beqohne \label{lamn}
\Lambda_{D,n}^-(y-|x) &=&
\su \int_{[0,y)}  \frac{d\left(\tilde W_i(x)\Ind{Z_i\leq t,
\Delta_i=1}\right)}{\sum_{j=1}^n \tilde W_j(x)\Ind{Z_j\geq t} }\\
&=& \su \int_{[0,y)} \frac{\tilde W_i(x)\Ind{Z_i\geq t}d\left(\Ind{Z_i\leq t, \Delta_i=1}\right)}{\sum_{j=1}^n \tilde W_j(x)\Ind{Z_j\geq t} }
\nonumber \\
&=& \su \int_{[0,y)}  C_i(x,t)\Ind{1-F_{Z,n}(t-|x)>0}dN_i(t)
\nonumber
\eeqohne
for the plug-in estimator of $\Lambda_D^-(.|x)$, where
\[
C_i(x,t) := \frac{\tilde W_i(x)\Ind{Z_i\geq t}}{\sum_{j=1}^n \tilde W_j(x)\Ind{Z_j\geq t}} = \frac{\tilde V_i(x)\Ind{Z_i\geq t}}{\sum_{j=1}^n
\tilde V_j(x)\Ind{Z_j\geq t}},
\]
and the quantity $N_i(t)$ is defined as $N_i(t) := \Ind{Z_i\leq t, \Delta_i=1}$.
In what follows, we will use the notation $G(A) = \int_A G(du)$ for a distribution function $G$ and a Borel set $A$.
With the definition
\[
\hat \Lambda_{D,n}^-(y-|x) := \su \int_{[0,y)} C_i(x,t)\Ind{1-F_{Z,n}(t-|x)>0}\Lambda^-_D(dt|X_i)
\]
we obtain the decomposition
\begin{eqnarray*}
|(\Lambda_{D,n}^- - \Lambda_D^-)((\sigma,t]|x)| &\leq & |(\Lambda_{D,n}^- - \hat
\Lambda_{D,n}^-)
((\sigma,U\wedge t]|x)| + |( \Lambda_{D,n}^- - \hat \Lambda_{D,n}^-)((U\wedge t,t]|x)|  \\
&+&
|( \hat \Lambda_{D,n} - \Lambda_D^-)
((\sigma,t]|x)|.
\end{eqnarray*}
Observing  that $\Lambda_{D,n}^-((U\wedge t,t]) = \hat \Lambda_{D,n}^-((U\wedge t,t])= 0$ it follows that
\bea
|( \Lambda_{D,n}^- - \hat \Lambda_{D,n}^-)((U\wedge t,t]|x)| &=& 0,
\\
|( \hat \Lambda_{D,n}^- - \Lambda_D^-)((\sigma,t]|x)| &\leq& |(\hat\Lambda_{D,n}^- - \Lambda_D^-)((\sigma,U\wedge t]|x)| + \Lambda_D^-((U\wedge t,t]|x),
\\
\sup_{\sigma \leq t < t_0}|(\hat \Lambda_{D,n}^- - \Lambda_D^-)((\sigma, t\wedge U]|x)| &\leq &\sup_{\sigma \leq t \leq U\wedge t_0}|(\hat \Lambda_{D,n}^- -
\Lambda_D^-)((\sigma, t]|x)|
\eea
where we set the supremum over the empty set to zero. Hence  assertion  (\ref{help1}) can be obtained from the statements
\begin{eqnarray}
\label{(A)}  &&\sqrt{nh^d} \sup_{\sigma \leq t < t_0} \Lambda_D^-((U\wedge t,t]|x) \pn 0 \\
\label{(B)} && \sqrt{nh^d} \sup_{\sigma \leq t \leq U\wedge t_0}|(\hat \Lambda_{D,n}^- - \Lambda_D^-)((\sigma, t]|x)| \pn 0 \\
\label{(C)}  &&\limsup_{\nti} P\left(\sqrt{nh^d}\sup_{\sigma\leq t < U\wedge t_0}|(\Lambda_{D,n}^- - \hat \Lambda_{D,n}^-)((\sigma,U\wedge t]|x)|>\delta \right) < \epsilon/2,
\end{eqnarray}
which will be shown separately.

\medskip


\underline{\textit{Proof of (\ref{(A)})}}
For a proof of (\ref{(A)}) note that
\[
\Lambda_D^-((U\wedge t,t]|x) = \left\{
\begin{array}{lcl}
0 &,& U \geq t \\
\Lambda_D^-((U,t]|x) &,& U< t
\end{array}
\right.
\]
and $\Lambda_D^-((U,t]|x)\leq \Lambda_D^-((U\wedge t_0,t_0]|x)$ whenever $U<t\leq t_0$. Hence, the supremum in (\ref{(A)}) can be bounded by
\[
(A) \quad\quad \quad \sup_{\sigma\leq t< t_0} \Lambda_D^-((U\wedge t,t]|x) \leq \Lambda_D^-((U\wedge t_0,t_0]|x). \hfill
\]

Observing \ref{1'} we have $F_D([t_0,\infty]|x)>0$ [note that $\Lambda_D^-(dt|x) = \frac{F_D(dt|x)}{1-F_D(t-|x)}$] and obtain
\[
(B) \quad \quad\quad\Lambda_D^-((U\wedge t_0,t_0]|x) \leq \int_{(U\wedge t_0,t_0]}\frac{F_D(dt|x)}{F_D([t_0,\infty]|x)} = \frac{F_D((U\wedge
t_0,t_0]|x)}{F_D([t_0,\infty]|x)}.
\]
Observing (A) and (B) it suffices to verify the convergence $\sqrt{nh^d}F_D((U\wedge t_0,t_0]|x)\pn 0$.
For this purpose we introduce the notation
\[
u_n^{\alpha} = u_n^{\alpha}(x) := \inf\left\{ s:\sqrt{nh^d}F_D((s,t_0]|x)\leq \alpha  \right\}
\]
[note that $u_n^{\alpha}\leq t_0$]. 
Then we obtain for any fixed $\alpha>0$ and sufficiently  large $n$
\bea
P\left( \sqrt{nh^d}F_D((U\wedge t_0,t_0]|x) > \alpha \right) &\leq&  \E\left[\Ind{U\wedge t_0<u_n^{\alpha}} \right]
=\E\left[\E\left[\Ind{U\wedge t_0<u_n^{\alpha}}\middle|X_1,...,X_n\right] \right]
\\
&\leq& \E\Bigl[\E\Bigl[\prod_{j=1}^n\left\{1-\Ind{Z_j\geq u_n^{\alpha}}\Ind{\tilde W_i(x)\neq 0}\right\}\Bigl|X_1,...,X_n\Bigr] \Bigr]
\\
&\leq& \E\Bigl[\prod_{j=1}^n\left\{1-\E\left[\Ind{Z_j\geq u_n^{\alpha}}\middle|X_j\right]\Ind{\|X_j-x\|\leq c_n}\right\} \Bigr]
\\
&=& \E\Bigl[\prod_{j=1}^n\left\{1-F_Z([u_n^{\alpha},\infty]|X_j)\Ind{\|X_j-x\|\leq c_n} \right\} \Bigr]
\\
&\leq& \E\Bigl[\prod_{j=1}^n\left\{1-F_Z([u_n^{\alpha},\infty]|X_j)\Ind{X_j\in U_{c_n}(x)\cap I} \right\} \Bigl]
\\
&\stackrel{(*)}{\leq}& \E\Bigl[\prod_{j=1}^n\left\{1-CF_Z([u_n^{\alpha},\infty]|x)\Ind{X_j\in U_{c_n}(x)\cap I} \right\} \Bigl]
\\
&=& \prod_{j=1}^n\left\{1-CF_D([u_n^{\alpha},\infty]|x)F_B([u_n^{\alpha},\infty]|x)\E\left[\Ind{X_j\in U_{c_n}(x)\cap I}\right] \right\}
\\
&\leq& \prod_{j=1}^n\left\{1-CF_D([u_n^{\alpha},t_0)|x)F_B([u_n^{\alpha},\infty]|x)\E\left[\Ind{X_j\in U_{c_n}(x)\cap I}\right] \right\}
\\
&\stackrel{(**)}{\leq}& \prod_{j=1}^n\left\{1-CF_D([u_n^{\alpha},t_0)|x)F_B([u_n^{\alpha},\infty]|x)ch^dO(1) \right\}
\\
&\leq& \prod_{j=1}^n\left\{1-C\frac{\alpha^2}{nh^d}\frac{F_B([u_n^{\alpha},\infty]|x)}{F_D([u_n^{\alpha},t_0)|x)}ch^dO(1) 
\right\}
\\
&=& \Bigl(1-C\frac{\alpha^2}{n}\frac{F_B([u_n^{\alpha},\infty]|x)}{F_D([u_n^{\alpha},t_0)|x)}cO(1) \Bigl)^n,
\eea
where the inequalities $(*),(**)$ follow from \ref{4'}, the last inequality follows from the definition of $u_n^{\alpha}$ and the $O(1)$ is independent of $j$ [it comes from the ratio $c/h$].
Now we have
\bea
\frac{F_D([u_n^{\alpha},t_0)|x)}{F_B([u_n^{\alpha},\infty]|x)}
&\leq& \int_{[u_n^{\alpha},t_0)}\frac{F_D(ds|x)}{F_B((s,\infty]|x)}
\leq \int_{[u_n^{\alpha},t_0)}\frac{F_D(ds|x)}{F_B((s,\infty]|x)F_D((s,\infty]|x)F_D([s,\infty]|x)}
\\
&=& \int_{[u_n^{\alpha},t_0)}\frac{\Lambda_D^-(ds|x)}{F_Z((s,\infty]|x)} \longrightarrow 0,
\eea
by \ref{1'} [note that $u_n^{\alpha} \rightarrow t_0$ if $\nti$] and hence the proof of (\ref{(A)}) is complete.  \\
\\
\underline{\textit{Proof of (\ref{(B)})}}
For fixed $\sigma\leq s \leq U\wedge t_0$ and sufficiently  small $h$  we have
\bea
&&|(\hat \Lambda_{D,n}^- - \Lambda_D^-)((\sigma,s]|x)|
= \Bigl| \int_{\sigma}^{s}\su C_i(x,t)(\lambda_D(t|X_i)-\lambda_D(t|x))dt \Bigr|
\\
&=& \Bigl| \int_{\sigma}^{s}\su C_i(x,t)\left((x-X_i)'\partial_x\lambda_D(t|x)
+ \frac{1}{2}(x-X_i)'\partial_x^2\lambda_D(t|\xi_i)(x-X_i)\right)dt \Bigr|
\\
&\leq& \Bigl| \int_{\sigma}^{s}\su C_i(x,t)(x-X_i)'\partial_x\lambda_D(t|x)dt\Big|
+ \int_{\sigma}^{s}\su C_i(x,t)\|x-X_i\|^2\frac{C}{2}dt,
\eea
with some positive constant $C$, where we used \ref{3'} in the last inequality. The second term in the above inequality can be bounded as follows
\[
\frac{C}{2}\int_{\sigma}^{s}\su C_i(x,t)\|x-X_i\|^2dt
\leq \frac{C}{2}\int_{\sigma}^{s}\su C_i(x,t)O(h^2)dt
\leq   \frac{C}{2}(t_0-\sigma)O(h^2)
= O(h^2) = o\left(\frac{1}{\sqrt{nh^d}}\right),
\]
where the last inequality holds uniformly in $s \in [\sigma, t_0]$. Thus it remains to consider the first term, which can be represented as follows
\bea
R_n &:=&
\Bigl| \int_{\sigma}^{s} \frac{\su\tilde V_i(x)\Ind{Z_i\geq t}(x-X_i)'}{\sum_{j=1}^n\frac{\tilde V_j(x)}{\sum_{k=1}^n\tilde
V_k(x)}\Ind{Z_j\geq t}}\frac{1}{\sum_{k=1}^n\tilde V_k(x)}\partial_x\lambda_D(t|x)dt\Bigr|
\\
&=& \Bigl| \frac{1}{\sum_{k=1}^n\tilde V_k(x)}\int_{\sigma}^{s} \su\tilde V_i(x)\Ind{Z_i\geq
t}(x-X_i)'\left(\frac{1-F_Z(t-|x)}{1-F_{Z,n}(t-|x)}\right)\frac{\partial_x\lambda_D(t|x)}{1-F_Z(t-|x)}dt\Bigr|.
\eea
Now, from condition \ref{W1c} and  \ref{W1d} $\frac{1}{\sum_{k=1}^n\tilde V_k(x)}=O_P(1)$, $\Big\|\su\tilde V_i(x)\Ind{Z_i\geq t}(x-X_i)\Big\| = o_P(1/\sqrt{nh^d})$
uniformly in $t \in (\sigma,U\wedge t_0)$, \ref{2'} and $\frac{1-F_Z(t-x)}{1-F_{Z,n}(t-|x)} = O_P(1)$ uniformly in $t \in (\sigma,U\wedge t_0)$ [see Lemma $\ref{le:1}$ in the Appendix B]  we obtain
\bea
R_n &=& o_P(1/\sqrt{nh^d})\Big\|\int_{\sigma}^{s} \frac{\partial_x\lambda_D(t|x)}{1-F_Z(t-|x)}dt\Big\| \leq o_P(1/\sqrt{nh^d})\int_{\sigma}^{t_0}
\frac{\|\partial_x\lambda_D(t|x)\|}{1-F_Z(t-|x)}dt = o_P(1/\sqrt{nh^d})
\eea
uniformly in $s \in [\sigma, t_0]$, and hence assertion (\ref{(B)}) is
established.
\\
\\
\underline{\textit{Proof of (\ref{(C)})}}
Observe that
$
|(\Lambda_{D,n}^- - \hat \Lambda_{D,n}^-)((\sigma,U\wedge t_0]|x)| \leq \left|D_1(U\wedge t_0) - D_1(\sigma)  \right|,
$
where we have used the notation $M_i(t) := N_i(t) - \int_0^t \Ind{Z_i\geq s}\Lambda_D^-(ds|X_i)$ and
\beq \label{d1}
D_1(t) := \su\int_{[0,t]} C_i(x,t)\Ind{1-F_{Z,n}(t-|x)>0}dM_i(t).
\eeq
Define $\mathcal{F}_t := \sigma(X_i,\Ind{Z_i\leq t,\Delta_i=1},\Ind{Z_i\leq t,\Delta_i=0}:i=1,...,n)$ and note that  $M_i$ are independent locally
bounded martingales with respect to $(\mathcal{F}_t)_t$  [see Theorem 2.3.2 p. 61 in
\cite{flemharr1991}]. Moreover, $\Ind{1-F_{Z,n}(t-|x)>0}$, $\Ind{Z_j\geq t}$ and $\tilde V_i(x)$ [and with them $C_i(x,t)$] are measurable with respect to
$\mathcal{F}_t$ and leftcontinuous, hence predictable. The structure of the 'weights' $C_i$ also implies their boundedness. \\
Thus for $t<t_0$ $D_1(t)$ is a locally bounded right continuous martingale with
predictable variation given by
\begin{eqnarray}
\label{d2} \left\langle D_1,D_1 \right\rangle(t) &=& \int_{[0,t]} \su C_i^2(x,s)\Ind{1-F_{Z,n}(t-|x)>0}d\left\langle M_i,M_i \right\rangle(s)
\\ \nonumber
&=& \int_{[0,t]} \su C_i^2(x,s)\Ind{1-F_{Z,n}(t-|x)>0}\Lambda_D^-(ds|X_i).
\end{eqnarray}
Note that with $D_1$, $D_1(t)-D_1(\sigma)$ is also a locally bounded martingale for $t \in [\sigma,t_0]$ with predictable variation $\left\langle D_1,D_1 \right\rangle(t)- \left\langle D_1,D_1 \right\rangle(\sigma)$. Hence from a version Lenglart's inequality [see \cite{shorwell1986}, p. 893, Example 1] we obtain
\beq \quad \quad \quad
P\Bigl(
\sup_{\sigma \leq t \leq U\wedge t_0}nh^d (D_1(t) - D_1(\sigma))^2\geq\epsilon\Bigr) \leq \frac{\eta}{\epsilon} +
P\left(\mathcal{D}_n \geq\eta \right) ,
\label{Th1.liq}
\eeq
where $\mathcal{D}_n = nh^d\left(\left\langle D_1,D_1 \right\rangle(U\wedge t_0) - \left\langle D_1,D_1 \right\rangle(\sigma) \right)$. If $\sigma$ is sufficiently close to $t_0$ it follows
\bea
\mathcal{D}_n &=& nh^d\int_{[\sigma,U\wedge t_0]} \su C_i^2(x,t)\Lambda_D^-(dt|X_i)
\\
&=& nh^d\int_{[\sigma,U\wedge t_0]} \su \frac{\tilde V_i^2(x)\Ind{Z_i\geq t}}{\left(\sum_{j=1}^n\tilde V_{j}(x)\Ind{Z_j\geq
t}\right)^2}\Lambda_D^-(dt|X_i)
\\
&\leq& nh^d\sup_j \tilde V_j(x)\int_{[\sigma,U\wedge t_0]} \su \frac{C_i(x,t)}{\left(1-F_{Z,n}(t-|x) \right)}\frac{1}{\sum_{k=1}^n\tilde V_k(x)}\Lambda_D^-(dt|X_i)
\\
&\stackrel{(*)}{=}& O_P(1)\int_{[\sigma,U\wedge t_0]} \su \frac{C_i(x,t)}{\left(1-F_{Z,n}(t-|x) \right)}\lambda_D(t|x)dt(1+o_P(1))
\\
&=& O_P(1)\int_{[\sigma,U\wedge t_0]} \frac{\lambda_D(t|x)}{1-F_{Z,n}(t-|x)}dt(1+o_P(1))
\\
&=& O_P(1)\int_{[\sigma,U\wedge t_0]} \frac{\lambda_D(t|x)}{1-F_Z(t-|x)}\frac{1-F_Z(t-|x)}{1-F_{Z,n}(t-|x)}dt(1+o_P(1))
\\
&=& O_P(1)\int_{[\sigma,U\wedge t_0]} \frac{\lambda_D(t|x)}{1-F_Z(t-|x)}dt
\eea
where we have used \ref{6'}, \ref{W1a} and \ref{W1c} in equality $(*)$ [note that the $(1+o_P(1))$ holds uniformly in $i$ and $t$] and Lemma \ref{le:1} in the last equality. Now we obtain from \ref{1'} the a.s. convergence $\int_{[\sigma,U\wedge t_0]}\frac{\lambda_D(t|x)}{1-F_Z(t-|x)}dt \stackrel{\sigma\rightarrow t_0}{\longrightarrow} 0$ and hence assertion $(\ref{(C)})$ ist established [first choose $\eta$ in $(\ref{Th1.liq})$ small enough to make $\eta/\epsilon$ small and then choose $\sigma$ close enough to $t_0$].\\
Summarizing these considerations, we have established (\ref{(A)})-(\ref{(C)}) and the proof of the theorem is complete. \hfill$\Box$\\


\section{Auxiliary results: technical details}
\def\theequation{B.\arabic{equation}}
\setcounter{equation}{0}
\begin{lemma} \label{le:LenglartsIneq}
Let $M$ be a locally bounded, rightcontinuous martingale on $[0,\infty)$ and denote by $\left\langle M,M\right\rangle$ the predictable
variation of $M$. Then we have for any stopping time $U$ with $P(U<\infty)=1$ and all
$\eta,\epsilon>0$
\[
P\Bigl(\sup_{t\leq U}M^2(t)\geq\epsilon \Bigr) \leq \frac{\eta}{\epsilon} +
P\left( \left\langle M,M \right\rangle (U) \geq \eta \right)
\]
\end{lemma}
\textbf{Proof:} In fact this Lemma is a specific version of Lenglart's inequality [see \cite{flemharr1991}, Theorem 3.4.1].
To be precise note that it suffices to prove that for any a.s.\ finite stopping time $T$
\begin{equation} \label{eqA3}
\E[M^2(T)]\leq\E[\langle M,M\rangle(T)].
\end{equation}
Let $\tau_k$ denote a localizing sequence such that $M(.\wedge \tau_k)\leq k$ and $M^2(t\wedge\tau_k)-\langle M,M\rangle(t\wedge\tau_k)$ is a
martingale. Define the processes
\[
X_k(t) := M^2(t\wedge\tau_k), \quad Y_k(t) := \langle M,M\rangle(t\wedge\tau_k).
\]
Note that by Theorem 2.2.2 in \cite{flemharr1991} $(X_k-Y_k)(t\wedge T)$ is a martingale and hence for all $t$:
\begin{equation}
\E[X_k(t\wedge T)] = \E[Y_k(t\wedge T)]. \label{le3i1}
\end{equation}
Moreover, $k \geq X_k(t\wedge T) \stackrel{t\rightarrow\infty}{\longrightarrow} X_k(T) \ a.s.$, and hence we obtain by the Dominated Convergence
Theorem  $$\E[X_k(T)]=\lim_{t\rightarrow\infty}\E[X_k(t\wedge T)].$$ Since the process $\langle M,M\rangle$ is increasing, we also have
$$\langle M,M\rangle(t\wedge T)\uparrow \langle M,M\rangle(T)\ a.s.$$ and by the Monotone Convergence Theorem  $$\E[Y_k(T)] =
\lim_{t\rightarrow\infty} \E[Y_k(t\wedge T)].$$ Combining this and $(\ref{le3i1})$ we obtain the identity $\E[X_k(T)] = \E[Y_k(T)]$ for all a.s.
finite stopping times $T$. Hence we can apply Lenglart's inequality to the process $X_k$ dominated by $Y_k$ which leads to:
\[
P_{1,k} := P\left(\sup_{t\leq U}M^2(t\wedge\tau_k)\geq\epsilon\right) \leq\frac{\eta}{\epsilon} + P\left(\langle
M,M\rangle(U\wedge\tau_k\right) \geq \epsilon) =: \frac{\eta}{\epsilon} + P_{2,k}.
\]
Finally, from $\sup_{t\leq U}M^2(t\wedge\tau_k) = \sup_{t\leq U\wedge\tau_k}M^2(t) \uparrow \sup_{t\leq U}M^2(t)$ and $\langle
M,M\rangle(U\wedge\tau_k)\uparrow \langle M,M\rangle(U)$ a.s. as $k$ tends to infinity we obtain the desired result. \hfill$\Box$

\begin{lemma} \label{le:PFL:1}
Assume that conditions \ref{D000} and \ref{D8} hold. Denote by $W_1(x,n),...,W_k(x,n)$ those values of $Y_1,...,Y_n$, whose weights fulfill
$W_i(x)\neq 0$ and by $W_{(1)}(x,n),...,W_{(k)}(x,n)$ the corresponding increasingly ordered values. Assume that the estimators $F_{L,n}$ and
$H_n$ are based on weights $W_i(x) = V_i(x)/\sum_j V_j(x)$ with $V_i(x)$ satisfying the conditions \ref{W1a}-\ref{W1b}, that $F_{S,n}(r|x) :=
H_n(r|x)/F_{L,n}(r|x)$ is consistent for some $r > t_{00}$ with $F_S(r|x)<1$ and that all the observations $Y_i$ are distinct. Then we have for
any $b<r$:
\[
\sup_{b\geq s\geq W_{(2)}(x,n)} \frac{1}{F_{L,n}(s-|x)- H_n(s-|x)} = O_P(1).
\]

\end{lemma}

\textbf{Proof:} As in the proof of Theorem \ref{Th1} we reverse the time and use the same notation. Write $V_x := a(W_{(2)}(x,n))$, $v=a(r)$, $w=a(b)$,
then the statement of the Lemma can be reformulated as
\[
\sup_{w \leq s\leq V_x} \frac{1}{1-F_{D,n}(s|x)- (1-F_{Z,n}(s|x))} = O_P(1).
\]
With the notation $F_{B,n}(s|x) := 1- (1- F_{Z,n}(s|x))/(1-F_{D,n}(s|x))$ the denominator in this expression can be rewritten as
\[
\frac{1}{1-F_{D,n}(s|x)- (1-F_{Z,n}(s|x))} = \frac{1}{(1-F_{D,n}(s|x))F_{B,n}(s|x)}
\]
[note that $F_{B,n}(v|x) = 1 - F_{S,n}(r-|x)$].
Since $F_{B,n}(s|x)$ is increasing in $s$ and consistent at some point $v \leq w$ with $F_{B,n}(v|x)>0$, we only need to worry about finding a bound in probability for the term  $1/(1-F_{D,n}(s|x))$. Such a bound can be derived by exploiting the underlying martingale structure of the
estimator $\Lambda_{D,n}^-(t)$ of the hazard measure. More precisely, using exactly the same arguments as given in the proof of Theorem \ref{Th1} and  the same notation we obtain $\Lambda_{D,n}^-(t\wedge V_x|x) - \hat \Lambda_{D,n}^-(t\wedge V_x|x) = D_1(t \wedge V_x)$, where $D_1(t)$ is defined in (\ref{d1}) and is a locally bounded continuous martingale on $[0,\infty)$ with predictable variation given in (\ref{d2}).
The martingale property of $D_1(t)$ implies that $|D_1(t)|$ is a nonnegative submartingale and from Doob's submartigale inequality we obtain
for any $\beta>0$ and sufficiently  large $n$
\[
\Pro{\sup_{t\leq V_x} |D_1(t)| \geq \frac{1}{\beta}} \leq \beta \E{|D_1(V_x)|} \leq \beta \sqrt{\E|D_1(V_x)|^2} \leq \beta
\sqrt{\E{\left\langle D_1,D_1 \right\rangle(V_x)}} \leq \beta \sqrt{\sup_{y\in U_\epsilon(x)} \Lambda_D^-(V_x|y)},
\]
where we have used the inequality (\ref{eqA3}) from the proof of Lemma \ref{le:LenglartsIneq} and the fact that  the weights $C_i$ are positive and sum up to one.
Note that the expression $\sqrt{\sup_{y\in U_\epsilon(x)} \Lambda_D^-(V_x|y)}$ is finite. This follows from condition \ref{D8}, which now reads $\sup_{y\in U_\eps(x)} 1-F_D(\tilde \tau_T(y)|y)<1$ since we have reversed time, and the relation $\Lambda_D^-(t|x) = -\log(1-F_D(t|x))$. Thus we have obtained the estimate $\sup_{t\leq V_x} |D_1(t)|=O_P(1).$\\
From the definition of $\hat \Lambda_{D,n}^-(t|x)$ we can derive the bound $\sup_t \hat \Lambda_{D,n}^-(t|x) \leq \sup_{y\in U_\epsilon(x)} \Lambda_D^-(V_x|y)$, and thus obtain
\begin{equation}\label{B:PFL:1}
\sup_{t\leq V_x} \Lambda_{D,n}^-(t|x) \leq \sup_{t\leq V_x} |D_1(t)| + \sup_{t\leq V_x} \hat \Lambda_{D,n}^-(t|x) = O_P(1).
\end{equation}
Finally, we note that the estimator $F_{D,n}(s|x)$ can be expressed in terms of the statistic $\Lambda_{D,n}^-(t|x)$ by using the product limit map as $1 - F_{D,n}(t|x) = \prod_{[0,t]}\left(1 - \Lambda_{D,n}^-(ds|x)\right)$. By exactly the same arguments as given in the proof of Lemma 6 in \cite{gilljoha1990} we obtain the inequality
\[
1 - F_{D,n}(t|x) \geq \exp\left(-c(\eta)\Lambda_{D,n}^-(t|x)\right) \qquad a.s.
\]
whenever $0< t \leq V_x$, where $1 - 2\eta$ is the size of the largest atom of $\Lambda_{D,n}^-$ on the interval $(0,V_x]$ and $c(\eta):= -\log(\eta)/(1-\eta)<\infty$ [note that, whenever all observations take distinct values, the size of the largest atom of $\Lambda_{D,n}^-$ on $(0,V_x]$ is less or equal to the largest possible
value of $\sum_i W_i(x)\Ind{Z_i=V_x,\Delta_i=1}/\sum_i W_i(x)\Ind{Z_i \geq V_x}$ which can in turn be bounded by
$\overline{c}/(\overline{c}+\underline{c})<1$ uniformly in $n$ and thus $\eta>0$]. The desired bound for $1/(1-F_{D,n}(s|x))$ now follows from
the above inequality together with (\ref{B:PFL:1}) and thus the proof is complete. \hfill$\Box$
\begin{lemma} \label{le:1}
Let $(X_1,Y_1),...,(X_n,Y_n)$ denote i.i.d.\ random variables with $F(y|x) := P(Y_1\leq y|X_1=x)$.
Define $\hat F(y|x) := \sum_i \frac{V_i(x)\Ind{Y_i\leq y}}{{\sum_j V_j(x)}}$, which is an estimator of the conditional distribution function $F(y|x)$ and assume that the weights weights $V_i(x)$ satisfy conditions \ref{W1a}-\ref{W1c}, the bandwidth $h$ fulfills $nh^{d}\rightarrow\infty$, $h\rightarrow 0$ and that additionally the following conditions hold
\begin{enumerate}
\item $F(t|x)$ is continuous at $(t_{0},x_0)$ \label{Le4.1}
\item $1-F_Z(t|y) \geq C(1-F_Z(t|x))$ for all $(t,y)\in (t_0-\eps,t_0]\times I$ where $I$ is a set with the property $\int_{I \cap U_\delta(x)} f_X(s) ds \geq c\delta^d$ for some $c>0$ and all $0<\delta\leq\eps$. \label{Le4.2}
\item $F(t_0-\delta|z)$ is continuous in the second component at the point $z=x$ \label{Le4.3}
\item The distribution function $G$ of the random variables $X_i$ has a continuous density $g$ with $g(x)>0$. \label{Le4.4}
\end{enumerate}
Then, with the notation $U:=\max\{Y_i: V_j(x)\neq 0\}$, we have for $\nti$
\[
\sup_{0\leq y\leq t_0\wedge U}\frac{1 - F(y-|x)}{1 - \hat F_n(y-|x)} = O_P(1).
\]
\end{lemma}
\textbf{Proof:}
Define
\[
\bar F_{n}(y|x) := \frac{\sum_{i=1}^n F(y|X_i)\Ind{ \|x-X_i\| \leq h }}{\sum_{i=1}^n \Ind{\|x-X_i\|\leq h}},
\]
and observe the representation
\[
\frac{1 - F(y-|x)}{1 - \hat F_n(y-|x)} = \frac{1 - \bar F_n(y-|x)}{1 - \hat F_n(y-|x)}\frac{1 - F(y-|x)}{1 - \bar F_n(y-|x)}.
\]
We now will derive bounds for both ratios on the right hand side. For the first factor we note 
for sufficiently small $h$ for all $t\in (t_0-\delta,t_0]$
\[
X_i\in I\cap U_h(x) \Rightarrow 1-F(t-|X_i) > C(1-F(t-|x))
\]
This implies
\bea
\sup_{t\in(t_0-\delta,t_0]} \frac{1-F(t-|x)}{1-\bar F_{n}(t-|x)}
&=& \sup_{t\in(t_0-\delta,t_0]}\frac{1-F(t-|x)\sum_i\Ind{\|x-X_i\|\leq h}}{\sum_i\Ind{X_i\in I\cap U_h(x)}(1-F(t-|X_i))}\frac{\sum_i \Ind{X_i\in I\cap U_h(x)}(1-F(t-|X_i))}{\sum_i\Ind{\|x-X_i\|\leq h}(1-F(t-|X_i))}
\\
&\leq& \frac{1}{C}\frac{\sum_i\Ind{\|x-X_i\|\leq h}}{\sum_i \Ind{X_i\in I\cap U_h(x)}}. 
\eea
A standard application of the Chebychef inequality yields for an arbitrary set $M$
\[
P\Big( \frac{1}{n}\Big|\sum_i \Big( I_{\{X_i \in M\}} - P(X_i \in M)\Big)\Big| > \eps \Big) \leq \frac{1}{\eps^2}\frac{1}{nP(X_1 \in M)},
\]
and a direct application of this result in combination with assumptions \ref{Le4.2},\ref{Le4.4} yield 
\[
P\Big(\frac{\sum_i\Ind{\|x-X_i\|\leq h}}{\sum_i \Ind{X_i\in I\cap U_h(x)}} > \frac{1}{c} + \eps \Big) \rightarrow 0
\]
for every $\eps>0$, which implies
\[
\Pro{ \sup_{t\in[t_0-\delta,t_0]} \frac{1-F(t-|x)}{1-\bar F_{n}(t-|x)} > \frac{1}{Cc}+\epsilon} \longrightarrow 0 \quad \alle \epsilon>0.
\]
It now remains to consider the interval $[0,t_0-\delta]$. Observe that condition $\ref{Le4.3}.$ implies $1- F(t_0-\delta-|X_i) \geq 0.5(1-F(t_0-\delta-|x))$ if $|X_i-x|$ is sufficiently small, which yields
\[
\frac{1-F(t-|x)}{1-\bar F_{n}(t-|x)} \leq \frac{1-F(t-|x)}{1-\bar F_{n}(t_0-\delta-|x)} \leq 2\frac{1-F(t-|x)}{1-F(t_0-\delta-|x)} < \infty
\]
for sufficiently large $n$. Summarizing, we have obtained the estimate
\[
\sup_{0\leq y\leq t_0}\frac{1 - \bar F_n(y-|x)}{1 - F(y-|x)} = O_P(1).
\]
Thus it remains to consider the ratio $(1 - \bar F_n(y-|x))/(1 - \hat F_n(y-|x))$. For this purpose note that
\begin{eqnarray}
1 - \hat F(y-|x) &=& \sum_i \frac{V_i(x)(1-\Ind{Y_i < y})}{\sum_j V_j(x)} =
\frac{1+o_P(1)}{C(x)} \sum_i V_i(x)(1-\Ind{Y_i< y}) \label{b:l3.g1}
\\
&\geq& \underline{c}\frac{1+o_P(1)}{C(x)} \frac{1}{nh^d}\sum_i \Ind{V_i(x)\neq 0}(1-\Ind{Y_i< y}) \nonumber
\\
&\geq& \underline{c}\frac{1+o_P(1)}{C(x)}\frac{1}{nh^d} \sum_i \Ind{\|x-X_i\|\leq h}(1-\Ind{Y_i< y}) \nonumber
\\
&=& \underline{c}f_X(x)\frac{1+o_P(1)}{C(x)}\frac{\sum_i \Ind{\|x-X_i\|\leq h}(1-\Ind{Y_i\leq y})}{\sum_j \Ind{\|x-X_j\|< h}}, \nonumber
\end{eqnarray}
uniformly in $y$. In (\ref{b:l3.g1}) the last equality follows from $\frac{1}{nh^d}\sum_j \Ind{\|x-X_j\|\leq h} = f_X(x)(1+o_P(1))$, the second equality is a consequence of \ref{W1c} and the two inequalities follow from \ref{W1a} and \ref{W1b}, respectively. Note that the quantity $\sum_i \Ind{\|x-X_i\|\leq h}(1-\Ind{Y_i< y})/\sum_j \Ind{\|x-X_j\|\leq h}$ equals $1- \hat F^{NW}(y-|x)$ where $\hat F^{NW}$ is the Nadaraya-Watson estimator of $F$ with rectangular kernel. Thus it remains to find a bound
for $(1-\bar F_n(y-|x))/(1- \hat F^{NW}(y-|x))$. Conditionally on $X_1,...,X_n$, this is simply the ratio between $1- F_n$ and $1-\bar F$ where
$F_n$ is the empirical distribution function of the sample $\{Y_i: \|x-X_i\|\leq h\}$ with sample size $\sum_j \Ind{\|x-X_j\|\leq h}$ and $\bar F$
is the averaged distribution function of the corresponding $Y_i$. Since the random variables $Y_i$ are independent conditionally on $X_i$, we can apply the results
from \cite{vanzuijlen1978} to obtain the bound
\[
\mbox{P}\left(1 - \hat F^{NW}(t-|x) < \beta(1-\bar F_n(t-|x))\ \forall t\leq U \middle| X_1,...,X_n \right) \leq
\frac{2\pi^2}{3}\frac{\beta^2}{(1-\beta)^4}.
\]
Since the right hand side of the last inequality does not depend on any random quantities or their distributions, this result also holds
unconditionally, and thus the proof is complete. \hfill$\Box$

\end{appendix}

\bigskip
\small

\bibliographystyle{apalike}
\bibliography{condest}
\end{document}